\documentclass[%
 reprint,
groupedaddress,
 amsmath,amssymb,
 aps,
prc,
]{revtex4-1}

\usepackage{graphicx}% Include Fig. files
\usepackage{dcolumn}% Align table columns on decimal point
\usepackage{bm}% bold math
\usepackage[modulo]{lineno}
\usepackage{float}

\usepackage{booktabs}
\newcolumntype{C}[1]{>{\centering\arraybackslash}p{#1}}
\AtBeginDocument{
\heavyrulewidth=.08em
\lightrulewidth=.05em
\cmidrulewidth=.03em
\belowrulesep=.65ex
\belowbottomsep=0pt
\aboverulesep=.4ex
\abovetopsep=0pt
\cmidrulesep=\doublerulesep
\cmidrulekern=.5em
\defaultaddspace=.5em
}

\begin{document}

\preprint{APS/123-QED}

\title{Total absorption $\gamma$-ray spectroscopy of the $\beta$-delayed neutron emitters $^{137}$I and $^{95}$Rb}% Force line breaks with \\

\author{V. Guadilla}
\email{guadilla@ific.uv.es}
\altaffiliation[\newline Present address: ]{Subatech, IMT-Atlantique, Universit\'e de Nantes, CNRS-IN2P3, F-44307, Nantes, France}
\affiliation{%
 Instituto de F\'isica Corpuscular, CSIC-Universidad de Valencia, E-46071, Valencia, Spain
}
\author{J. L. Tain}%
\affiliation{%
 Instituto de F\'isica Corpuscular, CSIC-Universidad de Valencia, E-46071, Valencia, Spain
}
\author{A. Algora}%
\altaffiliation{%
 Also at Institute of Nuclear Research of the Hungarian Academy of Sciences, Debrecen H-4026, Hungary.
}
\affiliation{%
 Instituto de F\'isica Corpuscular, CSIC-Universidad de Valencia, E-46071, Valencia, Spain
}
\author{J. Agramunt}  
\affiliation{%
 Instituto de F\'isica Corpuscular, CSIC-Universidad de Valencia, E-46071, Valencia, Spain
}
\author{D. Jordan} 
\affiliation{%
 Instituto de F\'isica Corpuscular, CSIC-Universidad de Valencia, E-46071, Valencia, Spain
}  
\author{M. Monserrate}  
\author{A. Montaner-Piz\'a}  
\affiliation{%
 Instituto de F\'isica Corpuscular, CSIC-Universidad de Valencia, E-46071, Valencia, Spain
}
\author{E. N\'acher}  
\altaffiliation{%
Also at Instituto de Estructura de la Materia, CSIC, E-28006, Madrid, Spain
}
\affiliation{%
 Instituto de F\'isica Corpuscular, CSIC-Universidad de Valencia, E-46071, Valencia, Spain
}
\author{S. E. A. Orrigo}  
\affiliation{%
 Instituto de F\'isica Corpuscular, CSIC-Universidad de Valencia, E-46071, Valencia, Spain
}
\author{B. Rubio}  
\affiliation{%
 Instituto de F\'isica Corpuscular, CSIC-Universidad de Valencia, E-46071, Valencia, Spain
}
\author{E. Valencia}  
\affiliation{%
 Instituto de F\'isica Corpuscular, CSIC-Universidad de Valencia, E-46071, Valencia, Spain
}

\author{M. Estienne}  
\affiliation{%
 Subatech, IMT-Atlantique, Universit\'e de Nantes, CNRS-IN2P3, F-44307, Nantes, France
}
\author{M. Fallot}  
\affiliation{%
 Subatech, IMT-Atlantique, Universit\'e de Nantes, CNRS-IN2P3, F-44307, Nantes, France
}
\author{L. Le Meur}  
\affiliation{%
 Subatech, IMT-Atlantique, Universit\'e de Nantes, CNRS-IN2P3, F-44307, Nantes, France
}
\author{J. A. Briz}  
\affiliation{%
 Subatech, IMT-Atlantique, Universit\'e de Nantes, CNRS-IN2P3, F-44307, Nantes, France
}
\author{A. Cucoanes}  
\affiliation{%
 Subatech, IMT-Atlantique, Universit\'e de Nantes, CNRS-IN2P3, F-44307, Nantes, France
}
\author{A. Porta}  
\affiliation{%
 Subatech, IMT-Atlantique, Universit\'e de Nantes, CNRS-IN2P3, F-44307, Nantes, France
}
\author{T. Shiba}  
\affiliation{%
 Subatech, IMT-Atlantique, Universit\'e de Nantes, CNRS-IN2P3, F-44307, Nantes, France
}
\author{A. -A. Zakari-Issoufou} 
\affiliation{%
 Subatech, IMT-Atlantique, Universit\'e de Nantes, CNRS-IN2P3, F-44307, Nantes, France
}

\author{A. A. Sonzogni}  
\affiliation{%
NNDC, Brookhaven National Laboratory, Upton, NY 11973-5000, USA
}

\author{J. \"Ayst\"o}  
\affiliation{%
 University of Jyv\"askyl\"a, FIN-40014, Jyv\"askyl\"a, Finland
}
\author{T. Eronen}  
\affiliation{%
 University of Jyv\"askyl\"a, FIN-40014, Jyv\"askyl\"a, Finland
}
\author{L. M. Fraile}  
\affiliation{%
Grupo de F\'isica Nuclear and IPARCOS, Universidad Complutense de Madrid, CEI Moncloa, E-28040 Madrid, Spain
}
\author{E. Ganio\u{g}lu}  
\affiliation{%
Department of Physics, Istanbul University, 34134, Istanbul, Turkey
}
\author{W. Gelletly}  
\affiliation{%
Department of Physics, University of Surrey, GU2 7XH, Guildford, UK
} 
\author{D. Gorelov}  
\author{J. Hakala} 
\author{A. Jokinen}
\affiliation{%
 University of Jyv\"askyl\"a, FIN-40014, Jyv\"askyl\"a, Finland
}
\author{A. Kankainen}  
\altaffiliation{%
 Also at University of Edinburgh, Edinburgh EH9 3JZ, United Kingdom
}
\affiliation{%
 University of Jyv\"askyl\"a, FIN-40014, Jyv\"askyl\"a, Finland
}
\author{V. S. Kolhinen}  
\author{J. Koponen}  
\affiliation{%
 University of Jyv\"askyl\"a, FIN-40014, Jyv\"askyl\"a, Finland
}
\author{M. Lebois}  
\affiliation{%
Institut de Physique Nucl\`eaire d'Orsay, 91406, Orsay, France
}
\author{T. Martinez}  
\affiliation{%
Centro de Investigaciones Energ\'eticas Medioambientales y Tecnol\'ogicas, E-28040, Madrid, Spain
}
\author{I. D. Moore}  
\affiliation{%
 University of Jyv\"askyl\"a, FIN-40014, Jyv\"askyl\"a, Finland
}

\author{H. Penttil\"a}  
\author{I. Pohjalainen}  
\affiliation{%
 University of Jyv\"askyl\"a, FIN-40014, Jyv\"askyl\"a, Finland
}
\author{J. Reinikainen}  
\author{M. Reponen}  
\author{S. Rinta-Antila}  
\affiliation{%
 University of Jyv\"askyl\"a, FIN-40014, Jyv\"askyl\"a, Finland
}
\author{K. Rytk\"onen}  
\affiliation{%
 University of Jyv\"askyl\"a, FIN-40014, Jyv\"askyl\"a, Finland
}

\author{V. Sonnenschein}  
\affiliation{%
 University of Jyv\"askyl\"a, FIN-40014, Jyv\"askyl\"a, Finland
}

\author{V. Vedia}  
\affiliation{%
Grupo de F\'isica Nuclear and IPARCOS, Universidad Complutense de Madrid, CEI Moncloa, E-28040 Madrid, Spain
}
\author{A. Voss} 
\affiliation{%
 University of Jyv\"askyl\"a, FIN-40014, Jyv\"askyl\"a, Finland
}
\author{J. N. Wilson}
\affiliation{%
Institut de Physique Nucl\`eaire d'Orsay, 91406, Orsay, France
}

\date{\today}% It is always \today, today,

\begin{abstract}

The decays of the $\beta$-delayed neutron emitters $^{137}$I and $^{95}$Rb have been studied with the total absorption $\gamma$-ray spectroscopy technique. The purity of the beams provided by the JYFLTRAP Penning trap at the ion guide isotope separator on-line facility in Jyv\"askyl\"a allowed us to carry out a campaign of isotopically pure measurements with the decay total absorption $\gamma$-ray spectrometer, a segmented detector composed of eighteen NaI(Tl) modules. The contamination coming from the interaction of neutrons with the spectrometer has been carefully studied, and we have tested the use of time differences between prompt $\gamma$-rays and delayed neutron interactions to eliminate this source of contamination. Due to the sensitivity of our spectrometer, we have found a significant amount of $\beta$-intensity to states above the neutron separation energy that de-excite by $\gamma$-rays, comparable to the neutron emission probability. The competition between $\gamma$ de-excitation and neutron emission has been compared with Hauser-Feshbach calculations, and it can be understood as a nuclear structure effect. In addition, we have studied the impact of the $\beta$-intensity distributions determined in this work on reactor decay heat and reactor antineutrino spectrum summation calculations. The robustness of our results is demonstrated by a thorough study of uncertainties, and with the reproduction of the spectra of the individual modules and the module-multiplicity gated spectra. This work represents the state-of-the-art of our analysis methodology for segmented total absorption spectrometers.
\end{abstract}

\keywords{Suggested keywords}%Use showkeys class option if keyword
\maketitle

\section{Introduction}

Neutron-rich nuclei far from stability may exhibit $\beta$-decay energy windows $Q_{\beta}$ larger than the neutron separation energy $S_n$ in the daughter nucleus. In those cases with  $Q_{\beta}>S_n$, neutron emission competes strongly with $\gamma$-ray emission in the de-excitation of excited levels populated above $S_n$ in the $\beta$-decay. This decay mode, known as $\beta$-delayed neutron emission, was discovered in 1939 by Roberts et al.~\cite{bneutron_discovery} and becomes dominant when the neutron excess is sufficiently large. 

The $\beta$-delayed neutron emission process plays an important role in stellar nucleosynthesis. Heavy nuclei beyond iron can be produced by means of the rapid-neutron-capture process, the so-called r-process~\cite{r-process_ref}. The main characteristic of the r-process is the availability of a large number of neutrons that are added in a short time interval to elements of the iron group in successive neutron capture processes followed by $\beta$-decays. Very neutron-rich  $\beta$-delayed neutron emitters up to very heavy nuclei are formed in this way. Core collapse supernovae of massive stars or neutron star mergers have been considered as possible astrophysical sites for the r-process. Recently, the combined detection of gravitational waves and electromagnetic radiation from the GW170817 neutron star merger, gave support to the idea that such mergers are important sources of r-process elements~\cite{NSM}.

The detailed study of the r-process requires nuclear data such as nuclear masses, half-lives ($T_{1/2}$), $\beta$-delayed neutron emission probabilities ($P_n$) and neutron capture (n,$\gamma$) reaction cross sections~\cite{Mumpower16} for nuclei far away from stability. In spite of considerable experimental effort, a large amount of data is still lacking, and are thus obtained theoretically from nuclear models. 

In the case of (n,$\gamma$) cross sections, when no experimental information is available, statistical calculations using the Hauser-Feschbach formalism (HFF) \cite{Hauser-Feshbach} are used. These calculations rely on parameters obtained close to the valley of $\beta$ stability for Nuclear Level Densities (NLD), Photon Strength Functions (PSF) and Neutron Transmission Coefficients (NTC)~\cite{ReactionRates_StatisticalModel}. In recent years the connection between the $\beta$-delayed neutron emission process and (n,$\gamma$) reactions as a possible source of experimental information has been highlighted~\cite{vTAS_PRL, vTAS_PRC, Tain_proceedings_bneu1, Tain_proceedings_bneu2}. In both processes resonances that decay either by $\gamma$ or neutron emission are populated, though they usually have different spin-parity values. The experimental difficulty when taking advantage of this connection is related to the accurate measurement of the $\beta$-intensity followed by $\gamma$ emission above $S_n$. Traditional high-resolution experiments with HPGe detectors have been shown to be limited in detecting $\beta$-intensity at high excitation energies. This is due to the so-called Pandemonium systematic error~\cite{Pandemonium}, associated with the limited efficiency of such detectors. The Total Absorption $\gamma$-ray Spectroscopy (TAGS) technique allows one to overcome this effect and it has proven to be capable of extracting the $\beta$ intensity followed by $\gamma$-rays above $S_n$ in previous works~\cite{Alkhazov_BDN,vTAS_PRL,Sun_betan,vTAS_PRC}. This technique uses large scintillator crystals covering almost the full solid angle in order to maximize the $\gamma$-detection efficiency. The sum of the $\gamma$-rays de-exciting each level fed in the daughter nucleus is detected, instead of the individual $\gamma$-rays. The TAGS technique allows one to obtain the $\beta$-intensity distribution followed by $\gamma$-ray emission, $I_{\beta \gamma}$, by means of a deconvolution process.

On the other hand, in the case of $T_{1/2}$ and $P_n$ values, predictions from QRPA $\beta$-strength calculations~\cite{Moller-r-process, Borzov-r-process} have been compared in recent years with experimental results to test the accuracy of the nuclear models. A more stringent cross-check implies a comparison of calculated and measured $\beta$-strength distributions, since they are particularly sensitive to the details of the nuclear model. A key ingredient to determine the $\beta$-strength distributions are the $\beta$ intensity probabilities, which can be obtained free from the Pandemonium effect with TAGS, as mentioned above.

The $\beta$-decay of fission fragments plays a crucial role in nuclear reactors, where on average six $\beta$-decays follow each fission reaction. A precise knowledge of the energy released by their radioactive decay, the so-called Decay Heat (DH), turns out to be important in order to maintain the safe operation of a reactor after shutdown. Furthermore it can help to understand the occurrence of accidents, as shown in the case of the Fukushima-Daiichi plant \cite{Fukushima}, a consequence of the non-effective dissipation of the DH in the reactor core and in the adjacent cooling pool for spent-fuel. In addition, predictions of the DH associated with innovative fuels and reactors are needed. 

Apart from its importance for the safe operation of reactors, accurate information on the $\beta$-decay of the resulting fission fragments can be used to improve our understanding of the reactor antineutrino spectrum, important for reactor-based antineutrino experiments on fundamental neutrino physics~\cite{DoubleChooz,DayaBay,Reno} and for reactor monitoring~\cite{non_proliferation_2017}. The standard approach used in antineutrino spectrum calculations is based on the conversion of integral $\beta$ spectra measurements for the main fissile isotopes \cite{ILL_3,238U_beta}. However, the recent observation of discrepancies between experimental data and calculations of the absolute flux \cite{Anomaly} and shape  \cite{RENO_shoulder,DayaBay_shoulder,DoubleChooz_shoulder} of the reactor antineutrino spectrum, has encouraged further improvements in the alternative summation approach, which relies on the information from nuclear databases. In this approach, the total antineutrino spectrum is calculated as the sum of the antineutrino spectra associated with the decay of each fission product weighted by the corresponding activity. The antineutrino spectrum for each decay is constructed by using the $\beta$-intensity probabilities. An improvement in the summation method from the point of view of decay data consists of the provision of data free from the Pandemonium systematic error~\cite{neutrinos_PRL}. The same applies for the calculation of the reactor DH, where a summation over the inventory of fission products provides an alternative to integral measurements. The DH as a function of time is computed by summing the energy released by the decay of each nucleus (average $\gamma$ and $\beta$ energies for $\beta$-decaying nuclei) weighted by the activity at this time. For the calculation of the average energies, the $\beta$-intensity distributions are needed, so that the use of $\beta$ intensities suffering from Pandemonium limits the accuracy of such calculations~\cite{DecayHeat}.

Here we study the decays of two important $\beta$-delayed neutron emitters by means of the TAGS technique: the decay of $^{137}$I ($Q_{\beta}$=6.027~MeV) into $^{137}$Xe ($S_n$=4.025~MeV), and the decay of $^{95}$Rb ($Q_{\beta}$=9.228~MeV) into $^{95}$Sr ($S_n$=4.348~MeV). 
The decay properties of their respective $\beta$-n branches (neutron emission probability, neutron branching to states in the final nucleus and neutron spectra) are well known, hence one can make a detailed study of the neutron/$\gamma$ competition. $^{137}$I is identified as an important contributor to the reactor DH and to the reactor antineutrino spectrum. In fact, a high priority has been assigned to measurements of this decay with the TAGS technique by the International Atomic Energy Agency (IAEA)~\cite{IAEA2015}. Although $^{95}$Rb has a smaller contribution to the total decay energy released in a reactor, its large $Q_{\beta}$ value makes it a good candidate for investigating the Pandemonium systematic error.

The paper is organized as follows. In Section~\ref{Exp} we describe the experimental measurements of these decays and in Section~\ref{TAGS} a detailed discussion of the TAGS analyses will be presented. The competition between neutron emission and $\gamma$-ray emission will be addressed in Section~\ref{Competition}. In Section~\ref{betaspec} we compare the $\beta$ energy spectra obtained from the results of this work with previous measurements, as well as the new average $\beta$ and $\gamma$ energies with previous values.  Finally, in Section~\ref{reactor} the impact of these results in reactor DH and reactor antineutrino spectrum summation calculations will be discussed.

\section{Experiment}\label{Exp}

\begin{figure}[!hbt]
\begin{center} 
\includegraphics[width=0.5 \textwidth]{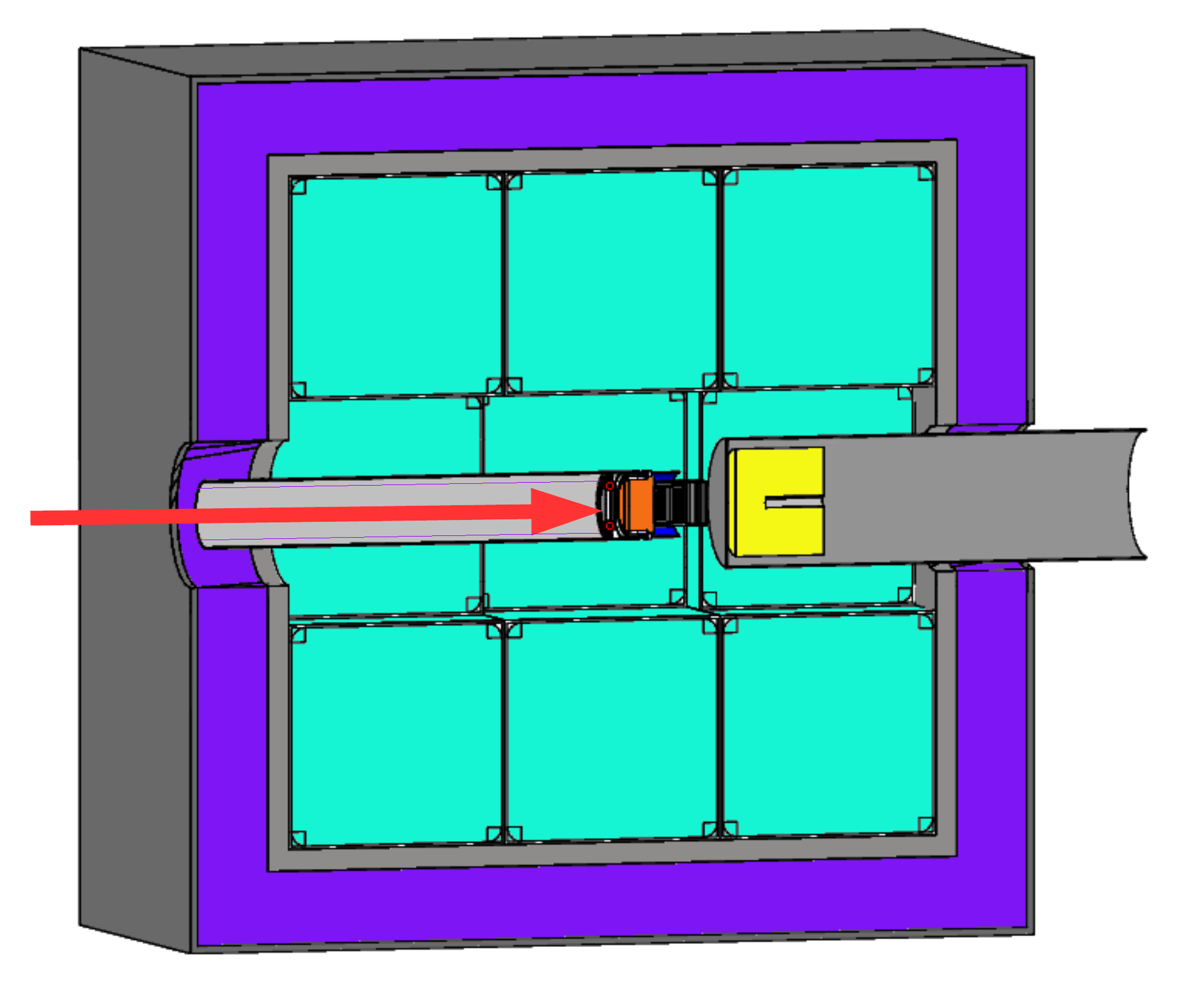} 
\caption{Partial view of a longitudinal cut in a drawing of the experimental set-up. The following elements are depicted: the NaI(Tl) crystals of DTAS (in green) surrounded by the lead shielding (in violet), the beam pipe to the left (in gray), the plastic detector with its light guide close to the center (in orange), and the HPGe detector to the right (in yellow). The red arrow represents the beam direction and indicates the implantation point.}
\label{setup}
\end{center}
\end{figure}

A campaign of measurements, including $^{137}$I and $^{95}$Rb decays, was carried out in 2014 at the upgraded Ion Guide Isotope Separator On-Line (IGISOL) facility at the University of Jyv\"askyl\"a~\cite{Moore_IGISOLIV}. We employed the new Decay Total Absorption $\gamma$-ray Spectrometer (DTAS), composed of 18 NaI(Tl) crystals~\cite{DTAS_design}. In the set-up a plastic $\beta$ detector of 3~mm thickness was located close to the center of DTAS, and a HPGe detector was placed behind the $\beta$ detector. A schematic picture of the set-up can be seen in Fig.~\ref{setup}. The fission ion guide was used to extract the nuclei produced by 25~MeV proton-induced fission on natural uranium. The IGISOL separator magnet was employed to separate the radioactive nuclei based on their mass to charge ratio before using the double Penning trap system JYFLTRAP~\cite{JYFLTRAP} for isobaric separation. The ions extracted from the trap were implanted on a tape placed in front of the plastic $\beta$ detector. A tape transport system was employed to remove the activity from DTAS during the measurements. The collection cycles of the tape transport system were selected to allow the reduction of the descendant activity in the measurements. For the decay of $^{137}$I ($T_{1/2}$=24.5~s) the collection cycle was $\sim$4$\times T_{1/2}$, while for $^{95}$Rb ($T_{1/2}$=377.7~ms) it was $\sim$7$\times T_{1/2}$. 

\subsection{Experimental spectra}

A coincidence between DTAS and the $\beta$ detector was required to get a spectrum free from environmental background. The total sum energy of DTAS was reconstructed off-line from the signals in the individual modules as described in~\cite{NIMA_DTAS}, with threshold values of $\sim$90 keV for DTAS modules and $\sim$70 keV for the $\beta$ detector. Standard calibration sources were used to obtain the energy and resolution calibration of DTAS ($^{22,24}$Na, $^{60}$Co, $^{137}$Cs and $^{152}$Eu-$^{133}$Ba)~\cite{NIMA_DTAS}. There are three main sources of contamination in our TAGS spectra that need to be corrected: 1) the summing-pileup distortion, 2) the activity of the descendants and 3) the contribution of the $\beta$-delayed neutrons interacting with the detector. The first two are discussed here, while the third one is discussed in subsection \ref{bnBkg}. 

The summing-pileup distortion was calculated as in previous works \cite{Zak_PRL,vTAS_PRC, Simon_PRC, 100Tc}, with a Monte Carlo (MC) procedure based on the random superposition of two stored events within the ADC gate length~\cite{TAS_pileup,NIMA_DTAS}. 

The decay of $^{137}$Xe ($T_{1/2}$=3.818~m) was measured and its contribution to the measurement of $^{137}$I was calculated using the $\gamma$ transition at 455.5~keV from the decay of $^{137}$Xe as normalization. In the case of $^{95}$Rb, both the daughter ($^{95}$Sr, with $T_{1/2}$=23.90~s) and the granddaughter ($^{95}$Y, with $T_{1/2}$=10.3~m) contaminate the measurement. Both decays were measured, and their contribution was estimated with the help of the Bateman equations. Contaminant fractions of 3.13$\%$ and 0.38$\%$ were calculated for daughter and granddaughter respectively. The contamination of $^{95}$Y in the measurement of $^{95}$Sr (3.96$\%$) was also taken into account in the same way.

\subsection{$\beta$-delayed neutron background}\label{bnBkg}

\begin{figure}[!hbt]
\begin{center}
\includegraphics[width=0.5 \textwidth]{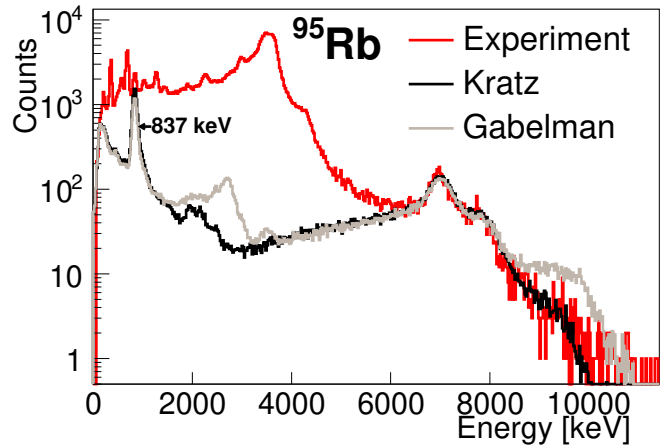}
\caption{Comparison of the effect in the simulation of the $\beta$-delayed neutron branch in the decay of $^{95}$Rb of two different $I_n$ distributions. See text for details. The MC spectra are normalized to the experimental $\beta$-gated DTAS spectrum around 6.8~MeV. The 837~keV $\gamma$-ray peak from the first excited state in $^{94}$Sr is highlighted.}
\label{95Rb_In_94Sr}
\end{center}
\end{figure}

$\beta$-delayed neutrons interact with the NaI(Tl) material of the detector producing $\gamma$-rays, either in an inelastic reaction or after neutron capture, that are easily detected in DTAS. The most clear evidence of these interactions is a structure in the spectra above 6.8~MeV, mainly due to neutron capture in the $^{127}$I of the crystals, that can be seen in Fig. \ref{95Rb_In_94Sr}. On the other hand, $\gamma$-rays from inelastic scattering, less evident, concentrate at low energy. This contamination was treated in two different ways. In the first method it is calculated using dedicated MC simulations. In the second method we exploit the fact that $\gamma$-rays from neutron interactions are delayed, due to the low velocity of the neutrons, with respect to prompt $\gamma$-rays emitted after the decay. 

The simulation of the contamination due to the $\beta$-n branch was done using the Geant4 simulation code \cite{GEANT4} and the DECAYGEN event generator, as described in~\cite{NIMA_DTAS}. The generator uses the $\beta$-intensity distribution followed by neutron emission, $I_{\beta n}$, that was reconstructed from the measured neutron spectra using the information on neutron branching ratios to the excited levels in the final nucleus, $I_n$. The neutron spectra are obtained from ENDF/B-VII.0, based on an evaluation of experimental data \cite{Brady_thesis}. In the case of $^{95}$Rb the experimental information is completed at high energies with QRPA and Hauser-Feshbach theoretical calculations~\cite{Kawano_neutron_spec}. For $^{137}$I the neutron spectrum directly provides $I_{\beta n}$ since neutron emission proceeds to the ground state (g.s.) only. For $^{95}$Rb there are several measurements of the neutron branching $I_n$ to the different levels in $^{94}$Sr~\cite{Kratz_bn94Sr,Hoff_bn94Sr,Thesis_Gabelmann}. The values quoted in ENSDF come from H. Gabelmann \cite{Thesis_Gabelmann}. However, we observe that our experimental spectrum is compatible with the simulation performed using the intensities coming from the experiment of K.-L. Kratz~\textit{et al.} \cite{Kratz_bn94Sr} but not with the one using $I_n$ from ENSDF, as shown in Fig.~\ref{95Rb_In_94Sr}. 

As discussed in \cite{NIMA_DTAS}, a 500~ns time window for accumulation of the energy deposited in DTAS was used for both the experimental spectra and in the MC simulations. This time is enough to allow the full energy deposition of $\beta$ delayed neutron-induced $\gamma$-rays. The simulated $\beta$-n branch has been normalized to match the low-energy tail of the experimental neutron capture peak (see Fig.~\ref{95Rb_In_94Sr}). In $^{95}$Rb this normalization matches at the same time the peak corresponding to the 837~keV $\gamma$-ray, emitted from the first excited state in $^{94}$Sr, which is populated in the $\beta$-n decay (see Fig.~\ref{95Rb_In_94Sr}).

\begin{figure}[!hbt]
\begin{center}
\includegraphics[width=0.5 \textwidth]{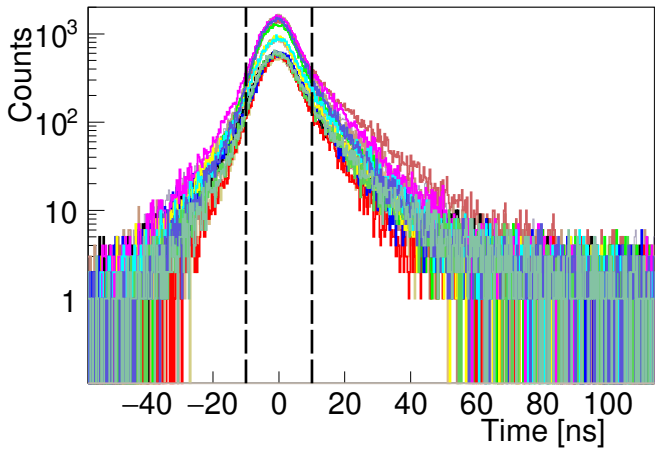}
\caption{Individual time correlation spectra between each module and the plastic detector for the decay of $^{137}$I. The prompt gate of 20~ns ($\pm 10$~ns)  is indicated with vertical dashed lines.}
\label{137I_neutron_discrimination}
\end{center}
\end{figure}

\begin{figure}[!hbt]
\begin{center} 
\includegraphics[width=0.5 \textwidth]{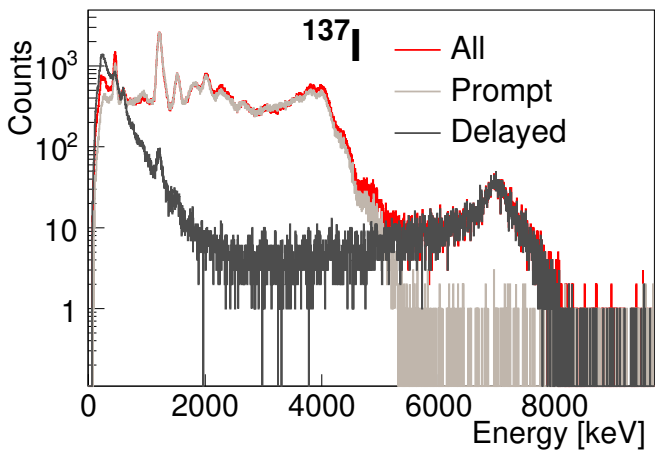} 
\caption{Effect of different time correlation windows $\Delta t = t_{DTAS} - t_{plastic}$ on the $\beta$-gated DTAS spectrum for the decay of $^{137}$I. Prompt gate ($\Delta t \leq  20$~ns) in light grey, delayed gate ($\Delta t > 10$~ns) in dark grey. The spectrum in red corresponds to a gate of $\Delta t = 500$~ns that covers both the prompt and the delayed signals.}
\label{neutron_gates}
\end{center}
\end{figure}

As discussed in \cite{DTAS_design}, one can use the timing information between $\gamma$-rays detected in DTAS and $\beta$-particles detected in the plastic scintillation detector in order to distinguish whether $\gamma$-rays are coming from neutron interactions or from the $\beta$ decay. Time correlation spectra $\Delta t = t_{DTAS} - t_{plastic}$ were reconstructed for the individual modules as shown in Fig.~\ref{137I_neutron_discrimination}. For convenience the peak positions of all spectra are aligned to zero applying an offset. We found that a time gate of 20~ns length (i.e. $\pm 10$~ns with respect to zero) is adequate to separate prompt contributions and delayed ones (those with $\vert \Delta_{t}\vert > 10$~ns). As shown in Fig. \ref{neutron_gates} the neutron capture peak disappears when we use the prompt gate. Unfortunately this gate impairs the reconstruction of the low energy part of the spectra (compare the light grey and the black spectra in Fig. \ref{neutron_gates}). This is related to the relatively poor individual timing resolution of around 20~ns, not properly optimized in the present measurements and much worse than the 5~ns reported in \cite{DTAS_design} for the DTAS prototype module. As a consequence, the effective energy threshold is increased, affecting the sum energy reconstruction. In fact, in the measurement of $^{95}$Rb the intense $\gamma$-ray of 204~keV energy is cut  with this procedure (as can be seen in Fig.~\ref{95Rb_fit}), and we estimate that the effective threshold is about 280~keV instead of 90~keV. As we will show later, this has an impact on the determination of the $\beta$ intensity distribution. In the future, a proper optimization of the individual timing resolution and the use of narrower gates, could make this a better method than the MC simulation method for the study of isotopes with very large neutron emission probabilities or unknown $\beta$-n decay properties.
 
\section{TAGS analyses}\label{TAGS}

In the analysis we follow the method developed by the Valencia group to determine the $\beta$-intensity distributions in TAGS experiments~\cite{TAS_MC,TAS_algorithms,TAS_decaygen}. For that, we have to solve the following inverse problem~\cite{TAS_algorithms}:

\begin{center}
\begin{equation}\label{inverse}
d_i=\sum\limits_{j}R_{ij}(B)f_j+C_i
\end{equation}
\end{center}

\noindent where $d_i$ represents the number of counts in channel $i$ of the experimental spectrum, $f_j$ is the number of events that feed level $j$ in the daughter nucleus, $C_i$ is the contribution of all contaminants to channel $i$, and $R_{ij}$ is the response function of the spectrometer, that depends on the branching ratios ($B$) between the states in the daughter nucleus. The branching ratio matrix is calculated using the known decay information for the levels at low excitation energy complemented with an estimate based on the nuclear statistical model at high excitation energy. 

\begin{figure}[!hbt]
\begin{center} 
\includegraphics[width=0.5 \textwidth]{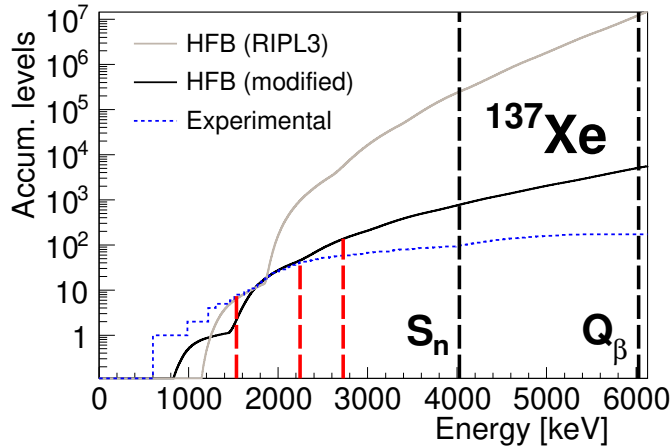} 
\caption{Accumulated number of levels as a function of excitation energy for $^{137}$Xe. The dotted-blue line is the experimental information obtained from ENSDF. The gray line corresponds to the HFB level density as obtained from RIPL-3. The black line comes from the HFB density modified to reproduce the experimental information both at low and high energies. The three vertical dotted red lines represent the three limits considered for the known part of the level scheme. See text for details.}
\label{137Xe_lev}
\end{center}
\end{figure}

According to the Reference Input Parameter Library (RIPL-3) \cite{RIPL-3}, the level scheme of $^{137}$Xe is complete up to a level at 2726.140~keV, whereas for $^{95}$Sr it is only considered complete up to a level at 1259.7~keV. These limits define the known parts of the branching ratio matrices for the two cases studied. For $^{137}$Xe we considered, in addition, two alternative known parts of the level scheme: up to the level at 2244.1~keV, where there is good agreement between  the experimental number of levels and the prediction of the statistical model (see Fig.~\ref{137Xe_lev}), and up to the level at 1534.32~keV, where there is a substantial gap to the next level at 1621.1~keV. %A recent TAGS study of $^{137}$I \cite{MTAS_137I} considered a known level scheme up to 4100~keV, but we did not find any justification for that. 

In all cases, from the last known level included in the known level scheme up to the $Q_{\beta}$ value (maximum decay energy window), a continuum region with 40~keV bins is defined. The branching ratios in this continuum region are determined with the statistical model, as presented in \cite{TAS_decaygen}. All parameters used for the statistical model calculations are extracted from RIPL-3 \cite{RIPL-3} and summarized in Table \ref{parameters}, with PSF and deformation parameters based on \cite{PSF} and \cite{DeformationPar_exp}, respectively. The level density parameter ``a'' at the neutron binding energy used to calculate the $E1$ $\gamma$-strength function is obtained from Enhanced Generalized Superfluid Model (EGSM) calculations for $^{95}$Sr~\cite{RIPL-3}, while for $^{137}$Xe it is taken from B. Fogelberg \textit{et al.} \cite{137I_Fog_fermi_gas}. The Hartree-Fock-Bogoliubov (HFB) plus combinatorial nuclear level densities \cite{Gorieli1,Gorieli2} have been used. For the level density of $^{95}$Sr the C and P correction parameters from RIPL-3 were used (0.0 and 0.78795, respectively). However, since the level density correction factors for $^{137}$Xe in RIPL-3 (C=2.96189 and P=1.09479) did not reproduce the available experimental information, we have calculated new corrections (C=-1.02 and P=0.69). In particular, the experimental number of resonances in $^{137}$Xe in the region 4.03-4.53~MeV, just above $S_n$, according to the experimental work of B. Fogelberg et al. \cite{137I_Fog} is $\leq$4 $1/2^+$ levels, (24 $\pm$ 8) $1/2^-$ levels and (16 $\pm$ 5) $3/2^-$ levels. With the original correction factors from RIPL-3 one obtains unrealistically large values: 618 $1/2^+$ levels, 12630 $1/2^-$ levels, and 24722 $3/2^-$ levels. The new correction factors have been calculated to obtain a more reasonable number of levels in the resonance region: 2.6 $1/2^+$ levels, 12 $1/2^-$ levels, and 24 $3/2^-$ levels. In addition, both sets of correction factors reproduce the accumulated number of levels at 1808.75~keV. A comparison of the original level density and the modified one for $^{137}$Xe is shown in Fig.~\ref{137Xe_lev}. We should mention that the analysis of the DTAS spectra using the original level density did not allow a good reproduction of the experimental spectrum (see Subsection \ref{137I_TAGS}).

\begin{table*}[!h]
\begin{ruledtabular}
\caption{\label{parameters} Parameters used in the statistical model calculation of the branching ratio matrix (B) of the daughter nuclei.}
  \begin{tabular}{@{}C{1.cm}C{2.5cm}C{2cm}C{1cm}C{1cm}C{1cm}C{1cm}C{1cm}C{1cm}C{1cm}C{1cm}C{1cm}@{}}
    Nucleus
    & Level-density parameter
    & Deformation parameter
    & \multicolumn{9}{c}{Photon strength function parameters}
    \\ \cmidrule(lr){4-12}
    & &  & \multicolumn{3}{c}{E1} & \multicolumn{3}{c}{M1} & \multicolumn{3}{c}{E2}   \\
    \cmidrule(r){2-2}\cmidrule(lr){3-3}\cmidrule(lr){4-6}\cmidrule(lr){7-9}\cmidrule(l){10-12}
     & a($S_n$) & $\beta$  & E & $\Gamma$ & $\sigma$ & E & $\Gamma$ & $\sigma$ & E & $\Gamma$ & $\sigma$  \\
    & $[$MeV$^{-1}$] &  & [MeV] & [MeV] & [mb] & [MeV] & [MeV] & [mb] & [MeV] & [MeV] & [mb]  \\
    \\ \cmidrule(lr){1-12}
    $^{137}$Xe & 12.3 &  -0.018
    & \begin{tabular}{@{}c@{}}15.279\\ 15.043\end{tabular} 
    & \begin{tabular}{@{}c@{}}4.621\\ 4.487\end{tabular} 
    & \begin{tabular}{@{}c@{}}94.649\\ 194.973\end{tabular} 
    & 7.966 & 4.000 & 0.329 & 12.241 & 4.466 & 2.851  
    \\ \cmidrule(lr){1-12}
    $^{95}$Sr & 15.802 &  0.31
    & \begin{tabular}{@{}c@{}}14.069\\ 18.236\end{tabular} 
    & \begin{tabular}{@{}c@{}}3.951\\ 6.467\end{tabular} 
    & \begin{tabular}{@{}c@{}}77.156\\ 94.264\end{tabular} 
    & 8.999 & 4.000 & 0.476 & 13.828 & 4.970 & 1.829 \\
  \end{tabular}
\end{ruledtabular}
\end{table*}

Once the branching-ratio matrix is constructed, the response function $R_{ij}(B)$ is calculated by means of MC simulations \cite{TAS_MC}. The detailed description of the geometry of the set-up and the nonproportionality of the light yield in NaI(Tl), as described in \cite{TAS_MC}, are included in the simulations. The MC simulations were validated by comparison with measurements of
well-known radioactive sources \cite{NIMA_DTAS}. The TAGS analysis is then carried out by applying the expectation maximization (EM) algorithm to extract the $\beta$-feeding distribution \cite{TAS_algorithms}.

The branching ratio matrix constructed combining the information from the statistical model and the known level scheme, provides a realistic guess of the true branching ratio matrix. Differences in model branching ratio matrices can appear not only because of the use of different nuclear statistical model parameters, but also because of ambiguities in the spin-parity values of levels in the known part of the level scheme. The impact of different choices of parameters can be used to estimate systematic uncertainties in the resulting $\beta$-intensity distribution, as will be shown later. In fact, some choices can be ruled out because they do not lead to a good reproduction of the total absorption spectrum. In this respect, a significant advance in the TAGS technique is the introduction of segmented spectrometers like DTAS. The model branching ratio matrix can now be subjected to more restrictive tests, using the reproduction of the individual-module spectra and the module-multiplicity gated total absorption spectra as additional analysis criteria (where the module-multiplicity of an event, $M_m$, is defined as the number of modules that fire above the threshold). All these tests improve significantly the reliability of the results, and they can provide a guide for empirical modification of the branching ratio matrix.

\subsection{Decay of $^{137}$I}\label{137I_TAGS}

\begin{figure}[!hbt]
\begin{center} 
\includegraphics[width=0.5 \textwidth]{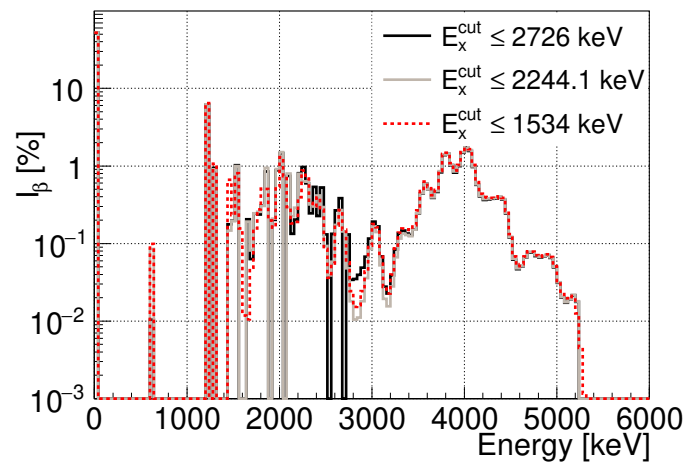} 
\caption{Comparison of the $\beta$-intensities obtained for the decay of $^{137}$I using three different excitation energy limits in the $^{137}$Xe known level scheme considered ($E_x^{\text{cut}}$).}
\label{CUTS}
\end{center}
\end{figure}

The spin-parity value of the ground state (g.s.) of $^{137}$I, according to ENSDF, is ($7/2^+$), based on systematics \cite{NDS_A137}. In the analysis we used the value $7/2^+$ as the primary choice and we considered decays by allowed transitions and first forbidden transitions to states in the known part of the level scheme, while only allowed transitions were considered to states in the continuum. Alternative g.s. spin-parity values of $7/2^-$ and $5/2^+$ were also used to construct alternative branching ratio matrices that also gave a reasonable reproduction of the total absorption spectrum. The associated $\beta$-intensities were considered in the evaluation of uncertainties. The three different choices for the known part of the level scheme in $^{137}$Xe mentioned above were also investigated. They were found to be equivalent, although the best reproduction of the total absorption spectrum was obtained with the known level scheme extending up to 2726.140~keV. Moreover, this choice was shown to reproduce better the module-multiplicity gated spectra, as will be discussed later. A comparison of the $\beta$-intensities obtained with the three level schemes is shown in Fig. \ref{CUTS}. 

\begin{figure}[H]
\begin{center}
\includegraphics[width=0.5 \textwidth]{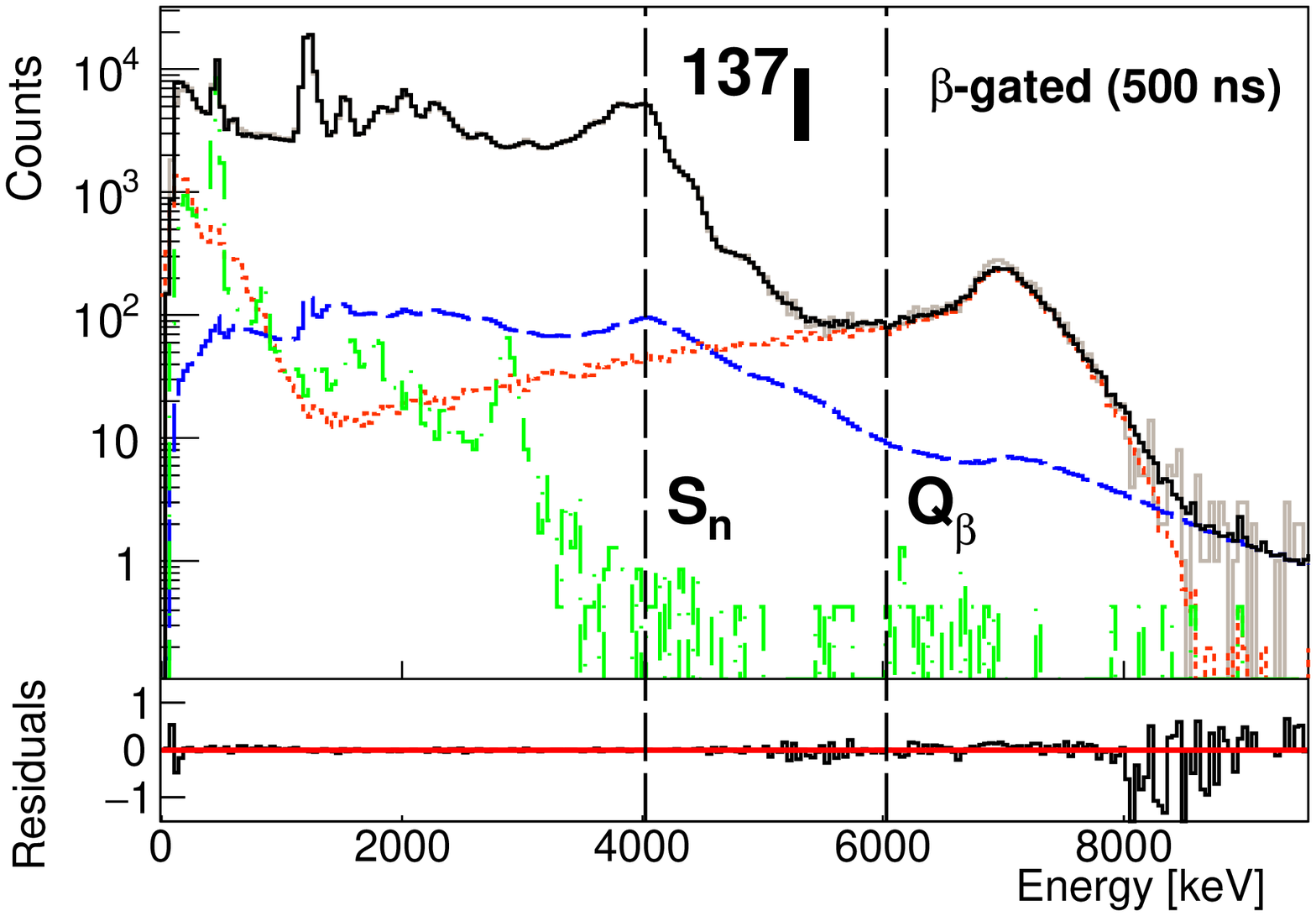} \\
\includegraphics[width=0.5 \textwidth]{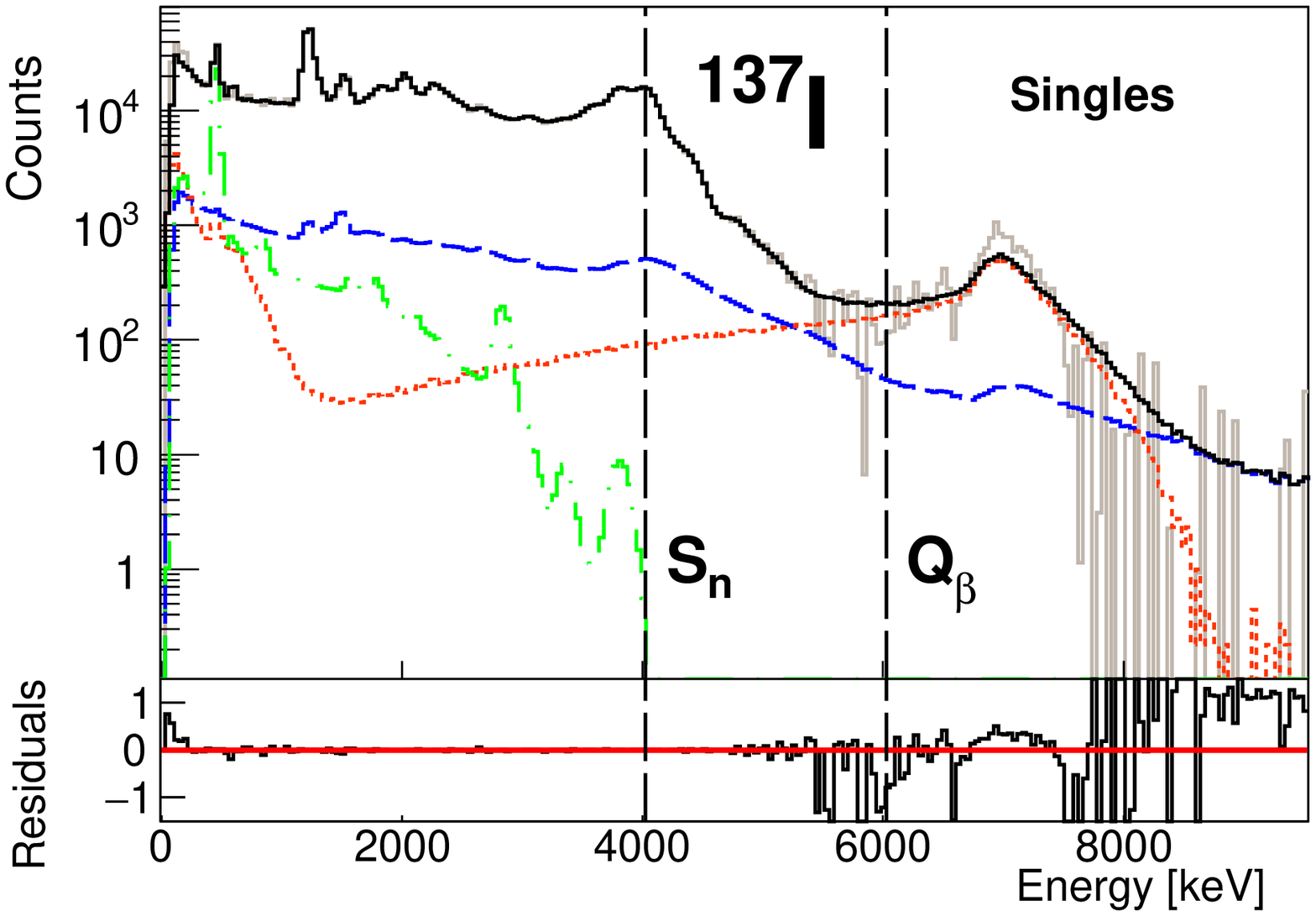} \\
\includegraphics[width=0.5 \textwidth]{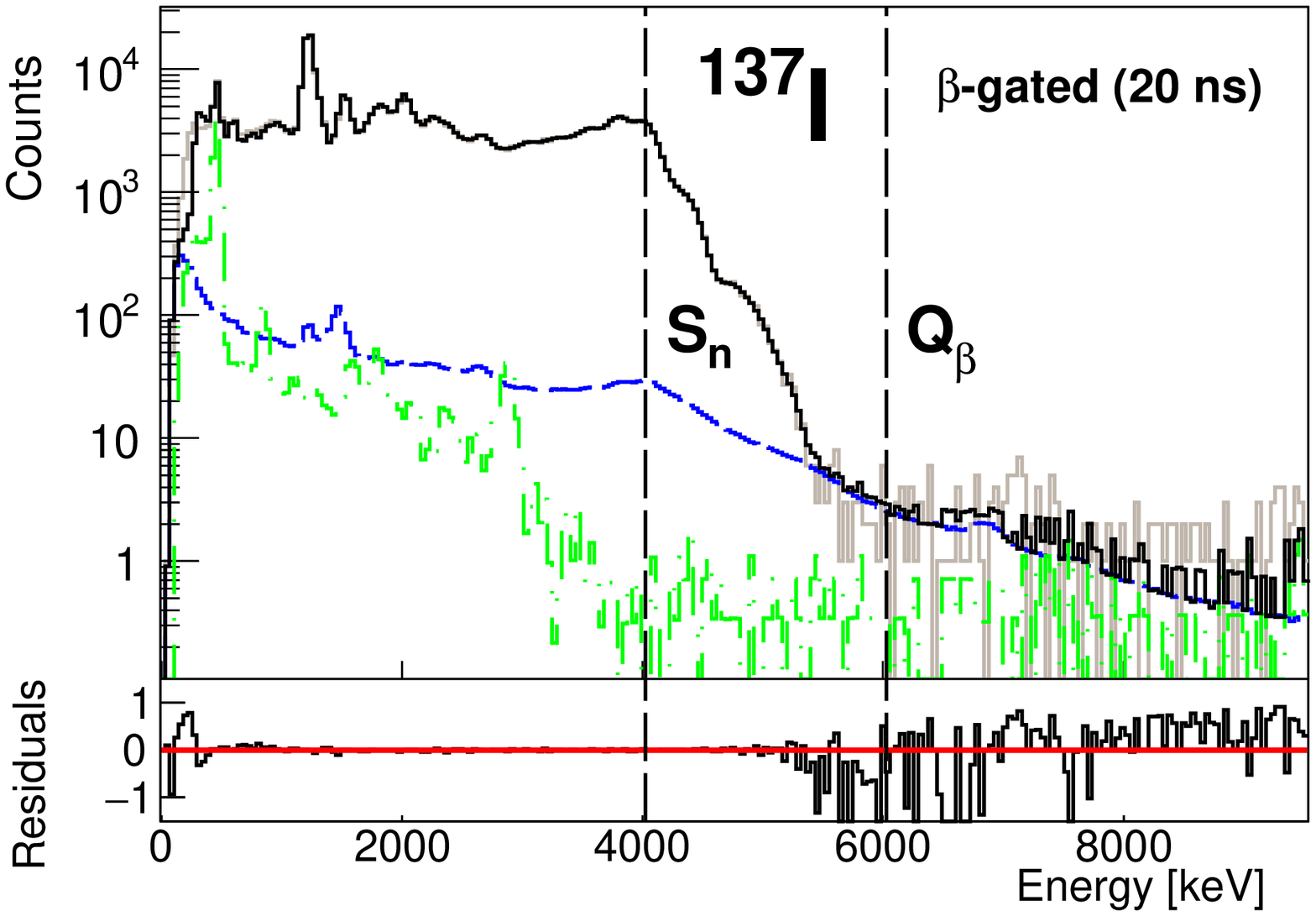}
\caption{Relevant histograms for the analysis of the decay of $^{137}$I: experimental total absorption spectrum (solid grey), summing-pileup contribution (dashed blue), daughter spectrum (dashed-dotted green), $\beta$-n decay branch (dotted orange) and reconstructed spectrum (solid black). The analyses of three different experimental spectra are shown: $\beta$-gated with a 500~ns gate (top), singles background subtracted (middle) and $\beta$-gated with a gate of 20~ns to cut neutron-induced $\gamma$-rays. See text for further details. The relative deviations between experimental and reconstructed spectra are shown for each case.} 
\label{137I_fit}
\end{center}
\end{figure}

\begin{figure}[!hbt]
\begin{center} 
\includegraphics[width=0.5 \textwidth]{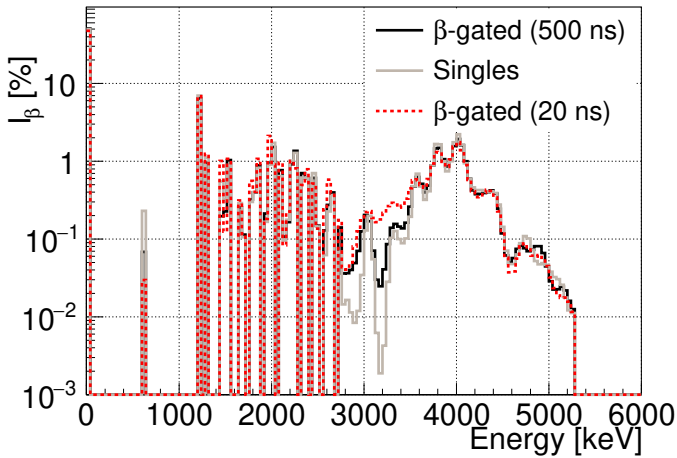} 
\caption{Comparison of the $\beta$-intensities obtained for the decay of $^{137}$I from the TAGS analyses of three experimental spectra with different $\beta$-gating conditions.}
\label{137I_comparisonI}
\end{center}
\end{figure}

In Fig.~\ref{137I_fit} we show the quality of the analysis by comparing experimental spectra with the spectra reconstructed using Eq.~\eqref{inverse} for different $\beta$-gating conditions. In the top panel of the figure we show the analysis of the $\beta$-gated spectrum with the 500~ns coincidence gate. In the central panel we show the analysis of the background subtracted singles spectrum (no $\beta$-gated). In the bottom panel we show the analysis of the $\beta$-gated spectrum with the 20~ns coincidence gate that eliminates neutron-induced $\gamma$-rays. The relative deviations between experimental and reconstructed spectra are shown in each case. They are small up to 5.2~MeV, except for the 20~ns gated spectrum in the first few hundred keV. This reflects the difficulty of reproducing the effective module threshold in the MC simulation. All three analyses were performed with the reference branching ratio matrix. It should be noted that the result of the analysis of the singles spectrum does not depend on the simulated $\beta$-efficiency of the $\beta$-detector, strongly varying close to $Q_{\beta}$, but is very sensitive to the proper background subtraction. On the other hand, the result of the analysis of the $\beta$-gated spectrum with the neutron background suppressed does not depend on the MC simulation of the $\beta$-n branch of the decay, but suffers from the higher threshold at low energies. A comparison of the $\beta$-intensity distributions obtained from these three analyses is presented in Fig. \ref{137I_comparisonI}. As can be seen the agreement is good except in the continuum region up to 3.5~MeV and for the weakly populated state at 601~keV. 

We should mention that in the analysis of the singles spectrum the contribution of the daughter decay (dashed-dotted green line in the central panel of Fig. \ref{137I_fit}) was obtained from MC simulations. For that, the information on this decay available in ENSDF~\cite{NDS_A137} was used as input for the DECAYGEN event generator \cite{TAS_decaygen}. This information is reliable according to our TAGS analysis of the $\beta$-gated $^{137}$Xe spectrum, and is in agreement with the recent TAGS result of \cite{MTAS_137I}. This procedure avoids the impact of the large statistical fluctuations of the experimental $^{137}$Xe singles spectrum after background subtraction. We would also like to point out that in the analysis of the $\beta$-gated spectrum with the 20~ns time window, the same time window was applied to obtain the spectrum of the daughter decay contamination and the summing-pileup contribution.

\begin{figure}[!hbt]
\begin{center} 
\includegraphics[width=0.5 \textwidth]{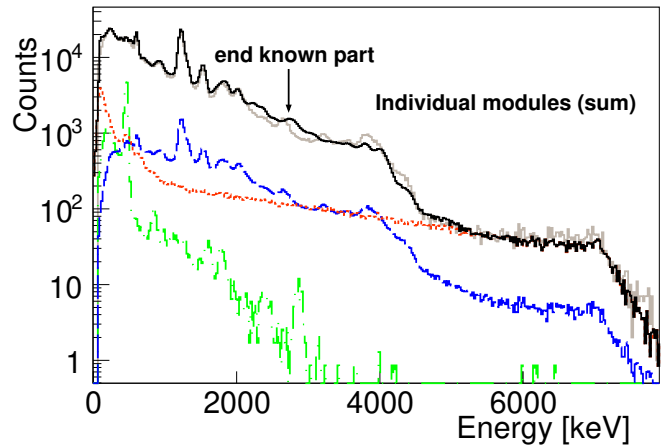} 
\caption{Comparison of the 18 individual experimental spectra summed (solid gray) with the reconstructed spectrum after the analysis (solid black) after taking into account the pileup contribution (dashed blue), the daughter contamination (dashed-dotted green) and the $\beta$-n branch (dotted orange).}
\label{137I_individuals}
\end{center}
\end{figure}

\begin{figure*}[!hbt]
\begin{center} 
\begin{tabular}{cc}
\includegraphics[width=0.5 \textwidth]{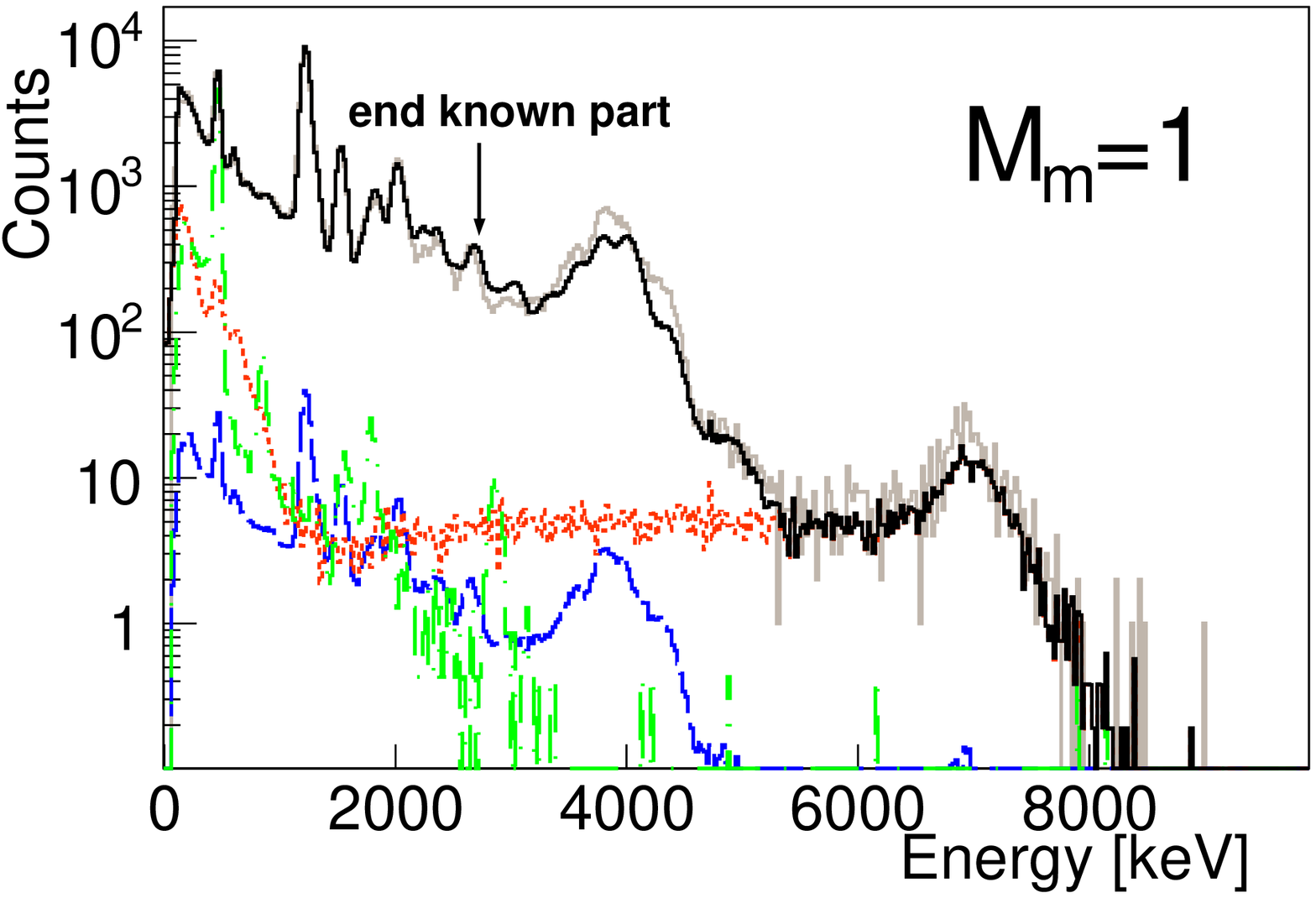} &
\includegraphics[width=0.5 \textwidth]{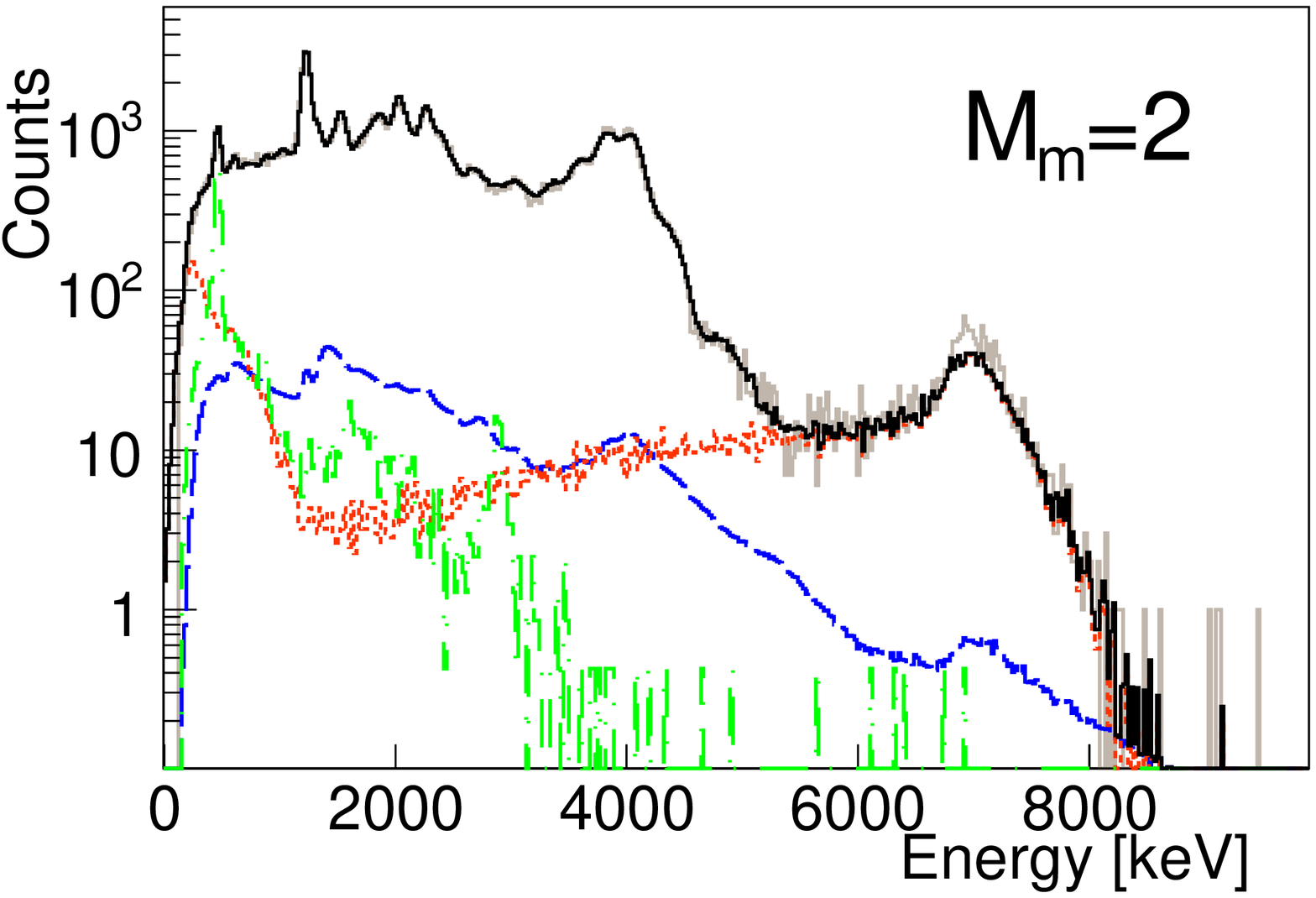} \\
\includegraphics[width=0.5 \textwidth]{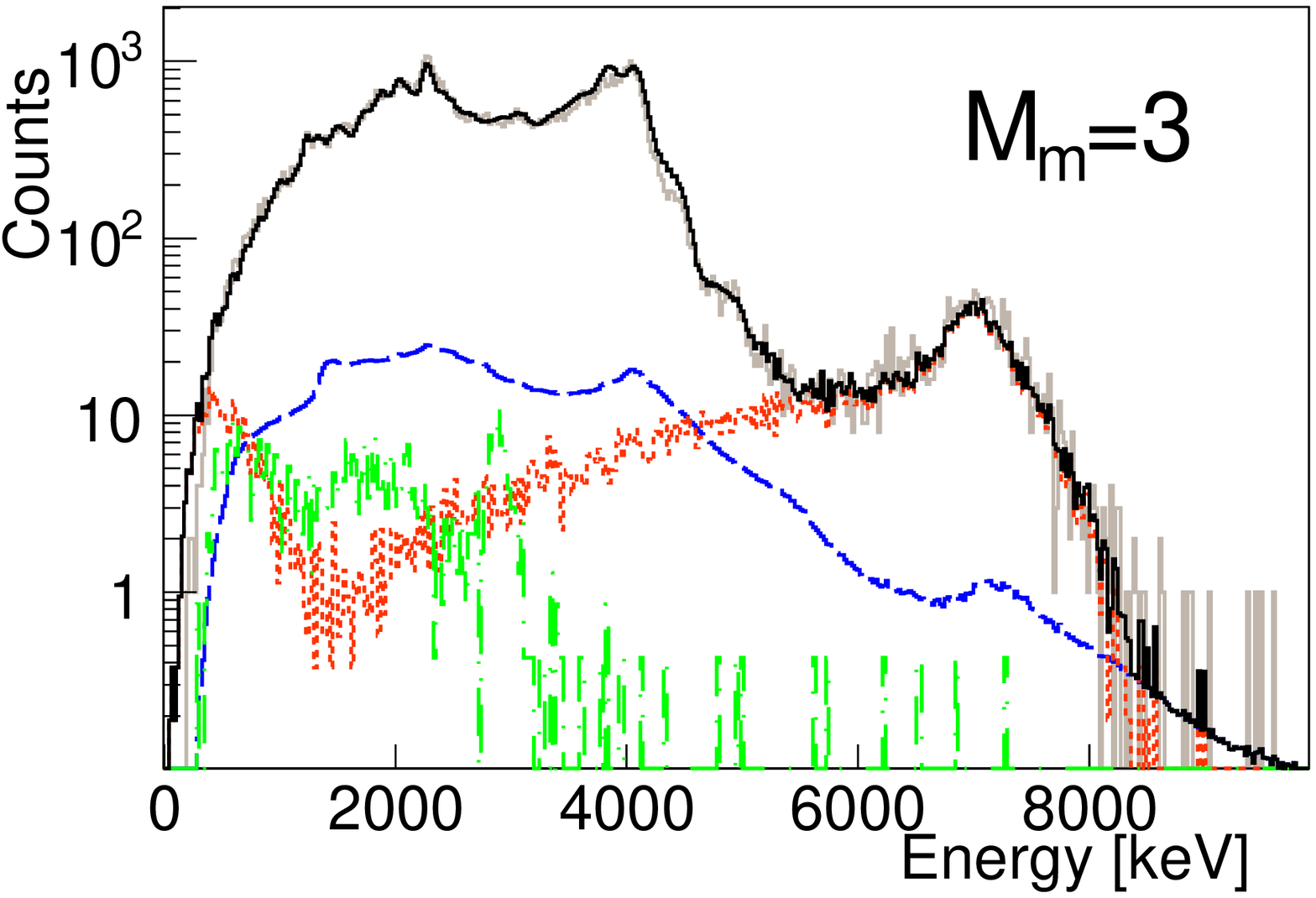} &
\includegraphics[width=0.5 \textwidth]{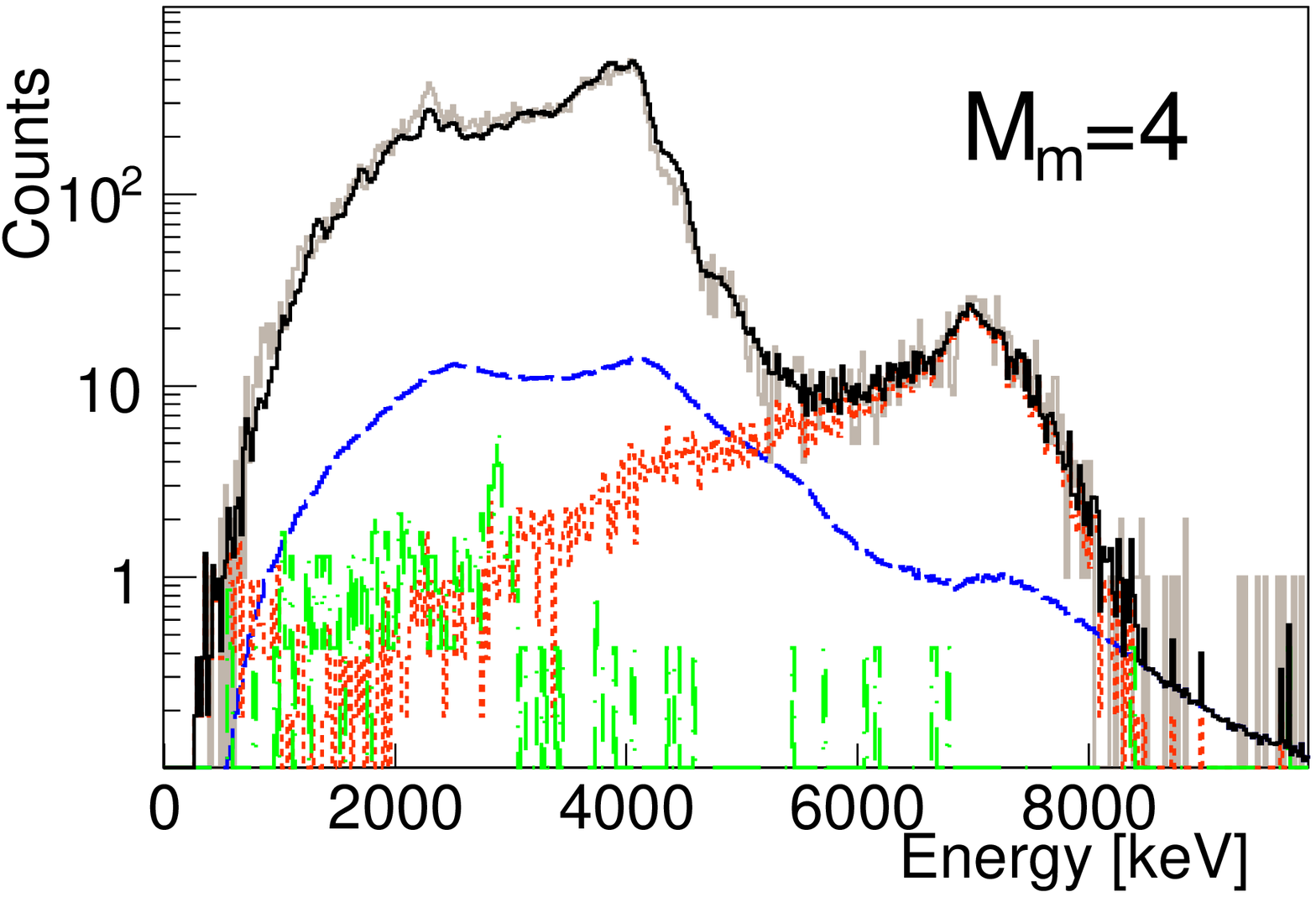} \\ 
\includegraphics[width=0.5 \textwidth]{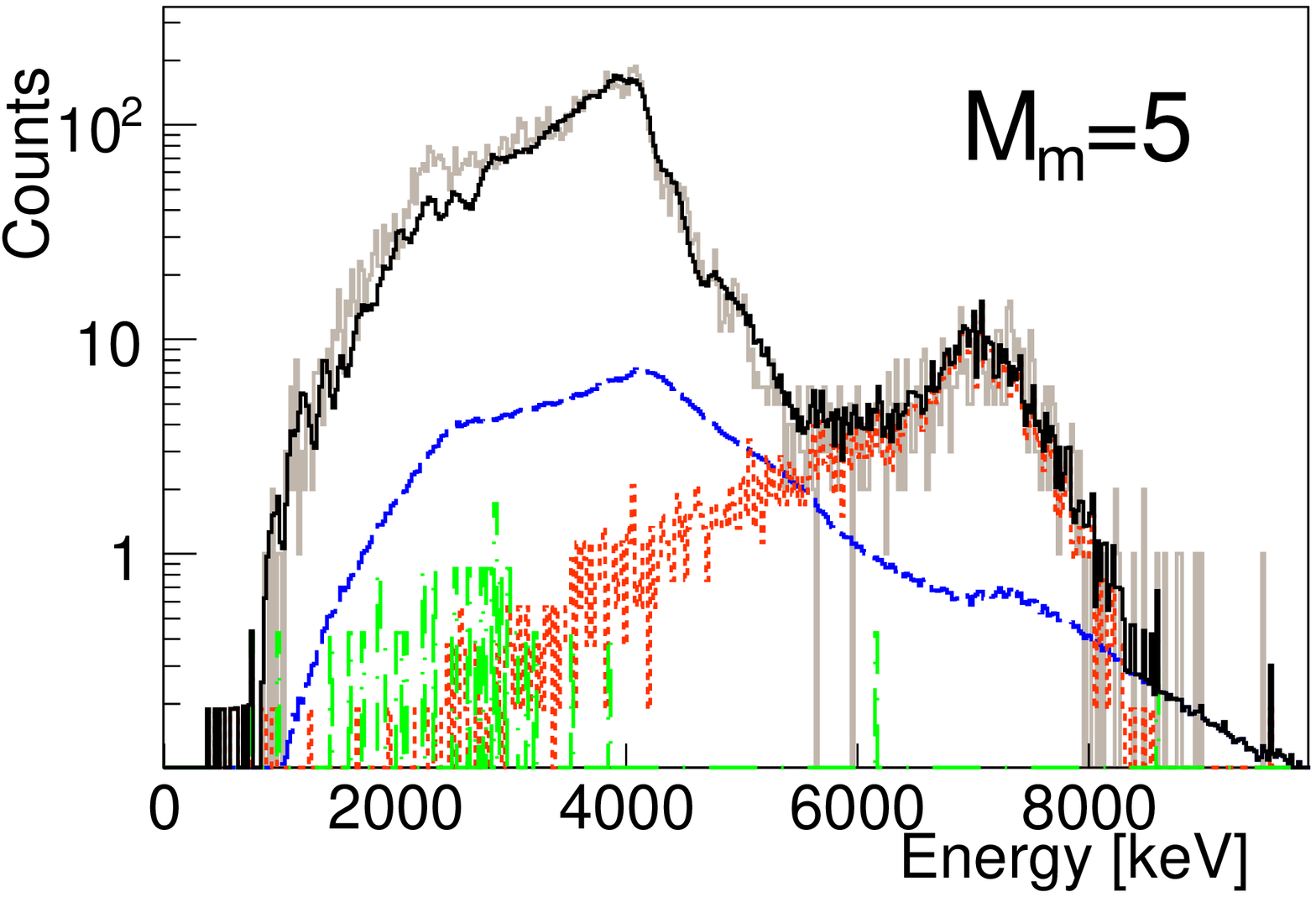} &
\includegraphics[width=0.5 \textwidth]{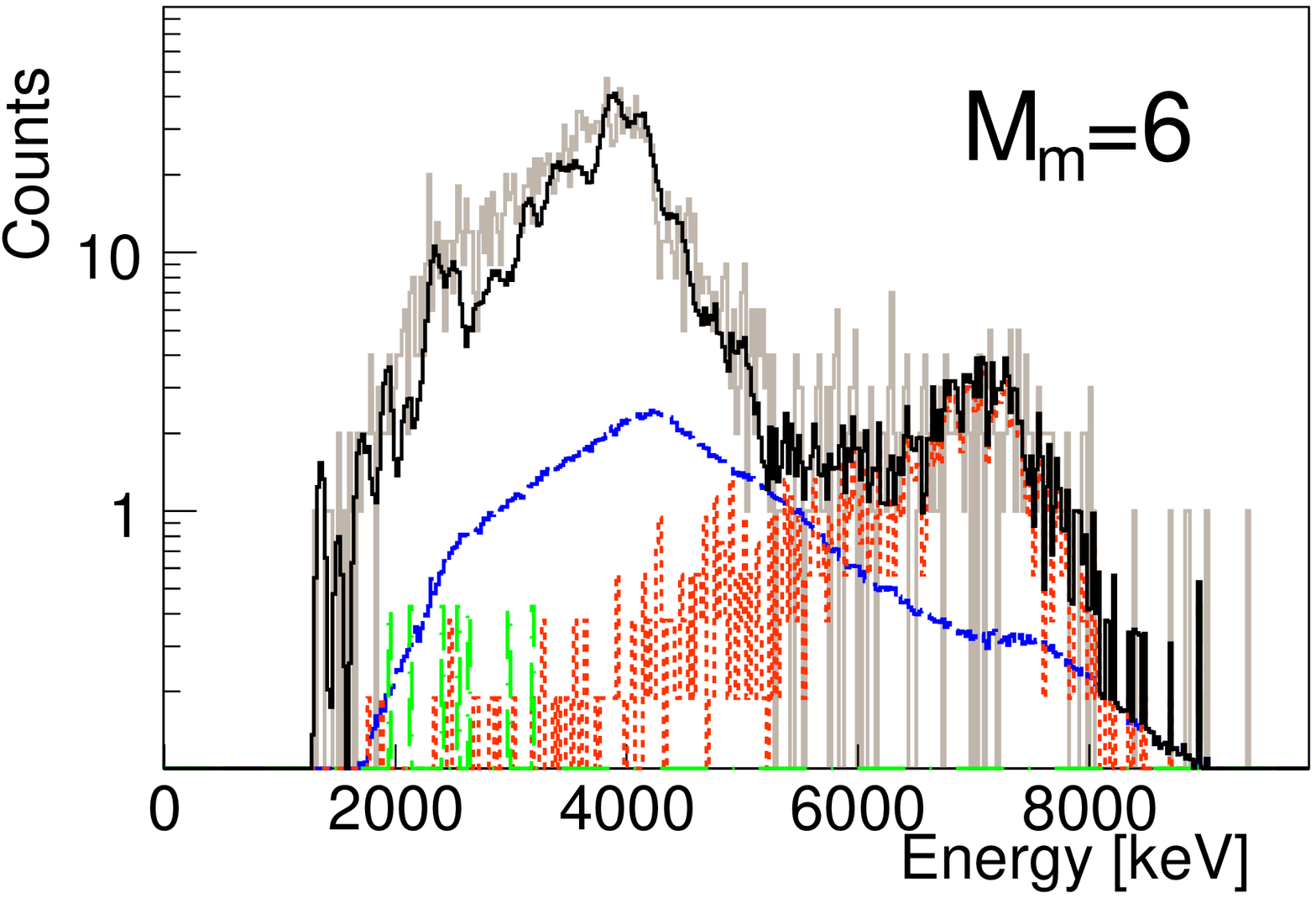} 
\end{tabular}
\caption{$^{137}$I $\beta$-gated spectra with a 500~ns $\beta$-gating time window and with a condition on module-multiplicity $M_m$ from 1 to 6 (solid grey) compared with the MC simulations (solid black) taking into account the summing-pileup contamination (dashed blue), the daughter activity (dashed-dotted green) and the $\beta$-n branch contribution (dotted orange).}
\label{137I_multiplicities}
\end{center}
\end{figure*}

\begin{figure}[!hbt]
\begin{center} 
\includegraphics[width=0.5 \textwidth]{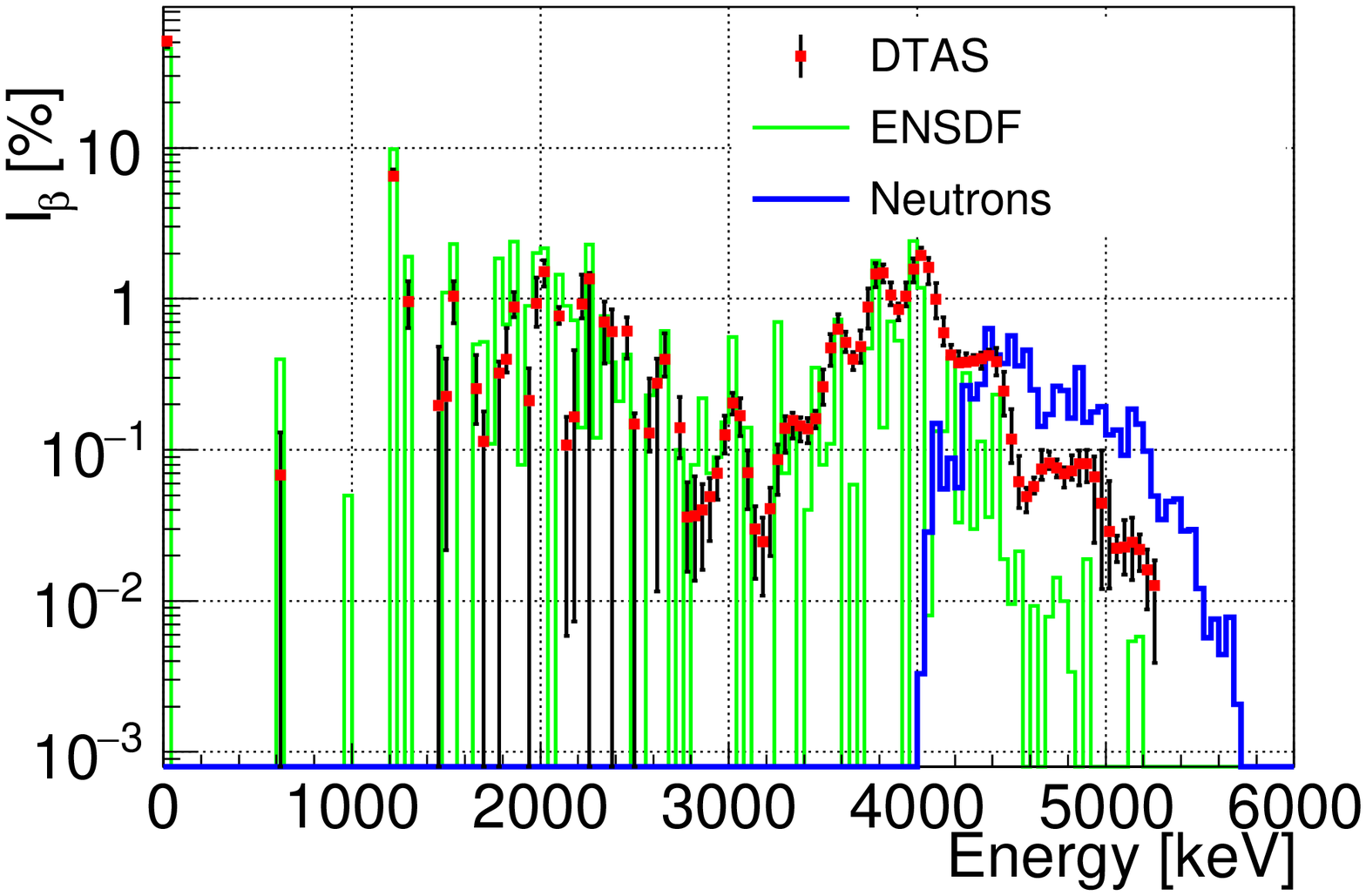} \\ 
\includegraphics[width=0.5 \textwidth]{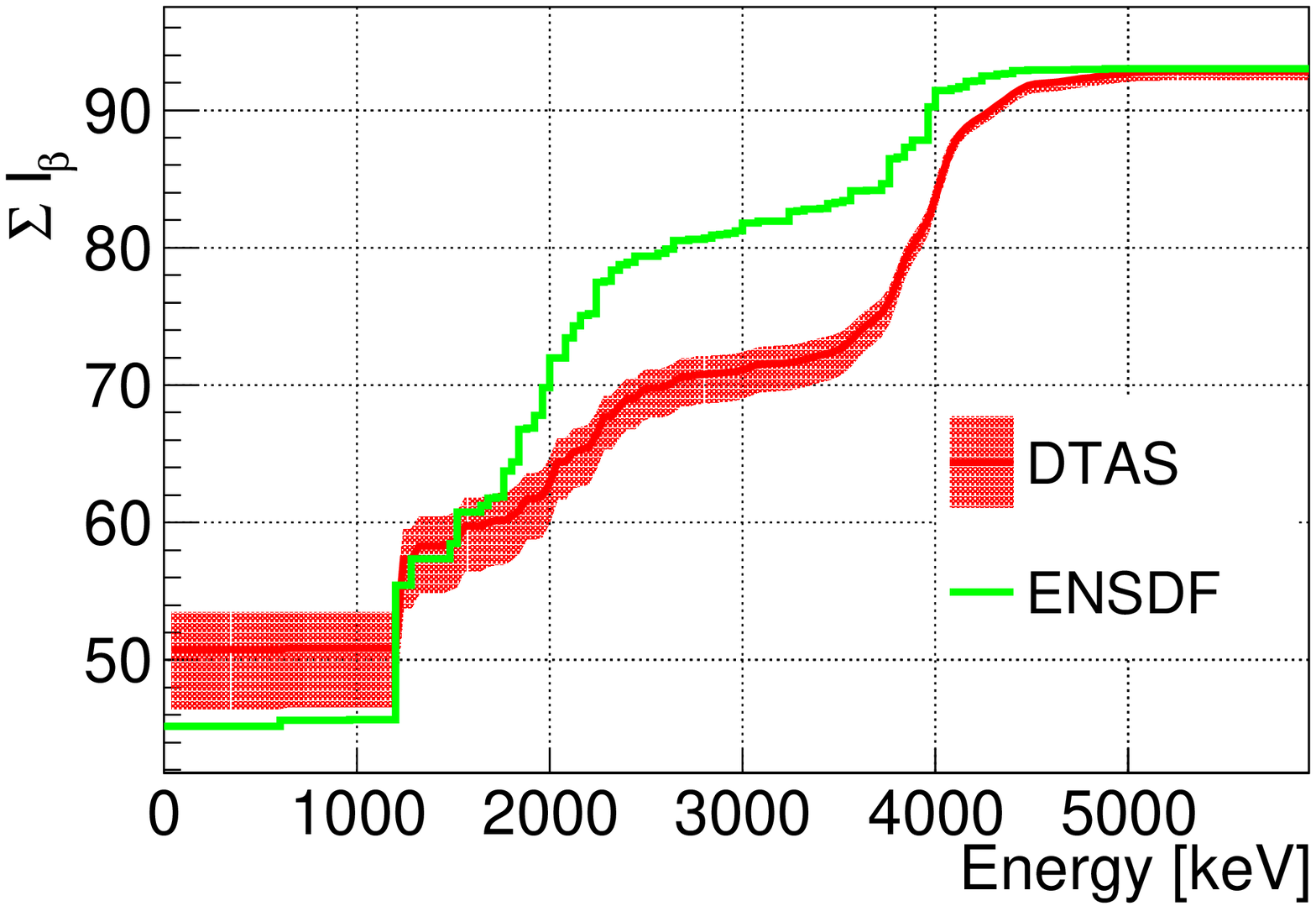} 
\caption{$\beta$-intensity distribution for the decay of $^{137}$I. Top panel: TAGS results (red dots with black error bars) and high-resolution $\gamma$-spectroscopy data from ENSDF (green line) are shown together with the $\beta$-n component (blue line). Bottom panel: accumulated $\beta$-intensity distribution obtained from the TAGS analysis (red line with error) and high-resolution $\gamma$-spectroscopy data from ENSDF (green line).}
\label{137I_I}
\end{center}
\end{figure}

As mentioned before, the segmentation of DTAS allows one to make more stringent tests of the branching ratio matrix used to construct the spectrometer response function for the decay of interest. The quality of the reproduction of the individual-module spectra and the module-multiplicity gated total absorption spectra was investigated for these purposes. The spectra of the individual modules is sensitive to the single $\gamma$-ray energy distribution from the whole decay. A more powerful test is provided by the total absorption spectra for different module-multiplicity conditions, which reflect the $\gamma$-cascade energy and multiplicity distribution as a function of excitation energy. The corresponding experimental spectra were generated with a 500~ns $\beta$-gating time window and are compared with MC simulations obtained with the DECAYGEN event generator \cite{TAS_decaygen} using the reference branching-ratio matrix and the $\beta$-intensity distribution from the analysis of the total absorption spectra (top panel in Fig.~\ref{137I_fit}) as input. The comparison for the sum of the 18 single-module spectra can be seen in Fig.~\ref{137I_individuals}. A reasonable reproduction of the spectrum is obtained, except for an excess of counts in the simulation in the range $2.2-3.2$~MeV and a deficit of counts in the interval $3.2-4.6$~MeV. The quality of the reproduction of the total absorption spectra for module-multiplicities from 1 to 6 is shown in Fig. \ref{137I_multiplicities}. We should emphasize that the same numerical factors used to normalize the contaminants for the total absorption spectrum (Fig.~\ref{137I_fit} top) were used to normalize the contaminants for all multiplicities. The agreement is excellent for $M_m=2$ to $M_m=6$. The most significant differences are found for $M_m=1$ and in fact the behavior is similar to that observed for single-module spectra, not surprising since the $M_m=1$ spectrum is mostly sensitive to single $\gamma$-ray de-excitations. None of the changes in the branching ratio matrix that we have tested (see discussion below) resulted in a better agreement, thus we adopted this $\beta$-intensity distribution as the reference solution. This distribution is obtained for the 500~ns time window and it is depicted with red points in the top panel of Fig~\ref{137I_I}. The black vertical bars indicate the uncertainty obtained with the procedure detailed below. For comparison we include in this figure the $\beta$-intensity obtained in high-resolution $\gamma$-ray spectroscopy from ENSDF~\cite{NDS_A137}, as well as the $\beta$-intensity followed by neutron emission, $I_{\beta n}$, discussed above. In the lower panel of Fig~\ref{137I_I} the accumulated $\beta$-intensity as a function of excitation energy from DTAS is compared with the one from ENSDF, showing clearly the effect of Pandemonium in the high-resolution data and the importance of using the TAGS technique.

%We like to point out also that the good reproduction of these spectra above 5.5~MeV proves the accuracy of our MC simulations of the $\beta$-n component.

For the evaluation of the systematic uncertainty in the $\beta$-intensity resulting from the TAGS analysis, we considered all possible solutions compatible with reproduction of the data. In comparison, the statistical uncertainty was found to be negligible. The envelope of all these $\beta$-intensity distributions defines the uncertainty bars shown in Fig. \ref{137I_I}. The $\beta$-intensity distribution obtained in the analysis of the $\beta$-gated spectrum with a 500~ns time window (shown in Fig. \ref{137I_fit} top) is taken as the reference result, and it can be found in the Supplemental Material~\cite{Suplement}. The other two $\beta$-intensity distributions plotted in Fig. \ref{137I_comparisonI} obtained from the same data with different gating conditions are considered as cross-checks of the reference solution. The influence of other experimental parameters in the results was investigated for the evaluation of the uncertainties of the reference solution. One of these parameters is the normalization factor of the contaminants. The corresponding normalization was varied until the reproduction of the total absorption spectrum was no longer considered acceptable. The normalization factor of the summing-pileup was changed by a factor of 30$\%$ and the ones for the daughter contribution and the $\beta$-n contribution were changed by 5$\%$. The resulting $\beta$-intensity distributions were included in the uncertainty estimation. We also changed the energy threshold in the simulation of the $\beta$-detector by $\pm$10~keV to evaluate the influence in the results of a change in the $\beta$-efficiency curve. The uncertainty in the experimental calibration of DTAS was also taken into account: the experimental energy calibration parameters have been changed by 0.5$\%$, while the width calibration parameters were changed by 15$\%$.

The influence of different branching ratio matrices was considered as well. This includes the $\beta$-intensity distributions obtained with the alternative spin-parity values for the g.s. level of $^{137}$I mentioned above. We tested the impact on the electric dipole PSF of alternative values of the level density parameter $a$ at $S_n$: 7.118, from EGSM calculations, and 16.461, from TALYS \cite{TALYS} (see Table~\ref{parameters}). We introduced by hand modifications in the branching ratio matrix of the continuum in order to reproduce better the experimental $\gamma$-intensities at low excitation energies obtained in high resolution studies \cite{137I_Fog,137I_Ohm,137I_Fog_fermi_gas} (see Table \ref{tableIg}). Although it worsened the reproduction of the sum of the individual modules and the module-multiplicity gated spectra, it gave a good reproduction of the total absorption spectrum and it was included in the evaluation of the uncertainties. 

\begin{table}[!hbt]
\centering
\begin{tabular}{cccc}
Energy [keV] & $I_{\gamma}$ ENSDF & $I_{\gamma}$ TAGS & $I_{\gamma}$ TAGS$^{*}$ \\ \hline
1218	&	0.128	&     0.071    & 0.127 \\      
\hline
352.01	&   0.490   &	 0.659  & 0.491 \\
556.06	&   0.151   &	 0.252  & 0.150 \\
680.7	&   0.244   &	 0.263  & 0.242 \\
\hline
\end{tabular}
\caption{\label{tableIg}Absolute $\gamma$-intensities per 100 decays de-exciting the main levels in the known part of the level scheme populated in the decays of $^{137}$I (first row) and $^{95}$Rb (rest). The second column corresponds to the intensities obtained from high resolution data \cite{NDS_A137,NDS_A95}. The third column gives the intensities obtained with TAGS for the reference analysis, whereas the intensities obtained with a modified branching ratio matrix are presented in the fourth column (TAGS$^{*}$).}
\end{table}

The impact of the first bin of the experimental spectrum included in the analysis was also evaluated. We found that it affects significantly the g.s. feeding intensity determination and we considered variations of $\pm$1 bin in the result. We have also used the maximum entropy (ME) algorithm \cite{TAS_algorithms} instead of the conventionally used EM algorithm to check the influence of the method of deconvolution in the analysis. The $\beta$ intensities determined in all cases were normalized to 100-$P_n$, where we take as a reference the value from ENSDF:  7.14$\%$~\cite{NDS_A137}. However, we also considered two alternative $P_n$ values: 7.76$\%$~\cite{NIM_BELEN} and 7.33$\%$~\cite{Pn_IAEA_2011}.

We obtained a g.s. feeding intensity of 50.8(43)$\%$. In the analysis with a 20~ns window, an intensity of 47.4$\%$ was obtained, while the analysis of the background subtracted singles spectrum gives a value of 49.9$\%$. This intensity was also calculated by means of a $\beta$-$\gamma$ counting method for TAGS data proposed by Greenwood et al. \cite{Greenwood_GS}, and we obtained a similar value of 51.2(6)$\%$. These two values are larger than the quoted number in ENSDF: 45.2(5)$\%$. It is also worth mentioning that B. Fogelberg \textit{et al.} \cite{137I_Fog} reported a $\beta$-intensity of 47$\%$ to the g.s., while a recent TAGS measurement determined a value of 49(1)$\%$ \cite{MTAS_137I}.

\subsection{Decay of $^{95}$Rb}

\begin{figure}[!hbt]
\begin{center}
\includegraphics[width=0.5 \textwidth]{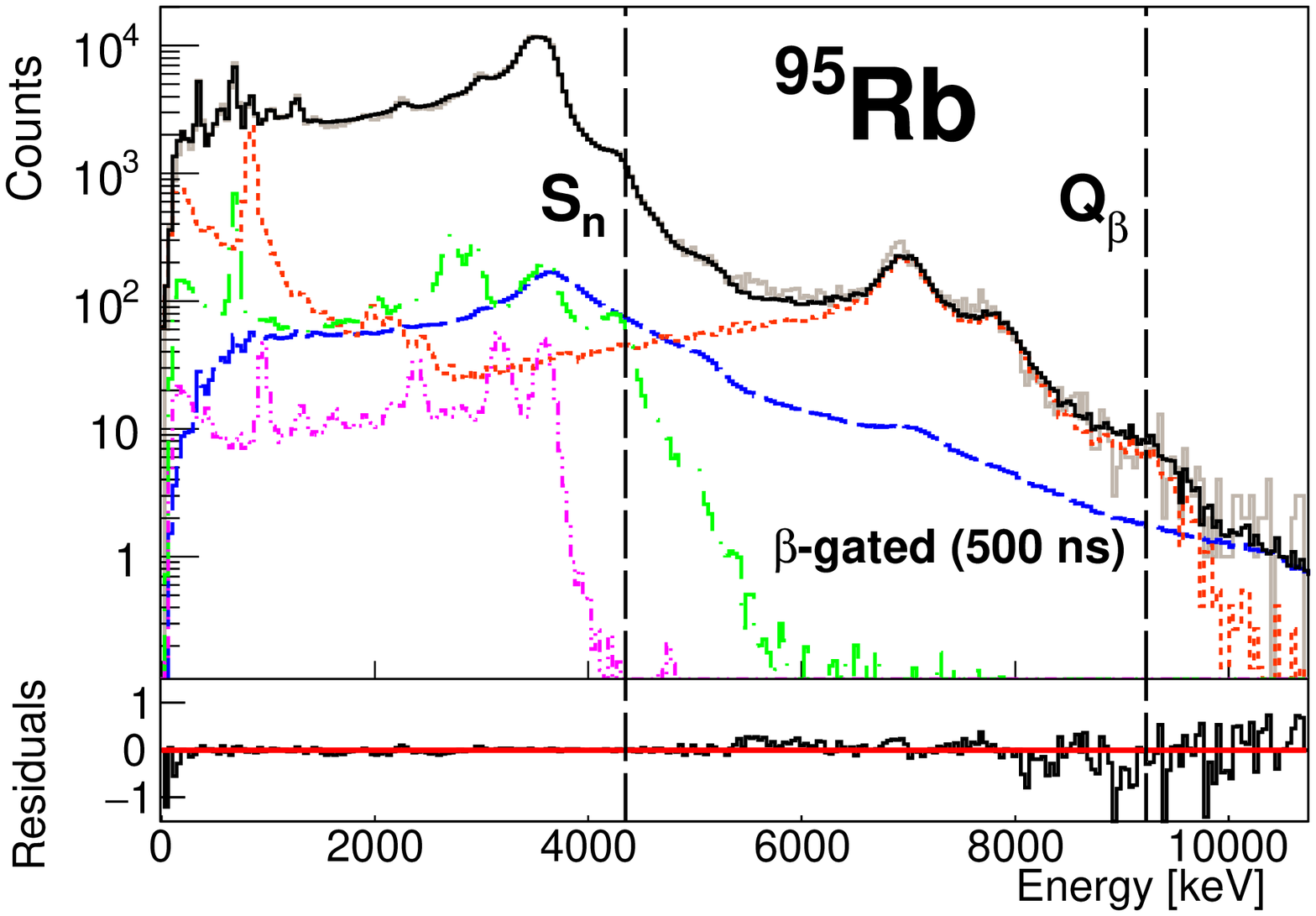} \\
\includegraphics[width=0.5 \textwidth]{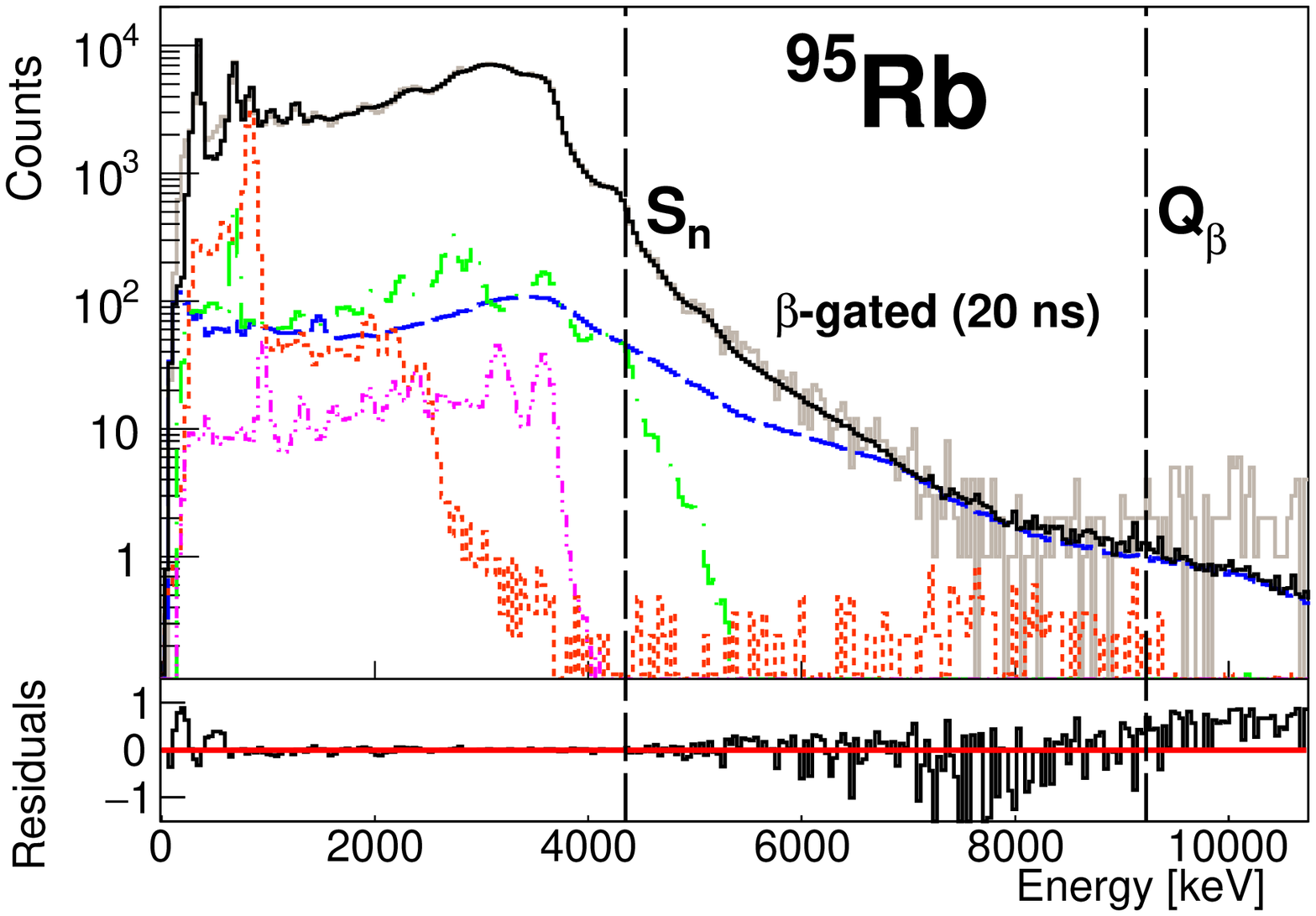}
\caption{Relevant histograms for the analysis of the decay of $^{95}$Rb: experimental $\beta$-gated spectrum (solid grey), summing-pileup contribution (dashed blue), daughter spectrum (dashed-dotted green), granddaughter spectrum (dashed-dotted-dotted pink) $\beta$-n branch (dotted orange) and reconstructed spectrum (solid black). The analyses of two different experimental spectra are shown: $\beta$-gated with 500~ns time gate (top) and $\beta$-gated with a gate of 20~ns to cut the $\beta$-delayed neutrons (bottom). See text for further details. The relative deviations between experimental and reconstructed spectra are shown for each case.} 
\label{95Rb_fit}
\end{center}
\end{figure}

\begin{figure}[!hbt]
\begin{center} 
\includegraphics[width=0.5 \textwidth]{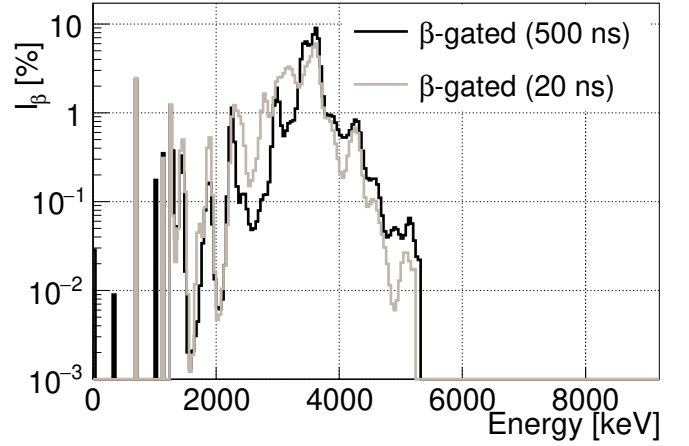} 
\caption{Comparison of the $\beta$-intensities obtained for the decay of $^{95}$Rb from the TAGS analysis of experimental spectra generated with two different time conditions.}
\label{95Rb_comparisonI}
\end{center}
\end{figure}

A spin-parity value of $5/2^-$~\cite{NDS_A95} is used for the g.s. of $^{95}$Rb. For the analysis, we considered allowed transitions and first forbidden transitions to states of the known part of the level scheme, and only allowed transitions to states in the continuum. Using the known level scheme and the nuclear statistical model parameters described at the beginning of Section \ref{TAGS}, we performed the analysis of the $\beta$-gated total absorption spectrum with the time window of 500~ns. The quality of the reproduction of the experimental spectrum is shown in the top panel of Fig.~\ref{95Rb_fit} and the corresponding $\beta$-intensity distribution is shown in Fig.~\ref{95Rb_comparisonI} (see also the Supplemental Material~\cite{Suplement}). We also performed the analysis of the $\beta$-gated spectrum with a condition of 20~ns to eliminate the delayed contribution associated with neutron interactions of the $\beta$-n branch (see bottom panel of Fig.~\ref{95Rb_fit}). Note that here there is a prompt contribution due to the $\gamma$-rays de-exciting $^{94}$Sr. This did not happen in $^{137}$I, where only the g.s. of $^{136}$Xe is populated. This contribution can be identified as an 837~keV $\gamma$-transition peak visible in the spectra. In the $\beta$-gated spectrum with the wide time coincidence window, the prompt contribution is partially added to the delayed contribution distorting the high energy side of the neutron capture bump, as shown in Section \ref{bnBkg}. As seen in the bottom panel of Fig. \ref{95Rb_fit}, the use of a 20~ns time coincidence window in the MC simulation of the $\beta$-n branch eliminates the neutron capture bump leaving a prominent peak at 837~keV used for normalization of this contamination.  It should be noted (compare top and bottom panels in  Fig.~\ref{95Rb_fit}) that the increase of effective DTAS threshold associated with the narrow time window, strongly affects the shape of the total absorption spectrum in this case. The reason is that we are cutting some low-energy $\gamma$-rays that take part in many cascades de-exciting high energy levels. In particular, the intense $\gamma$-ray of 204~keV, which comes from a $7/2^+$ level at 556~keV, is clearly cut, thus shifting and distorting the spectrum. This explains the large differences in the $\beta$-intensity distributions obtained from the analysis of both spectra, as shown in Fig. \ref{95Rb_comparisonI}.

\begin{figure}[!hbt]
\begin{center} 
\includegraphics[width=0.5 \textwidth]{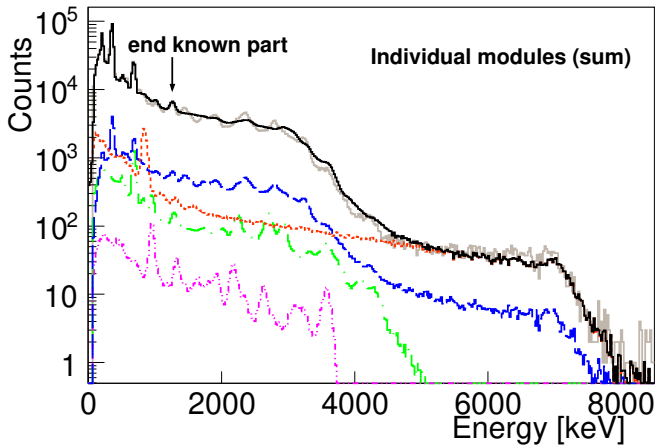} 
\caption{Comparison of the sum of the 18 individual experimental spectra (solid gray) with the reconstructed spectrum after the analysis (solid black) taking into account the pileup contribution (dashed blue), the daughter contamination (dashed-dotted green), the granddaughter contamination (dashed-dotted-dotted pink) and the $\beta$-n branch (dotted orange).}
\label{95Rb_individuals}
\end{center}
\end{figure}

\begin{figure*}[!hbt]
\begin{center} 
\begin{tabular}{cc}
\includegraphics[width=0.5 \textwidth]{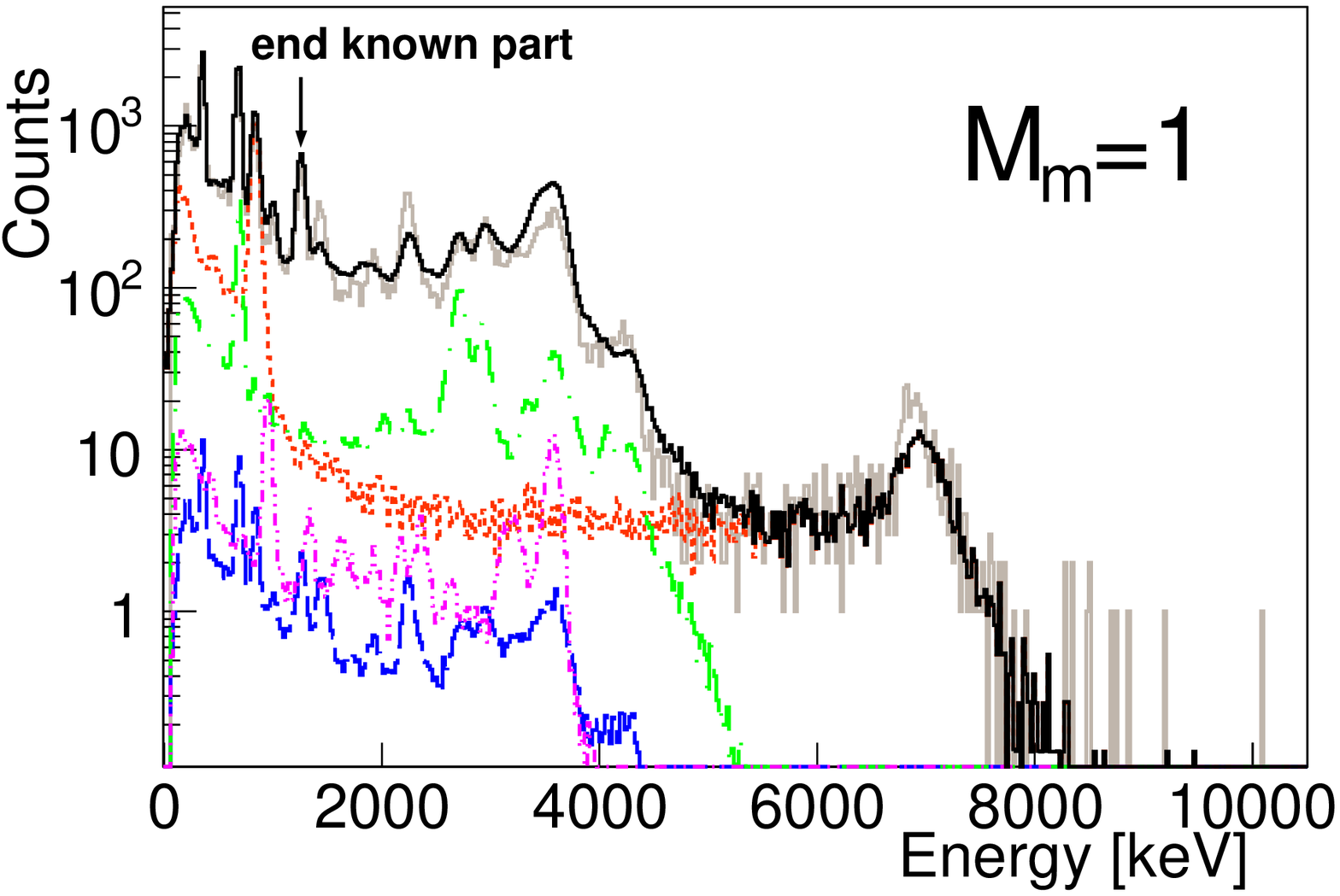} &
\includegraphics[width=0.5 \textwidth]{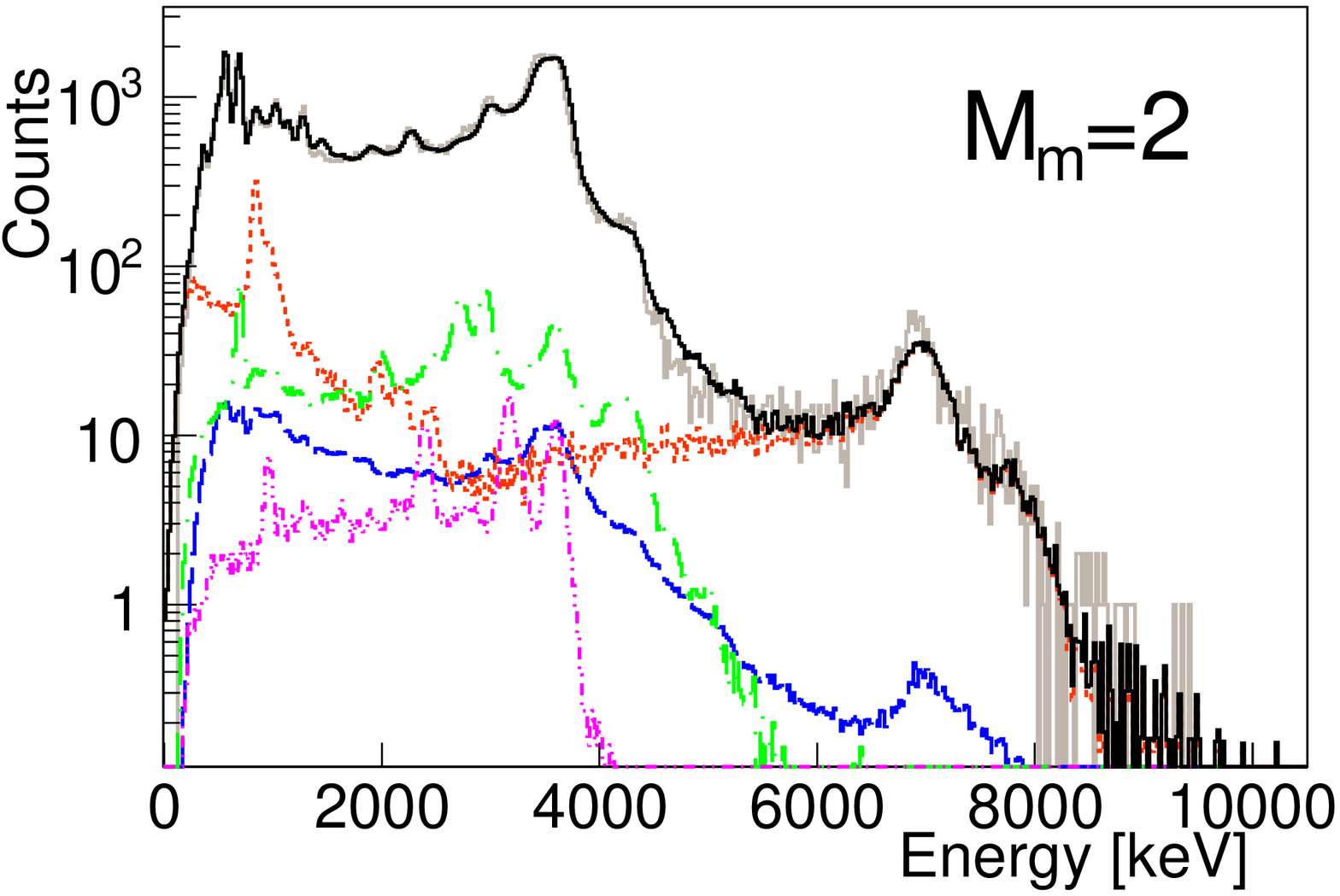} \\
\includegraphics[width=0.5 \textwidth]{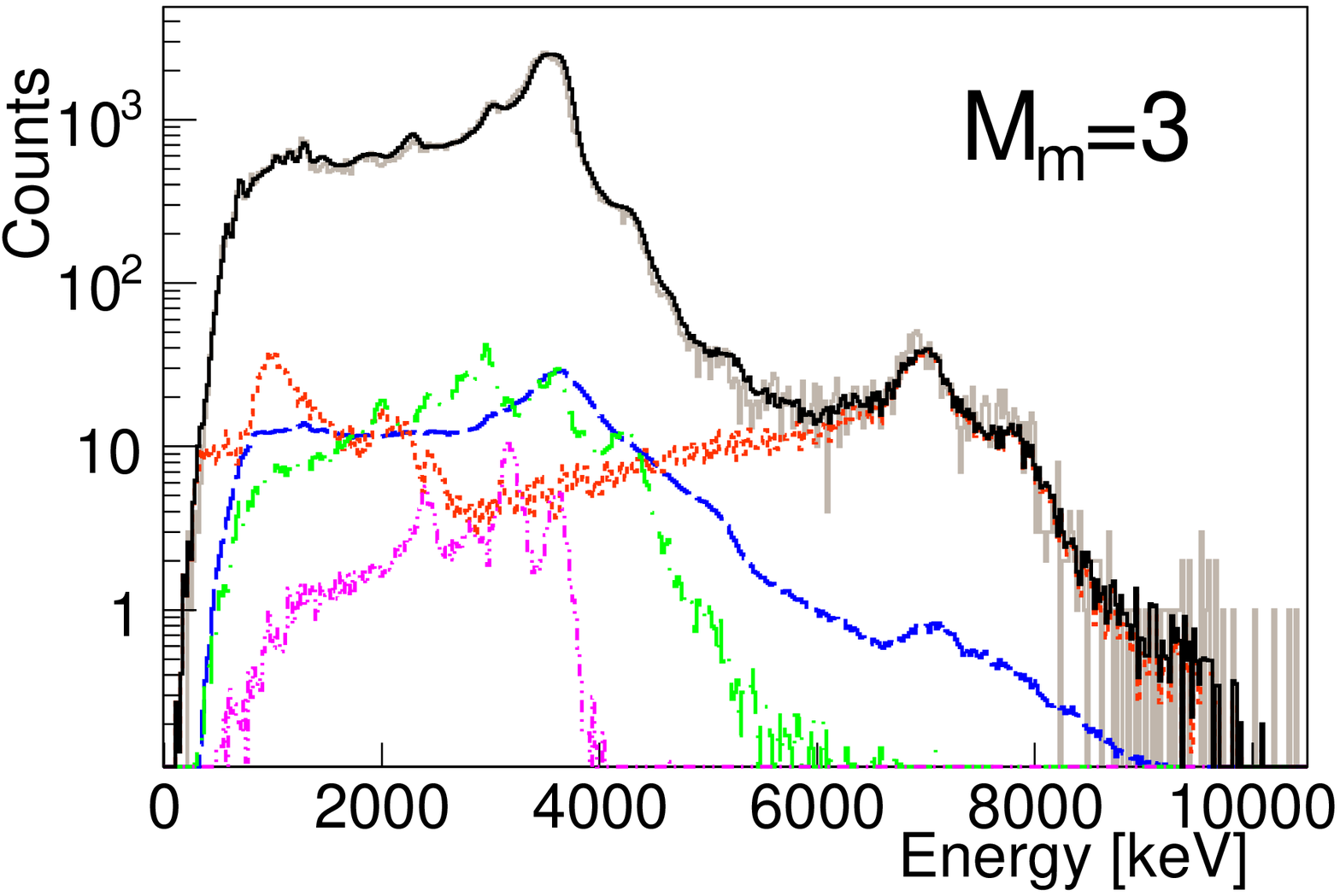} &
\includegraphics[width=0.5 \textwidth]{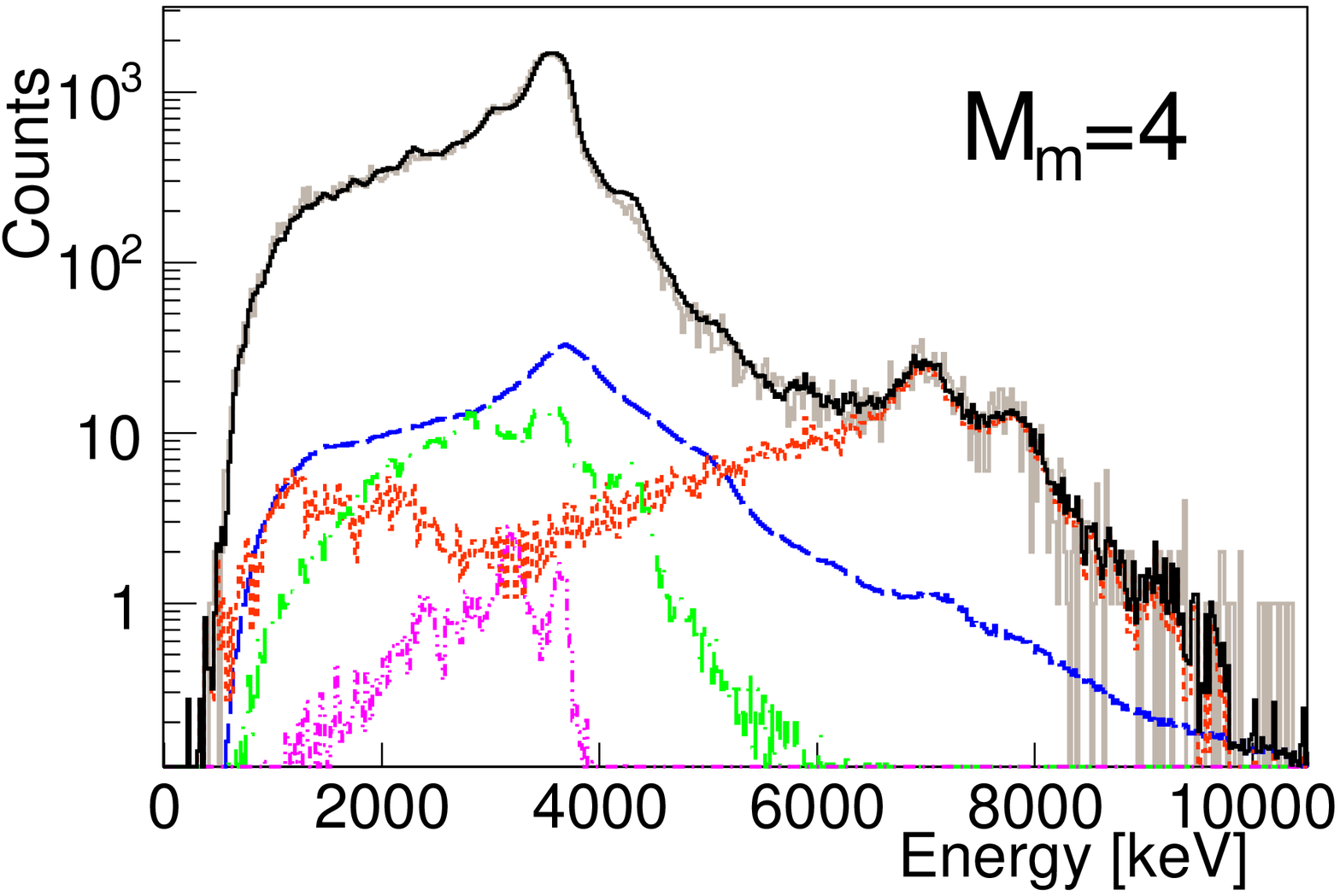} \\ 
\includegraphics[width=0.5 \textwidth]{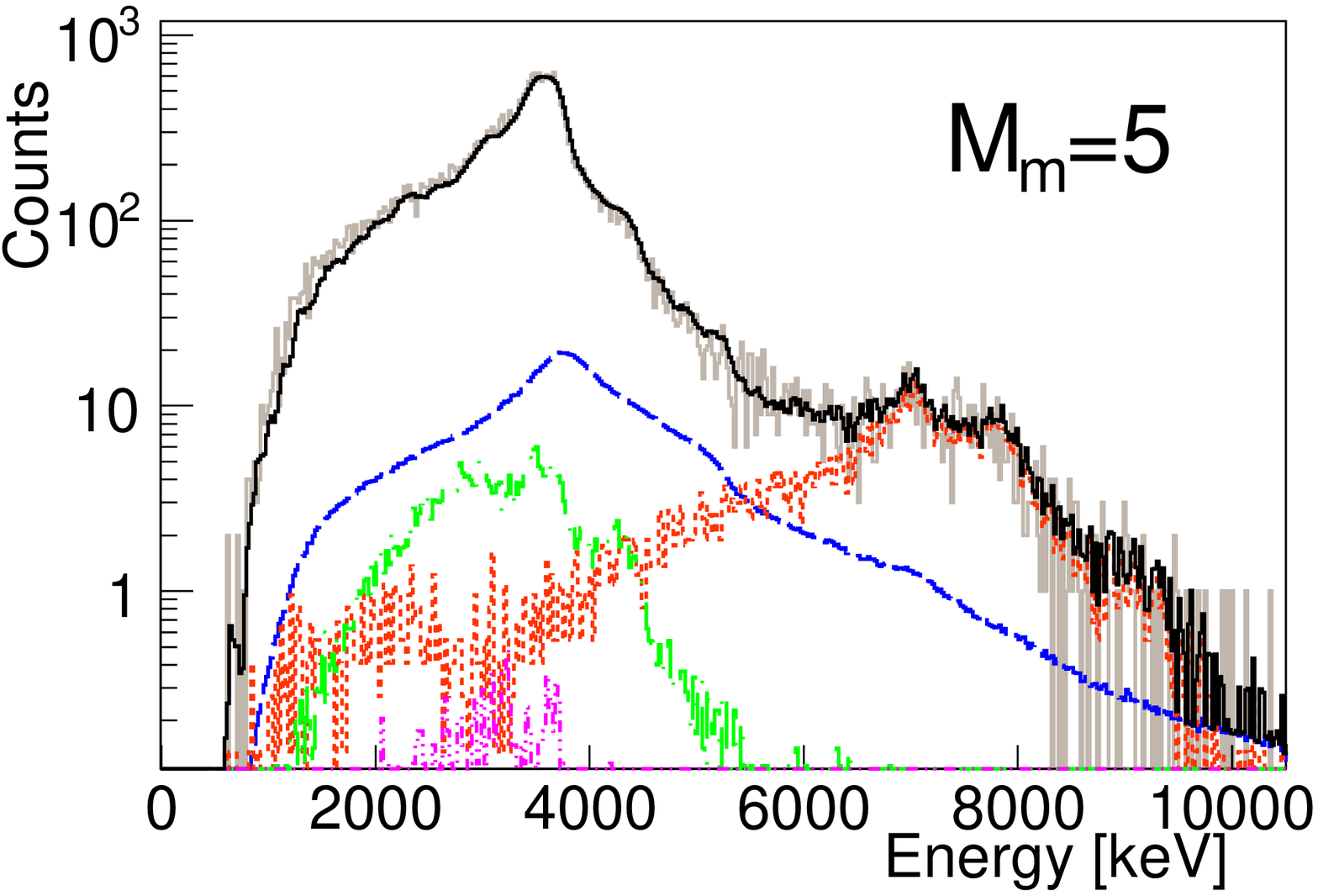} &
\includegraphics[width=0.5 \textwidth]{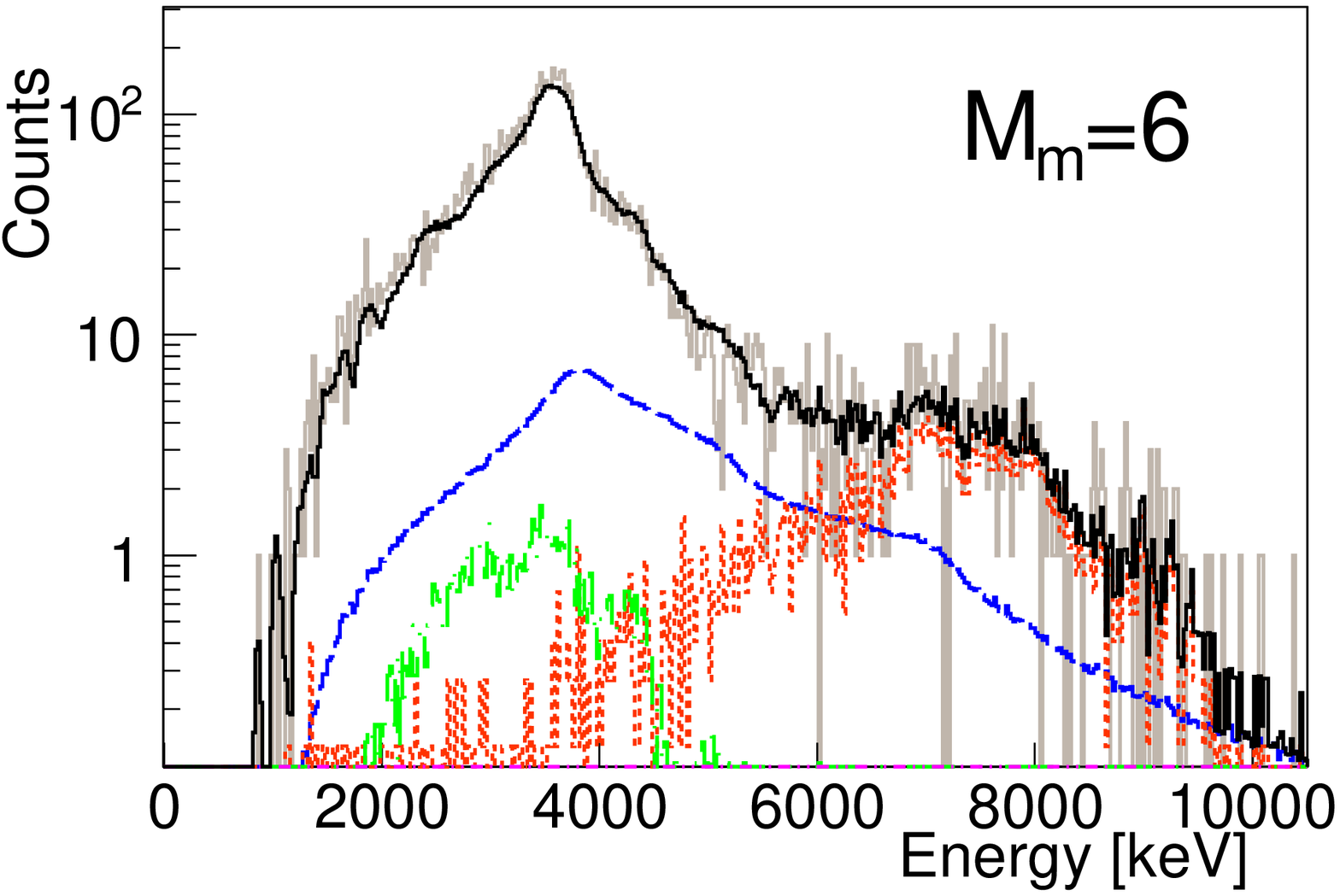} 
\end{tabular}
\caption{$^{95}$Rb $\beta$-gated spectra with a 500~ns $\beta$-gating time window and with a condition on module-multiplicity $M_m$ from 1 to 6 (solid grey) compared with the MC simulations (solid black) taking into account the summing-pileup contamination (dashed blue), the daughter activity (dashed-dotted green), the granddaughter activity (dashed-dotted-dotted pink) and the $\beta$-n branch contribution (dotted orange).}
\label{95Rb_multiplicities}
\end{center}
\end{figure*}

As in the case of $^{137}$I the branching ratio matrix used in these analyses was tested investigating the reproduction of the sum of the individual-module spectra (see Fig.~\ref{95Rb_individuals}) and the module-multiplicity gated spectra (see Fig. \ref{95Rb_multiplicities}). The result of the analysis of the $\beta$-gated spectrum with the 500~ns time gate was used as the reference one. As for $^{137}$I, the overall agreement is excellent and the larger differences are found for the $M_m = 1$ multiplicity gated total absorption spectrum and the sum of the individual-module spectra. The discrepancies in the reproduction of the fine structure in these spectra above the last discrete level in the known level scheme, seem to reflect the difficulty in the statistical model of reproducing variations in the branching to the g.s for individual levels. The impact of these discrepancies in the $\beta$-intensity distribution is small, given the limited contribution of $M_m = 1$ to the total spectrum.

Similarly to the $^{137}$I case, we have considered different sources of systematic error for the reference solution obtained with a 500~ns time window. The maximum variation of the normalization factor for the $\beta$-delayed neutron branch compatible with the reproduction of the total absorption spectrum  was 10$\%$. Due to the large contribution of the $\beta$-delayed neutron branch, especially at high energies, the summing-pileup normalization could be changed by a factor 100$\%$, and the activities of daughter and granddaughter were changed by a factor 50$\%$ without noticing a distortion in the reproduction of the spectrum. Apart from that, we have considered the same possible sources of uncertainty as in $^{137}$I for the threshold of the $\beta$-detector, the energy calibration and the width calibration. We also considered the solution obtained by applying the ME deconvolution method. Apart from the ENSDF $P_n$ value of 8.7$\%$~\cite{NDS_A95}, alternative values have been used to normalize the $\beta$-intensities to 100-$P_n$: 8.87$\%$~\cite{Pn_IAEA_2011} and 9.08$\%$~\cite{NIM_BELEN}. A modified branching ratio matrix reproducing the known $\gamma$-intensities for low-excitation levels, coming from high-resolution studies \cite{Kratz_95Rb_decay}, was also considered (see Table~\ref{tableIg}). Although it gave acceptable results, it worsened the reproduction of the total spectrum and the reproduction of the module-multiplicity gated spectra. The $\beta$ intensity distribution obtained with a narrow gate of 20~ns is considered just as a cross-check (see Fig. \ref{95Rb_comparisonI}) and it was not included in the error budget.

\begin{figure}[!hbt]
\begin{center} 
\includegraphics[width=0.5 \textwidth]{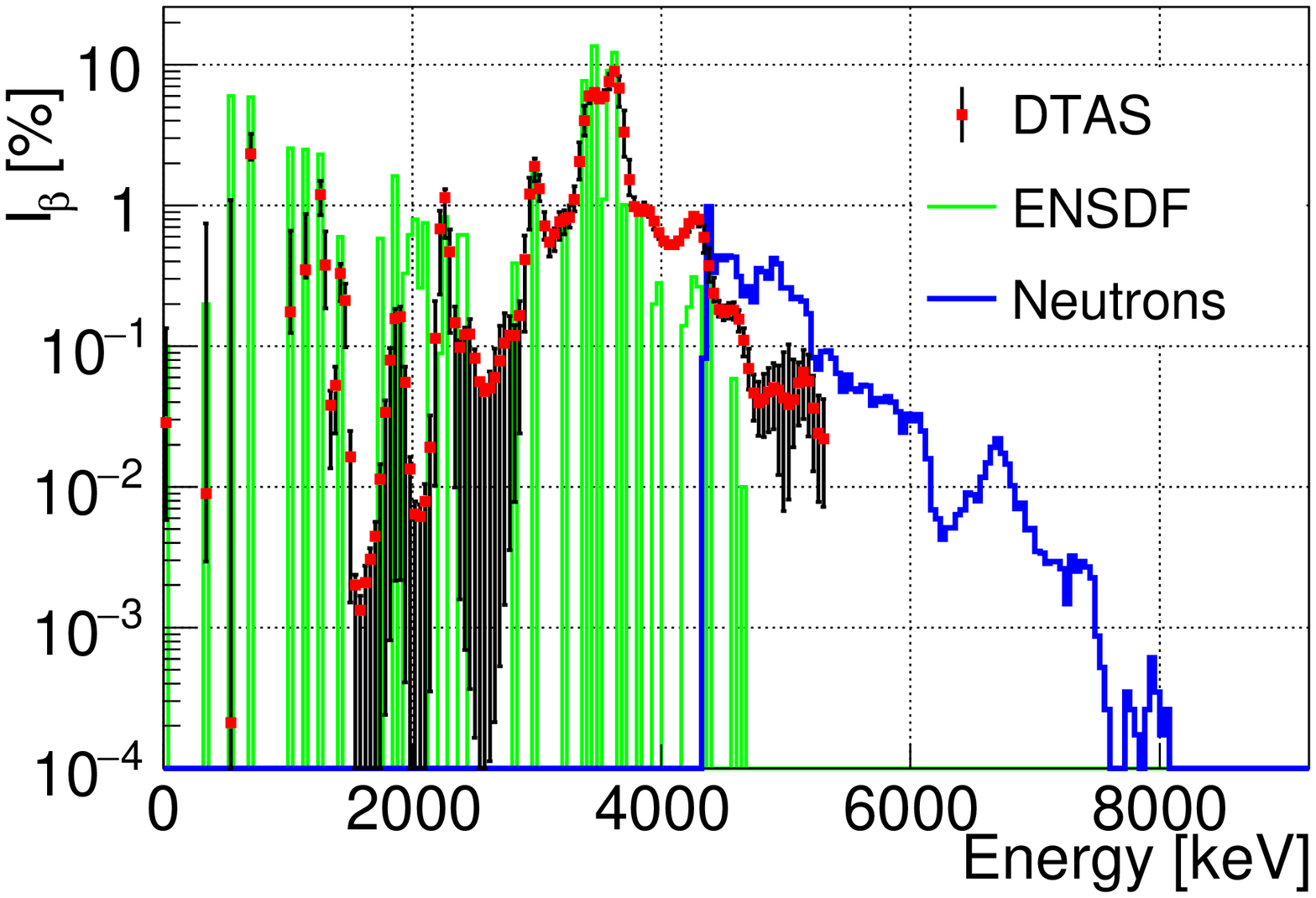} \\ 
\includegraphics[width=0.5 \textwidth]{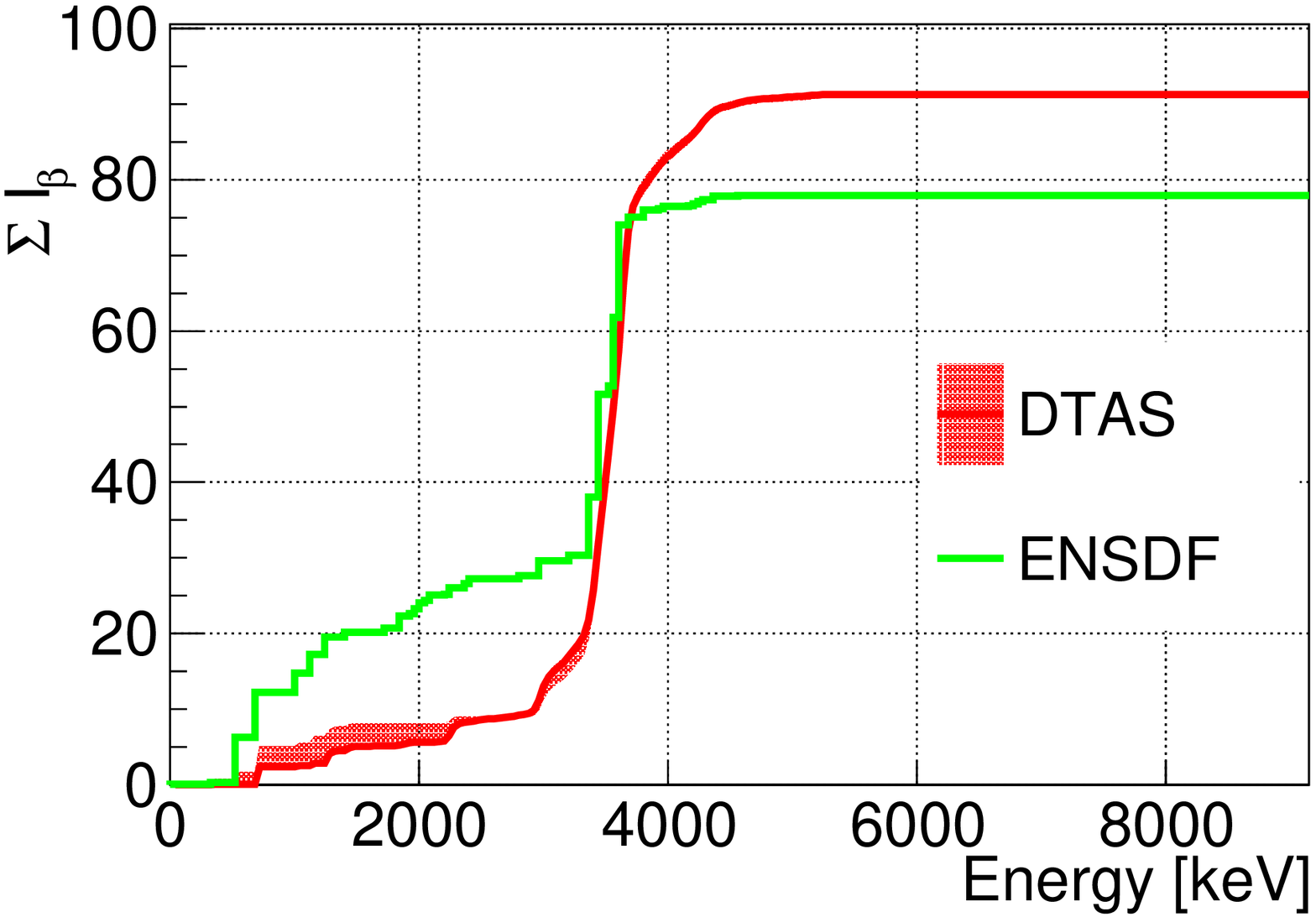} 
\caption{$\beta$-intensity distribution for the decay of $^{95}$Rb. Top panel: TAGS results (red dots with error bars) and high-resolution $\gamma$-spectroscopy data from ENSDF (green line) are shown together with the $\beta$-n component (blue line). Bottom panel: accumulated $\beta$-intensity distribution obtained from the TAGS analysis (red line with error) and high-resolution $\gamma$-spectroscopy data from ENSDF (green line).}
\label{95Rb_I}
\end{center}
\end{figure}

In Fig.~\ref{95Rb_I}, the $\beta$-intensity distribution including the systematic uncertainty is compared with the high-resolution result from ENSDF \cite{NDS_A95}. The $\beta$-intensity distribution obtained with DTAS is normalized to 100-$P_n$, whereas the intensity in ENSDF is normalized to 77.941$\%$ \cite{NDS_A95} since the evaluators recognized the incompleteness of the experimental information.

\section{$\gamma$-neutron competition}\label{Competition}

In the two $\beta$-delayed neutron emitters studied in this work a significant amount of $\beta$ intensity de-exciting by means of $\gamma$-rays, $I_{\beta \gamma}$, is observed above $S_n$. In order to compare with the neutron emission probability ($P_n$) we can define $P_{\gamma}$ as the integrated $I_{\beta \gamma}$ above $S_n$:

\begin{equation}\label{Pgamma_Def}
P_{\gamma}=\int_{S_n}^{Q_{\beta}}I_{\beta\gamma}dE_x
\end{equation}

In Table \ref{Pgamma} the $P_{\gamma}$ values obtained in this work are compared with the $P_n$ values. The $\beta$-intensity connecting to levels that de-excite by means of $\gamma$-rays represents 56$\%$ of the total $\beta$-intensity above $S_n$ for $^{137}$I and 25$\%$ for $^{95}$Rb. The situation is similar to that found in the decays of $^{87,88}$Br \cite{vTAS_PRL, vTAS_PRC} and can be understood as a nuclear structure effect, as discussed below. Compared to the $\gamma$ intensity observed above $S_{n}$ in high-resolution experiments, retrieved from the ENSDF data base, we observe 3 and 5 times higher values for $^{137}$I and $^{95}$Rb, respectively, indicating a sizable Pandemonium effect.

\begin{table}[!hbt]
\centering
  \begin{tabular}{|c|c|c|c|}
  Nucleus & $P_{\gamma}$ ENSDF & $P_{\gamma}$ TAGS & $P_n$ \\
          & [$\%$] & [$\%$] & [$\%$]  \\ \hline  
  $^{137}$I  & 2.76 & 9.25$_{-2.23}^{+1.84}$ & 7.14(23) \\
  $^{95}$Rb  & 0.57 & 2.92$_{-0.83}^{+0.97}$ & 8.7(3)\\
  \hline
  \end{tabular}
  \caption{Integral $I_{\beta \gamma}$ above $S_n$ ($P_{\gamma}$) obtained with TAGS in comparison with the value from ENSDF and with the neutron emission probability, $P_n$. }
  \label{Pgamma}
\end{table}

The uncertainties quoted in Table \ref{Pgamma} are the quadratic sum of two terms. One term is evaluated from the dispersion of $P_{\gamma}$ values calculated from the different $\beta$ intensity distributions obtained in our analysis under different assumptions, as explained in previous Sections. The second term arises from the uncertainty in the first bin considered in the integration of Eq.~\eqref{Pgamma_Def}, due to the uncertainty of our energy calibration. This term amounts to 20$\%$ for $^{137}$I, and 26$\%$ for $^{95}$Rb and dominates the upper limit of the uncertainty given in Table \ref{Pgamma}.

Similar to the previous works on $^{87,88}$Br and $^{94}$Rb~\cite{vTAS_PRL, vTAS_PRC}, we have evaluated the ratio $I_{\beta \gamma}/(I_{\beta \gamma}+I_{\beta n})$ as a function of the excitation energy above $S_n$. This ratio is equivalent to the average ratio of  widths: $\langle \Gamma_{\gamma}/(\Gamma_{\gamma}+\Gamma_n)\rangle$, that is calculated in the Hauser-Feshbach formalism as described in detail in \cite{vTAS_PRC, Tain_proceedings_bneu1}. The ingredients for these calculations are the NLD and the PSF in the daughter nucleus, and the NTC into the $\beta$-delayed neutron descendant. The first two are the same used for the construction of the branching ratio matrix in the TAGS analysis, whereas NTC are obtained from optical model calculations performed with TALYS-1.8 \cite{TALYS}. In the case of $^{137}$I we only need to consider the neutron transmission to the 0$^{+}$ g.s. of $^{136}$Xe, while in the case of $^{95}$Rb several levels are populated in $^{94}$Sr. Spin-parity values from RIPL-3 \cite{RIPL-3} have been selected for those levels in $^{94}$Sr that have no experimentally assigned values. 

\begin{figure}[!hbt]
\begin{center}
\includegraphics[width=0.5 \textwidth]{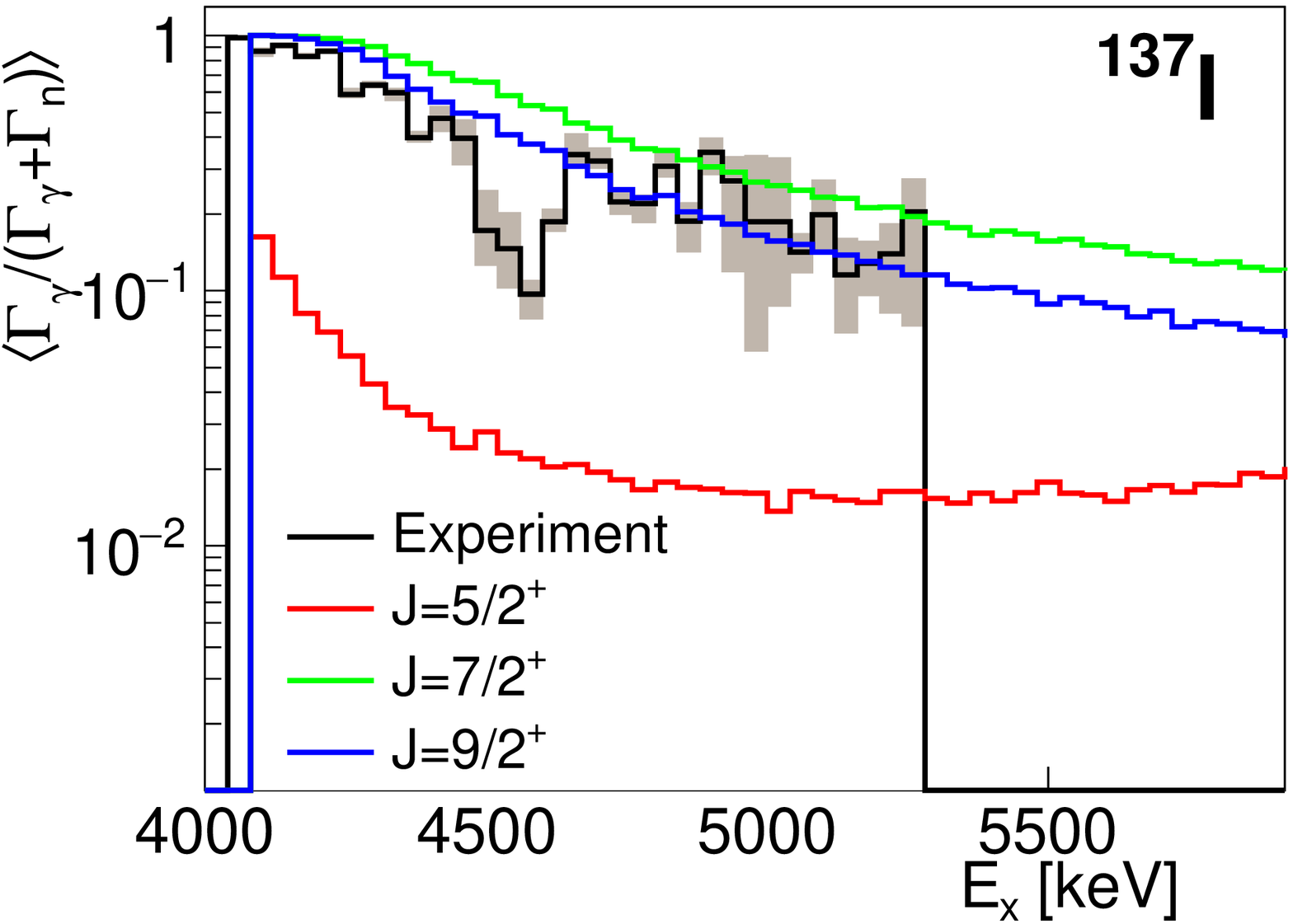} \\
\includegraphics[width=0.5 \textwidth]{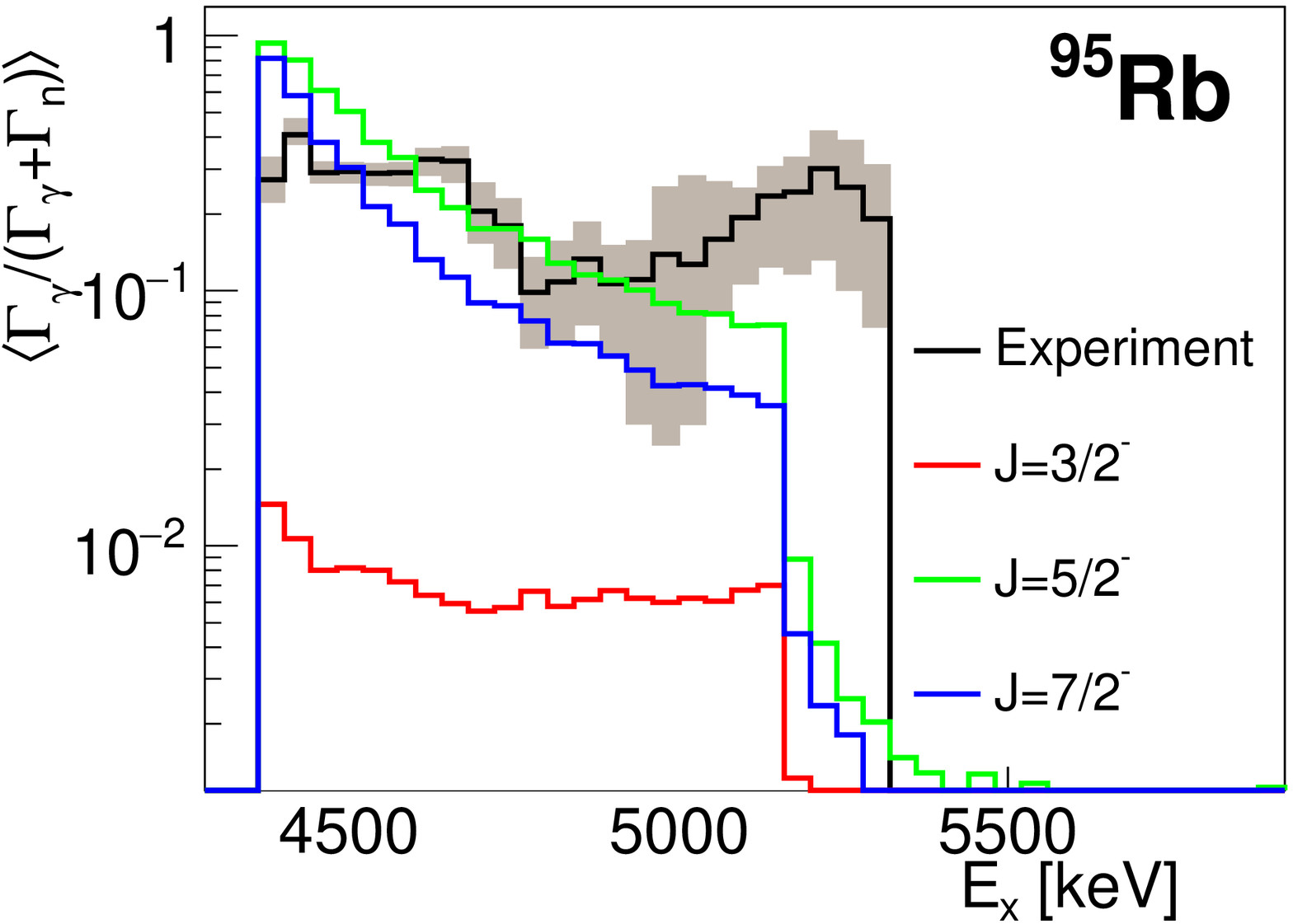}
\caption{Experimental average $\gamma$ to total width compared with Hauser-Feshbach calculations for allowed $\beta$-decays in the case of $^{137}$I (top) and $^{95}$Rb (bottom).}
\label{Average_widths}
\end{center}
\end{figure}

For the calculation of the average ratio of widths we used the MC method explained in \cite{vTAS_PRC}. This allows one to take into account Porter-Thomas fluctuations on neutron and $\gamma$ widths. A direct comparison between the experimental ratio of $\beta$-intensities and the average ratio of widths is meaningful, since the average is taken over all levels of a given $J^{\pi}$ in the daughter nucleus within the experimental energy bin of 40~keV. The allowed decay of $^{137}$I populates positive parity states with $J$=5/2, 7/2, 9/2, while in the decay of $^{95}$Rb negative parity states with $J$=3/2, 5/2, 7/2 are populated. In Fig. \ref{Average_widths} the comparison between the experimental ratio and the calculation for each $J^{\pi}$ is shown. The uncertainty band around the experimental result (grey filled area) corresponds to the envelope of the ratios calculated for all possible $\beta$ intensity distributions compatible with the data, discussed in the previous Section. As seen in the figure, the large $P_{\gamma}$ in $^{137}$I is coming from $7/2^{+}$ and $9/2^{+}$ states  that need to emit $l \geq 4$ neutrons to populate the $0^{+}$ g.s. in $^{136}$Xe. This emission is hindered by the centrifugal barrier. Analogously, in the decay of $^{95}$Rb, the emission of $l=3$ neutrons from $5/2^{-}$ and $7/2^{-}$ states to populate the g.s. in $^{94}$Sr is very suppressed up to $E_{x}-S_{n} = 837$~keV, the energy of the first excited state in $^{94}$Sr, with $J^{\pi} = 2^{+}$. Above that energy neutron emission can proceed via $l=1$ and $\gamma$ emission can no longer compete. Experimentally we observe still a sizable competition, but we ascribe it to the uncertainty in the subtraction of the contaminants. Note that this region contributes little to $P_{\gamma}$.

\section{$\beta$ energy spectra and mean energies}\label{betaspec}

The $\beta$-intensity distributions obtained in this work were also used to calculate the $\beta$ energy spectra by means of subroutines from the $\log ft$ program of NNDC \cite{logftNNDC}. In the calculations we have assumed allowed shapes for all decay branches.
 
\begin{figure}[!hbt]
\begin{center}
\includegraphics[width=0.5 \textwidth]{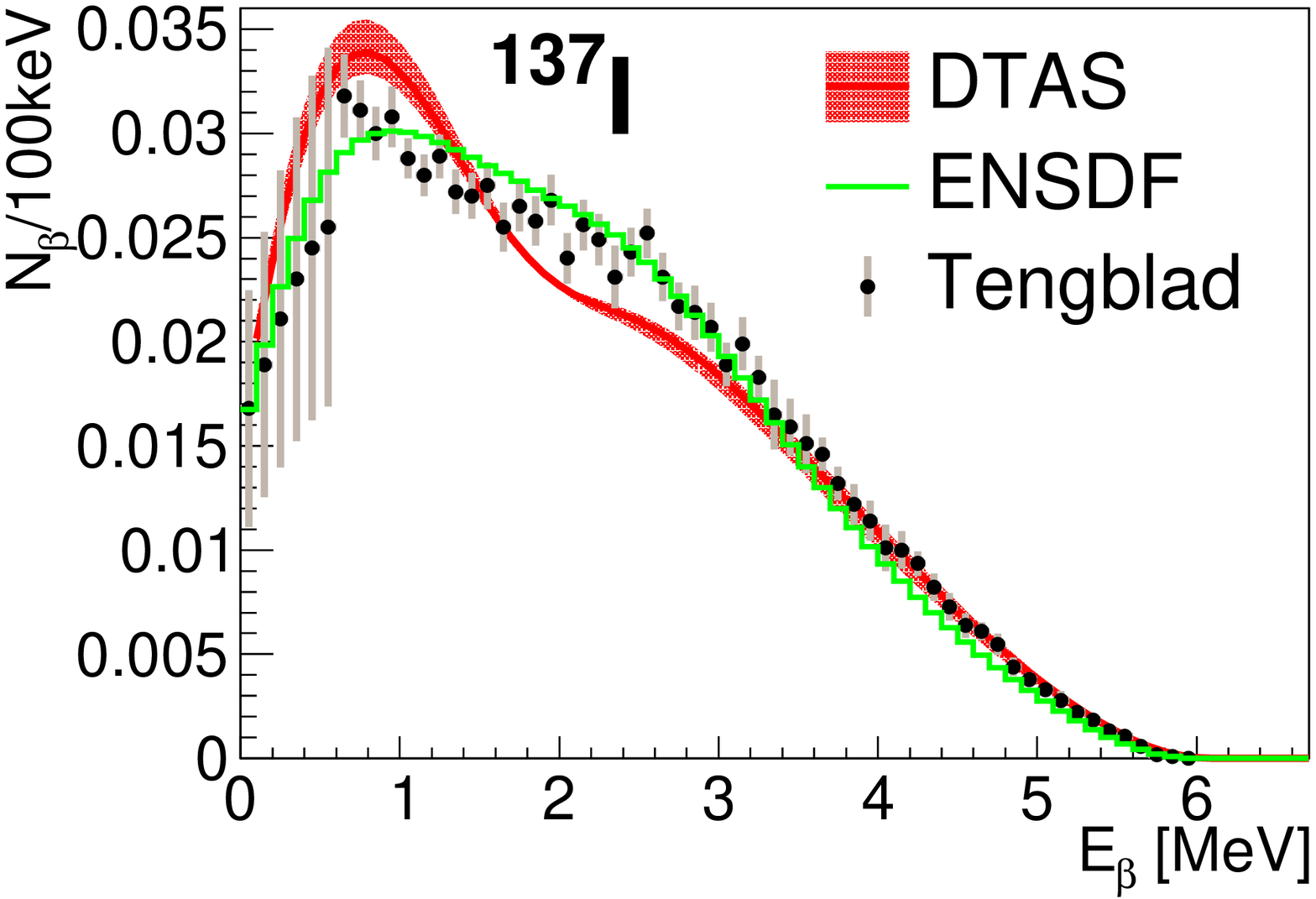} \\
\includegraphics[width=0.5 \textwidth]{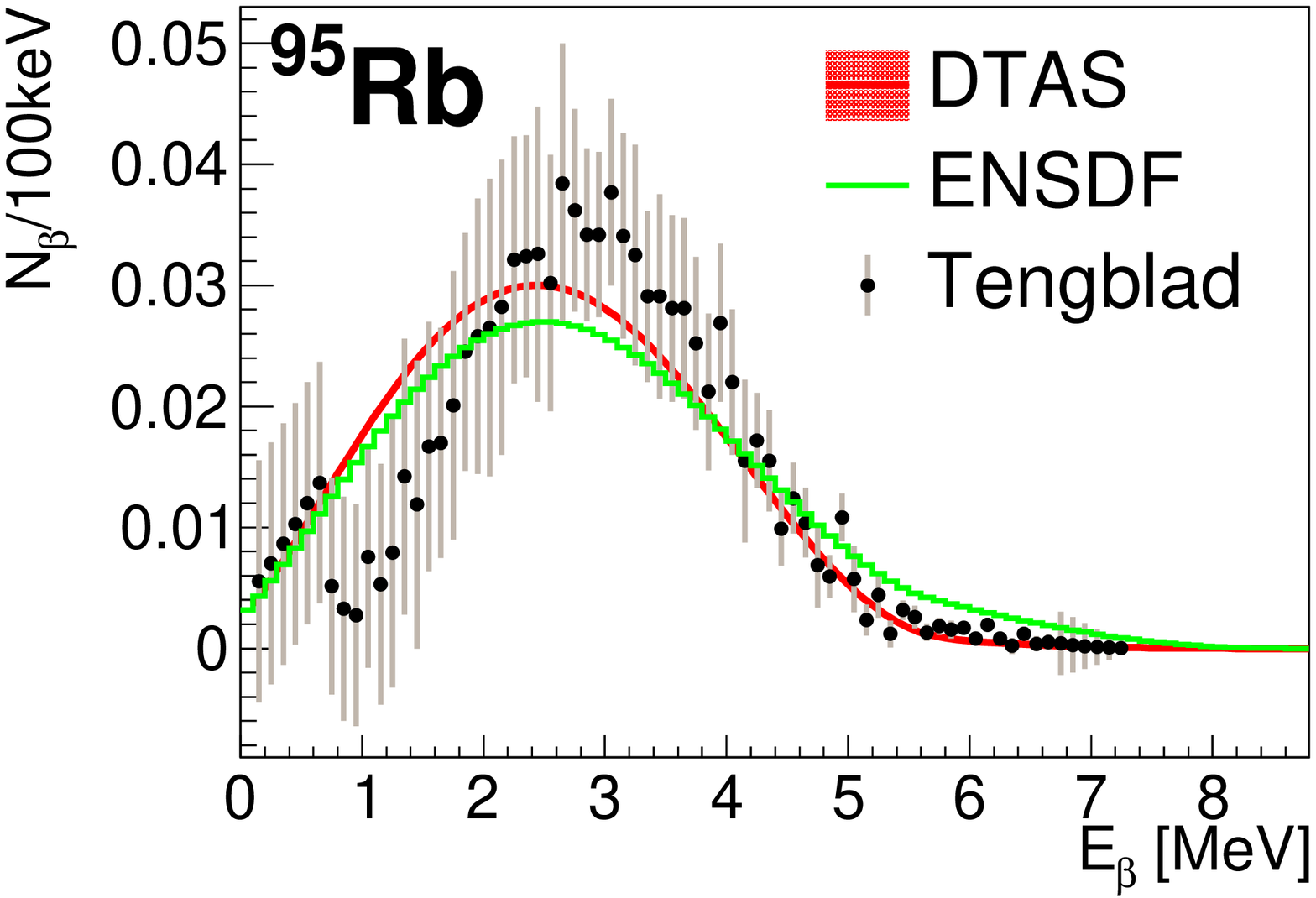}
\caption{$\beta$ spectra for the decays of $^{137}$I (top) and $^{95}$Rb (bottom) calculated with the $\beta$ intensity distributions obtained with DTAS (red line with error) compared to the spectra calculated with the data from ENSDF (green line), and with the experimental data of Tengblad \textit{et al.} (black points). 
}
\label{beta_spectra}
\end{center}
\end{figure}

In Fig.~\ref{beta_spectra} the deduced $\beta$ spectra for the decays of $^{137}$I and $^{95}$Rb are presented. The distributions calculated with the data from ENSDF, based on high-resolution $\gamma$-ray spectroscopy measurements~\cite{NDS_A137,NDS_A95}, are also included for comparison. They show a shift to higher energies with respect to the TAGS data, characteristic of the Pandemonium systematic error. The $\beta$ spectra include the contribution of the $\beta$-delayed neutron branch. For both nuclei the $\beta$ spectra were measured by Tengblad \textit{et al.} at OSIRIS-ISOLDE \cite{Olof_beta} using a $\beta$ spectrometer. This method is also Pandemonium free and a meaningful comparison between our calculated $\beta$ spectra and their experimental data can be made, in line with \cite{vTAS_PRC, Simon_PRC}. Differences in shape are observed for both isotopes, as shown in Fig. \ref{beta_spectra}. This is especially clear in the case of $^{95}$Rb. For both nuclei we observe agreement at high energies, while our spectra is lower at intermediate energies and higher at lower energies. The jump observed in Tengblad \textit{et al.} data below 0.8~MeV is due to the use of a different detector. 

The average $\gamma$ and $\beta$ energies obtained with the present TAGS results are listed in Table \ref{Mean_exp}, where uncertainties correspond to the evaluation of the mean energies for all the solutions compatible with a good analysis result mentioned in Section~\ref{TAGS}. For the mean $\beta$ energies we have summed the contribution from the beta-delayed neutron branch, taken from ENSDF~\cite{NDS_A137,NDS_A95}. In the mean $\gamma$-energy calculation of $^{95}$Rb the contribution of the $\gamma$ emission in the de-excitation of $^{94}$Sr has also been taken into account. The average $\gamma$ energies measured at OSIRIS by Rudstam \textit{et al.} \cite{Rudstam}, and the $\beta$ energies obtained by Tengblad \textit{et al.} \cite{Olof_beta} are also listed for comparison, taken from~\cite{Rudstam}. A similar comparison has been done in two recent publications \cite{vTAS_PRC, Simon_PRC} for a number of nuclei. A problem with the normalization of the average decay $\gamma$ energies was pinpointed in \cite{Simon_PRC}: all average  $\gamma$ energies in \cite{Rudstam} should be scaled up by 14$\%$. In our case, however, even though Rudstam mean $\gamma$ energies are larger than the TAGS values, both sets of numbers are compatible within the quoted errors. It is not the case for the mean $\beta$ energies, where discrepancies beyond the quoted errors are found, following the trend shown in~\cite{vTAS_PRC} and in line with the shape discrepancies of Fig.~\ref{beta_spectra}.

%\begin{table*}[!hbt]
%\centering
%  \begin{tabular}{|c|c|c|c|c|c|c|c|c|}
%  Nucleus & \multicolumn{2}{c|}{ENDF/B-VII.1} & \multicolumn{2}{c|}{JEFF-3.1.1} & \multicolumn{2}{c|}{Rudstam \cite{Rudstam}} & \multicolumn{2}{c|}{TAGS}  \\ 
%  & $\overline{E}_{\gamma}$ [keV] & $\overline{E}_{\beta}$ [keV]& $\overline{E}_{\gamma}$ [keV] & $\overline{E}_{\beta}$ [keV] & $\overline{E}_{\gamma}$ [keV] & $\overline{E}_{\beta}$ [keV]  & $\overline{E}_{\gamma}$ [keV] & $\overline{E}_{\beta}$ [keV]  \\  \hline
%  $^{137}$I &1135(20) & 1920(26) & 1212 & 1861 & 1230(150) & 2050(40) & 1220$_{-74}^{+121}$ & 1934$_{-56}^{+35}$   \\
%  $^{95}$Rb & 2162(42) & 2296(110) & 2629 & 2824 & 3370(220) & 2850(150) & 3110$_{-38}^{+17}$ & 2573$_{-8}^{+18}$ \\
%  \hline
%  \end{tabular}
%  \caption{Comparison of our average $\gamma$ and $\beta$ energies including the $\beta$-delayed branch with the average energies from \cite{Rudstam}. The values from the ENDF/B-VII.1~\cite{ENDF} and JEFF-3.1.1~\cite{JEFF} databases are also included for comparison.}
%  \label{Mean_exp}
%\end{table*}

\begin{table*}[!hbt]
\centering
  \begin{tabular}{|c|c|c|c|c|c|c|c|c|}
  Nucleus & \multicolumn{4}{c|}{$\overline{E}_{\gamma}$ [keV]} & \multicolumn{4}{c|}{$\overline{E}_{\beta}$ [keV]} \\
  & ENDF & JEFF & Rudstam & TAGS & ENDF & JEFF & Rudstam & TAGS \\ \hline
  $^{137}$I & 1135(20) & 1212 & 1230(150) & 1220$_{-74}^{+121}$ & 1920(26) & 1861 & 2050(40)  & 1934$_{-56}^{+35}$   \\
  $^{95}$Rb & 2162(42) & 2629 & 3370(220) & 3110$_{-38}^{+17}$ & 2296(110) & 2824 & 2850(150) & 2573$_{-8}^{+18}$ \\
  \hline
  \end{tabular}
  \caption{Comparison of our average $\gamma$ and $\beta$ energies including the $\beta$-delayed branch with the average energies from Rudstam \textit{et al.}~\cite{Rudstam}. The values from the ENDF/B-VII.1~\cite{ENDF} and JEFF-3.1.1~\cite{JEFF} databases are also included for comparison.}
  \label{Mean_exp}
\end{table*}

In Table \ref{Mean_exp} we also include for comparison the $\beta$ and $\gamma$ average energies from the ENDF/B-VII.1 \cite{ENDF} and JEFF-3.1 \cite{JEFF} databases. Although a clear Pandemonium effect was observed when comparing the present TAGS $\beta$-intensity distributions and the previous results from high-resolution measurements in Figs. \ref{137I_I} and \ref{95Rb_I} (especially evident for the comparison of the accumulated $\beta$ intensities), the average energies for $^{137}$I listed in Table \ref{Mean_exp} do not differ significantly. This is due to the redistribution of the $\beta$-intensity consequence of the larger g.s. feeding probability that we determined. This is not the case for $^{95}$Rb, where the average $\gamma$ energy of the databases is clearly underestimated, whereas the average $\beta$ energy is overestimated, as is normally found with the Pandemonium effect \cite{DecayHeat}.

\section{Reactor summation calculations}\label{reactor}

\begin{figure}[!hbt]
\begin{center}
\includegraphics[width=0.5 \textwidth]{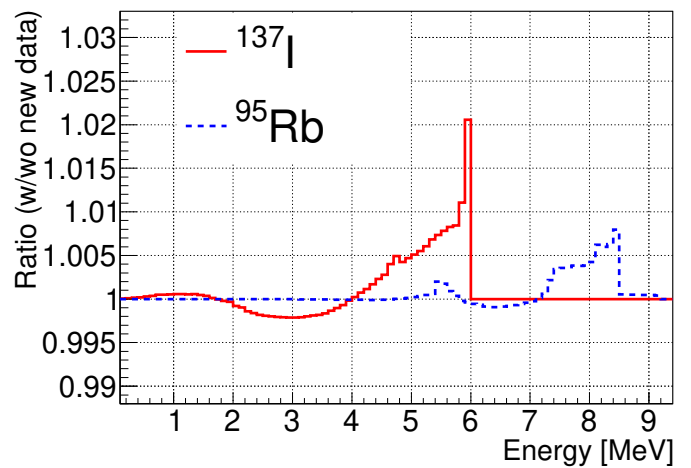}
\caption{Ratio of reactor antineutrino spectra as a function of energy for $^{239}$Pu when the results obtained in the present work replace the previous knowledge of the decays studied. The effect of $^{137}$I (solid red) and $^{95}$Rb (dotted blue) are presented.}
\label{antineutrino}
\end{center}
\end{figure}

The impact of the present results on reactor antineutrino summation calculations has been evaluated. For this, the summation method developed by the group of Nantes~\cite{neutrinos_PRL} has been employed assuming allowed shapes. The impact of the present results in the calculation for each of the four main fissile isotopes in a pressurized water reactor (PWR) - $^{235}$U, $^{239}$Pu, $^{241}$Pu, and $^{238}$U - has been evaluated. For $^{239}$Pu the ratio between the antineutrino spectrum calculated with the inclusion of the present TAGS $\beta$-intensity distributions, and the original calculation, where Rudstam data were taken, is shown in Fig.~\ref{antineutrino}. Similar figures are obtained for the other three fissile isotopes. The effect of the results for $^{137}$I is to increase the ratio by up to 1$\%$ in the region 4-6 MeV for uranium and plutonium isotopes. On the other hand, the new results for $^{95}$Rb increase the ratio up to 0.5$\%$ in the region of 7-8.5 MeV in the four fissile isotopes. 

A summation method was also employed for the calculation of the reactor DH. The impact of replacing the average $\gamma$ and $\beta$ energies available at ENDF/B-VII.1 by the present TAGS values has been studied. The effect in the electromagnetic component of $^{235}$U and $^{239}$Pu is an increase of $<1\%$ and $<0.5\%$, respectively, for times shorter than 1~s. The light particle component is reduced by less than 0.5$\%$ in both cases below 1~s. Such a modest impact can be understood on the one hand due to the similarity between the TAGS average energies and the values available at ENDF/B-VII.1 for $^{137}$I, and on the other hand because of the low cumulative fission yield of $^{95}$Rb.

\section{Conclusions}

In this work we reported the results of the TAGS measurements of two important $\beta$-delayed neutron emitters. The sensitivity of our technique made it possible to determine the $\beta$-intensity to states above $S_n$ followed by $\gamma$-rays. This $\beta$-intensity was found to be larger than in previous measurements affected by the Pandemonium effect. Moreover, it represents 56$\%$ and 25$\%$ of the $\beta$-intensity above $S_n$ in $^{137}$I and $^{95}$Rb, respectively. The competition between neutron emission and $\gamma$ de-excitation can be understood as an effect of nuclear structure, due to the large neutron angular momentum required to populate the granddaughter levels, because of their spin-parity values. 

The presence of the Pandemonium effect in previous high-resolution data was deduced when comparing the present TAGS $\beta$-intensity distributions and the values from ENSDF, as well as the average $\gamma$ and $\beta$ energies with the reference values from the databases. The $\beta$ spectra constructed with the $\beta$-intensity distributions of this work were compared with the measured spectra free from Pandemonium from Tengblad \textit{et al.}. Discrepancies in the shape of the spectra were found, in line with recent works~\cite{vTAS_PRC, Simon_PRC}.

A careful study of the systematic uncertainties was performed for each case in order to verify our results. We also considered the results obtained from the analyses of spectra constructed with a $\gamma$-neutron discrimination condition. Even though the set-up was not optimized for such a procedure, the reasonable quality of the results reinforces the interest of this methodology for future TAGS measurements of $\beta$-delayed neutron emitters. In addition, for $^{137}$I, the analysis of the background subtracted singles spectrum was found to be in good agreement with the analysis of the $\beta$-gated spectra. 

Two stringent cross-checks of the branching ratio matrices for $^{137}$I and $^{95}$Rb were carried out by means of MC simulations based on our results: reproduction of the individual crystal spectra and reproduction of the module-multiplicity gated spectra. Both were reasonably well reproduced with the results of our TAGS analysis, showing two important features of our analysis procedure: the quality of the MC simulations (including the reproduction of the $\beta$-n branch) and the validity of our branching ratio matrices. For low-lying levels, we have also reproduced the known absolute $\gamma$-intensities obtained with our $\beta$-intensity distributions after modifying the branching ratio matrix. Although it did not lead to the best reproduction of the spectra, which was interpreted as a consequence of the incomplete knowledge of these decays, it was included in the estimation of systematic uncertainties. 

Finally, the impact of these new results in reactor summation calculations has been evaluated. The effect of replacing previous database values with our new results was found to be less than 1$\%$ for both antineutrino spectrum calculations and DH calculations.

%%%%%%%%%%%%%%%%%%%%%%%%%%%%%%%%%%%%%%%%%%%%%%%%%%%%%%%%%%%%%%%%%%%%%%%%%%%%%%%%%%%%%%%%%%%%%%%%
\begin{acknowledgments}

This work has been supported by the Spanish Ministerio de Econom\'ia y Competitividad under Grants No. FPA2011-24553, No. AIC-A-2011-0696, No. FPA2014-52823-C2-1-P, No. FPA2015-65035-P, No. FPI/BES-2014-068222, No. FPA2017-83946-C2-1-P and the program Severo Ochoa (SEV-2014-0398), by the Spanish Ministerio de Educaci\'on under the FPU12/01527 Grant, by the European Commission under the FP7/EURATOM contract 605203 and the FP7/ENSAR contract 262010, and by the $Junta~para~la~Ampliaci\acute{o}n~de~Estudios$ Programme (CSIC JAE-Doc contract) co-financed by FSE. We acknowledge the support of the UK Science and Technology Facilities Council (STFC) Grant No. ST/P005314/1. This work
was also supported by the Academy of Finland under the Finnish Centre of Excellence Programme (Project No. 213503, Nuclear and Accelerator-Based Physics Research at JYFL). 

\end{acknowledgments}

\bibliography{BDN}% Produces the bibliography via BibTeX.

\begin{thebibliography}{68}%
\makeatletter
\providecommand \@ifxundefined [1]{%
 \@ifx{#1\undefined}
}%
\providecommand \@ifnum [1]{%
 \ifnum #1\expandafter \@firstoftwo
 \else \expandafter \@secondoftwo
 \fi
}%
\providecommand \@ifx [1]{%
 \ifx #1\expandafter \@firstoftwo
 \else \expandafter \@secondoftwo
 \fi
}%
\providecommand \natexlab [1]{#1}%
\providecommand \enquote  [1]{``#1''}%
\providecommand \bibnamefont  [1]{#1}%
\providecommand \bibfnamefont [1]{#1}%
\providecommand \citenamefont [1]{#1}%
\providecommand \href@noop [0]{\@secondoftwo}%
\providecommand \href [0]{\begingroup \@sanitize@url \@href}%
\providecommand \@href[1]{\@@startlink{#1}\@@href}%
\providecommand \@@href[1]{\endgroup#1\@@endlink}%
\providecommand \@sanitize@url [0]{\catcode `\\12\catcode `\$12\catcode
  `\&12\catcode `\#12\catcode `\^12\catcode `\_12\catcode `\%12\relax}%
\providecommand \@@startlink[1]{}%
\providecommand \@@endlink[0]{}%
\providecommand \url  [0]{\begingroup\@sanitize@url \@url }%
\providecommand \@url [1]{\endgroup\@href {#1}{\urlprefix }}%
\providecommand \urlprefix  [0]{URL }%
\providecommand \Eprint [0]{\href }%
\providecommand \doibase [0]{http://dx.doi.org/}%
\providecommand \selectlanguage [0]{\@gobble}%
\providecommand \bibinfo  [0]{\@secondoftwo}%
\providecommand \bibfield  [0]{\@secondoftwo}%
\providecommand \translation [1]{[#1]}%
\providecommand \BibitemOpen [0]{}%
\providecommand \bibitemStop [0]{}%
\providecommand \bibitemNoStop [0]{.\EOS\space}%
\providecommand \EOS [0]{\spacefactor3000\relax}%
\providecommand \BibitemShut  [1]{\csname bibitem#1\endcsname}%
\let\auto@bib@innerbib\@empty
%</preamble>
\bibitem [{\citenamefont {Roberts}\ \emph {et~al.}(1939)\citenamefont {Roberts}
  \emph {et~al.}}]{bneutron_discovery}%
  \BibitemOpen
  \bibfield  {author} {\bibinfo {author} {\bibfnamefont {R.}~\bibnamefont
  {Roberts}} \emph {et~al.},\ }\href@noop {} {\bibfield  {journal} {\bibinfo
  {journal} {Phys. Rev.}\ }\textbf {\bibinfo {volume} {55}},\ \bibinfo {pages}
  {510} (\bibinfo {year} {1939})}\BibitemShut {NoStop}%
\bibitem [{\citenamefont {{Margaret Burbidge}}\ \emph
  {et~al.}(1957)\citenamefont {{Margaret Burbidge}} \emph
  {et~al.}}]{r-process_ref}%
  \BibitemOpen
  \bibfield  {author} {\bibinfo {author} {\bibfnamefont {E.}~\bibnamefont
  {{Margaret Burbidge}}} \emph {et~al.},\ }\href@noop {} {\bibfield  {journal}
  {\bibinfo  {journal} {Rev. Mod. Phys.}\ }\textbf {\bibinfo {volume} {29}},\
  \bibinfo {pages} {547} (\bibinfo {year} {1957})}\BibitemShut {NoStop}%
\bibitem [{\citenamefont {Kasen}\ \emph {et~al.}(2017)\citenamefont {Kasen}
  \emph {et~al.}}]{NSM}%
  \BibitemOpen
  \bibfield  {author} {\bibinfo {author} {\bibfnamefont {D.}~\bibnamefont
  {Kasen}} \emph {et~al.},\ }\href@noop {} {\bibfield  {journal} {\bibinfo
  {journal} {Nature}\ }\textbf {\bibinfo {volume} {551}},\ \bibinfo {pages}
  {80} (\bibinfo {year} {2017})}\BibitemShut {NoStop}%
\bibitem [{\citenamefont {Mumpower}\ \emph {et~al.}(2016)\citenamefont
  {Mumpower} \emph {et~al.}}]{Mumpower16}%
  \BibitemOpen
  \bibfield  {author} {\bibinfo {author} {\bibfnamefont {M.}~\bibnamefont
  {Mumpower}} \emph {et~al.},\ }\href@noop {} {\bibfield  {journal} {\bibinfo
  {journal} {Prog. Part. Nucl. Phys.}\ }\textbf {\bibinfo {volume} {86}},\
  \bibinfo {pages} {86} (\bibinfo {year} {2016})}\BibitemShut {NoStop}%
\bibitem [{\citenamefont {Hauser}\ and\ \citenamefont
  {Feshbach}(1952)}]{Hauser-Feshbach}%
  \BibitemOpen
  \bibfield  {author} {\bibinfo {author} {\bibfnamefont {W.}~\bibnamefont
  {Hauser}}\ and\ \bibinfo {author} {\bibfnamefont {H.}~\bibnamefont
  {Feshbach}},\ }\href@noop {} {\bibfield  {journal} {\bibinfo  {journal}
  {Phys. Rev.}\ }\textbf {\bibinfo {volume} {87}},\ \bibinfo {pages} {366}
  (\bibinfo {year} {1952})}\BibitemShut {NoStop}%
\bibitem [{\citenamefont {Rauscher}\ and\ \citenamefont
  {Thielemann}(2000)}]{ReactionRates_StatisticalModel}%
  \BibitemOpen
  \bibfield  {author} {\bibinfo {author} {\bibfnamefont {T.}~\bibnamefont
  {Rauscher}}\ and\ \bibinfo {author} {\bibfnamefont {F.-K.}\ \bibnamefont
  {Thielemann}},\ }\href@noop {} {\bibfield  {journal} {\bibinfo  {journal}
  {At. Data Nucl. Data Tables}\ }\textbf {\bibinfo {volume} {75}},\ \bibinfo
  {pages} {1} (\bibinfo {year} {2000})}\BibitemShut {NoStop}%
\bibitem [{\citenamefont {Tain}\ \emph
  {et~al.}(2015{\natexlab{a}})\citenamefont {Tain} \emph {et~al.}}]{vTAS_PRL}%
  \BibitemOpen
  \bibfield  {author} {\bibinfo {author} {\bibfnamefont {J.~L.}\ \bibnamefont
  {Tain}} \emph {et~al.},\ }\href@noop {} {\bibfield  {journal} {\bibinfo
  {journal} {Phys. Rev. Lett.}\ }\textbf {\bibinfo {volume} {115}},\ \bibinfo
  {pages} {062502} (\bibinfo {year} {2015}{\natexlab{a}})}\BibitemShut
  {NoStop}%
\bibitem [{\citenamefont {Valencia}\ \emph {et~al.}(2017)\citenamefont
  {Valencia} \emph {et~al.}}]{vTAS_PRC}%
  \BibitemOpen
  \bibfield  {author} {\bibinfo {author} {\bibfnamefont {E.}~\bibnamefont
  {Valencia}} \emph {et~al.},\ }\href@noop {} {\bibfield  {journal} {\bibinfo
  {journal} {Phys. Rev. C}\ }\textbf {\bibinfo {volume} {95}},\ \bibinfo
  {pages} {024320} (\bibinfo {year} {2017})}\BibitemShut {NoStop}%
\bibitem [{\citenamefont {Tain}\ \emph
  {et~al.}(2017{\natexlab{a}})\citenamefont {Tain} \emph
  {et~al.}}]{Tain_proceedings_bneu1}%
  \BibitemOpen
  \bibfield  {author} {\bibinfo {author} {\bibfnamefont {J.~L.}\ \bibnamefont
  {Tain}} \emph {et~al.},\ }\href@noop {} {\bibfield  {journal} {\bibinfo
  {journal} {EPJ Web of Conferences}\ } (\bibinfo {year}
  {2017}{\natexlab{a}})}\BibitemShut {NoStop}%
\bibitem [{\citenamefont {Tain}\ \emph
  {et~al.}(2017{\natexlab{b}})\citenamefont {Tain} \emph
  {et~al.}}]{Tain_proceedings_bneu2}%
  \BibitemOpen
  \bibfield  {author} {\bibinfo {author} {\bibfnamefont {J.~L.}\ \bibnamefont
  {Tain}} \emph {et~al.},\ }\href@noop {} {\bibfield  {journal} {\bibinfo
  {journal} {JPS Conf. Proc.}\ }\textbf {\bibinfo {volume} {14}},\ \bibinfo
  {pages} {010607} (\bibinfo {year} {2017}{\natexlab{b}})}\BibitemShut
  {NoStop}%
\bibitem [{\citenamefont {Hardy}\ \emph {et~al.}(1977)\citenamefont {Hardy}
  \emph {et~al.}}]{Pandemonium}%
  \BibitemOpen
  \bibfield  {author} {\bibinfo {author} {\bibfnamefont {J.}~\bibnamefont
  {Hardy}} \emph {et~al.},\ }\href@noop {} {\bibfield  {journal} {\bibinfo
  {journal} {Phys. Lett. B}\ }\textbf {\bibinfo {volume} {71}},\ \bibinfo
  {pages} {307} (\bibinfo {year} {1977})}\BibitemShut {NoStop}%
\bibitem [{\citenamefont {Alkhazov}\ \emph {et~al.}(1989)\citenamefont
  {Alkhazov} \emph {et~al.}}]{Alkhazov_BDN}%
  \BibitemOpen
  \bibfield  {author} {\bibinfo {author} {\bibfnamefont {G.~D.}\ \bibnamefont
  {Alkhazov}} \emph {et~al.},\ }\href@noop {} {\bibfield  {journal} {\bibinfo
  {journal} {Leningrad Nuclear Physics Institute Report No. 1497}\ } (\bibinfo
  {year} {1989})}\BibitemShut {NoStop}%
\bibitem [{\citenamefont {Spyrou}\ \emph {et~al.}(2016)\citenamefont {Spyrou}
  \emph {et~al.}}]{Sun_betan}%
  \BibitemOpen
  \bibfield  {author} {\bibinfo {author} {\bibfnamefont {A.}~\bibnamefont
  {Spyrou}} \emph {et~al.},\ }\href@noop {} {\bibfield  {journal} {\bibinfo
  {journal} {Phys. Rev. Lett.}\ }\textbf {\bibinfo {volume} {117}},\ \bibinfo
  {pages} {142701} (\bibinfo {year} {2016})}\BibitemShut {NoStop}%
\bibitem [{\citenamefont {Moeller}\ \emph {et~al.}(2003)\citenamefont {Moeller}
  \emph {et~al.}}]{Moller-r-process}%
  \BibitemOpen
  \bibfield  {author} {\bibinfo {author} {\bibfnamefont {P.}~\bibnamefont
  {Moeller}} \emph {et~al.},\ }\href@noop {} {\bibfield  {journal} {\bibinfo
  {journal} {Phys. Rev. C}\ }\textbf {\bibinfo {volume} {67}},\ \bibinfo
  {pages} {055802} (\bibinfo {year} {2003})}\BibitemShut {NoStop}%
\bibitem [{\citenamefont {Borzov}\ \emph {et~al.}(2003)\citenamefont {Borzov}
  \emph {et~al.}}]{Borzov-r-process}%
  \BibitemOpen
  \bibfield  {author} {\bibinfo {author} {\bibfnamefont {I.~N.}\ \bibnamefont
  {Borzov}} \emph {et~al.},\ }\href@noop {} {\bibfield  {journal} {\bibinfo
  {journal} {Phys. Rev. C}\ }\textbf {\bibinfo {volume} {67}},\ \bibinfo
  {pages} {025802} (\bibinfo {year} {2003})}\BibitemShut {NoStop}%
\bibitem [{\citenamefont {Okumura}\ \emph {et~al.}()\citenamefont {Okumura}
  \emph {et~al.}}]{Fukushima}%
  \BibitemOpen
  \bibfield  {author} {\bibinfo {author} {\bibfnamefont {K.}~\bibnamefont
  {Okumura}} \emph {et~al.},\ }\href@noop {} {\bibinfo  {journal} {Proceedings
  of the 2012 Symposium on Nuclear Data, Kyoto, JAEA-Conf. 2013-002,
  INDC(JPN)-198 (Japan Atomic Energy Agency, Tokaimura, 2013), p. 15.}\
  }\BibitemShut {NoStop}%
\bibitem [{\citenamefont {Abe}\ \emph {et~al.}(2012)\citenamefont {Abe} \emph
  {et~al.}}]{DoubleChooz}%
  \BibitemOpen
\bibfield  {journal} {  }\bibfield  {author} {\bibinfo {author} {\bibfnamefont
  {Y.}~\bibnamefont {Abe}} \emph {et~al.},\ }\href@noop {} {\bibfield
  {journal} {\bibinfo  {journal} {Phys. Rev. Lett.}\ }\textbf {\bibinfo
  {volume} {\textbf{108}}},\ \bibinfo {pages} {131801} (\bibinfo {year}
  {2012})}\BibitemShut {NoStop}%
\bibitem [{\citenamefont {An}\ \emph {et~al.}(2012)\citenamefont {An} \emph
  {et~al.}}]{DayaBay}%
  \BibitemOpen
  \bibfield  {author} {\bibinfo {author} {\bibfnamefont {F.~P.}\ \bibnamefont
  {An}} \emph {et~al.},\ }\href@noop {} {\bibfield  {journal} {\bibinfo
  {journal} {Phys. Rev. Lett.}\ }\textbf {\bibinfo {volume} {\textbf{108}}},\
  \bibinfo {pages} {171803} (\bibinfo {year} {2012})}\BibitemShut {NoStop}%
\bibitem [{\citenamefont {Ahn}\ \emph {et~al.}(2012)\citenamefont {Ahn} \emph
  {et~al.}}]{Reno}%
  \BibitemOpen
  \bibfield  {author} {\bibinfo {author} {\bibfnamefont {J.~K.}\ \bibnamefont
  {Ahn}} \emph {et~al.},\ }\href@noop {} {\bibfield  {journal} {\bibinfo
  {journal} {Phys. Rev. Lett.}\ }\textbf {\bibinfo {volume} {\textbf{108}}},\
  \bibinfo {pages} {191802} (\bibinfo {year} {2012})}\BibitemShut {NoStop}%
\bibitem [{\citenamefont {Kim}(2017)}]{non_proliferation_2017}%
  \BibitemOpen
  \bibfield  {author} {\bibinfo {author} {\bibfnamefont {Y.}~\bibnamefont
  {Kim}},\ }\href@noop {} {\bibfield  {journal} {\bibinfo  {journal} {J. Phys.:
  Conf. Ser.}\ }\textbf {\bibinfo {volume} {888}},\ \bibinfo {pages} {012010}
  (\bibinfo {year} {2017})}\BibitemShut {NoStop}%
\bibitem [{\citenamefont {Hahn}\ \emph {et~al.}(1988)\citenamefont {Hahn} \emph
  {et~al.}}]{ILL_3}%
  \BibitemOpen
  \bibfield  {author} {\bibinfo {author} {\bibfnamefont {A.~A.}\ \bibnamefont
  {Hahn}} \emph {et~al.},\ }\href@noop {} {\bibfield  {journal} {\bibinfo
  {journal} {Phys. Lett. B}\ }\textbf {\bibinfo {volume} {218}},\ \bibinfo
  {pages} {365} (\bibinfo {year} {1988})}\BibitemShut {NoStop}%
\bibitem [{\citenamefont {Haag}\ \emph {et~al.}(2014)\citenamefont {Haag} \emph
  {et~al.}}]{238U_beta}%
  \BibitemOpen
  \bibfield  {author} {\bibinfo {author} {\bibfnamefont {N.}~\bibnamefont
  {Haag}} \emph {et~al.},\ }\href@noop {} {\bibfield  {journal} {\bibinfo
  {journal} {Phys. Rev. Lett.}\ }\textbf {\bibinfo {volume} {112}},\ \bibinfo
  {pages} {122501} (\bibinfo {year} {2014})}\BibitemShut {NoStop}%
\bibitem [{\citenamefont {Mention}\ \emph {et~al.}(2011)\citenamefont {Mention}
  \emph {et~al.}}]{Anomaly}%
  \BibitemOpen
  \bibfield  {author} {\bibinfo {author} {\bibfnamefont {G.}~\bibnamefont
  {Mention}} \emph {et~al.},\ }\href@noop {} {\bibfield  {journal} {\bibinfo
  {journal} {Phys. Rev. D}\ }\textbf {\bibinfo {volume} {\textbf{83}}},\
  \bibinfo {pages} {073006} (\bibinfo {year} {2011})}\BibitemShut {NoStop}%
\bibitem [{\citenamefont {Choi}\ \emph {et~al.}(2016)\citenamefont {Choi} \emph
  {et~al.}}]{RENO_shoulder}%
  \BibitemOpen
  \bibfield  {author} {\bibinfo {author} {\bibfnamefont {J.~H.}\ \bibnamefont
  {Choi}} \emph {et~al.},\ }\href@noop {} {\bibfield  {journal} {\bibinfo
  {journal} {Phys. Rev. Lett.}\ }\textbf {\bibinfo {volume} {116}},\ \bibinfo
  {pages} {211801} (\bibinfo {year} {2016})}\BibitemShut {NoStop}%
\bibitem [{\citenamefont {An}\ \emph {et~al.}(2016)\citenamefont {An} \emph
  {et~al.}}]{DayaBay_shoulder}%
  \BibitemOpen
  \bibfield  {author} {\bibinfo {author} {\bibfnamefont {F.~P.}\ \bibnamefont
  {An}} \emph {et~al.},\ }\href@noop {} {\bibfield  {journal} {\bibinfo
  {journal} {Phys. Rev. Lett.}\ }\textbf {\bibinfo {volume} {116}},\ \bibinfo
  {pages} {061801} (\bibinfo {year} {2016})}\BibitemShut {NoStop}%
\bibitem [{\citenamefont {Abe}\ \emph {et~al.}(2016)\citenamefont {Abe} \emph
  {et~al.}}]{DoubleChooz_shoulder}%
  \BibitemOpen
  \bibfield  {author} {\bibinfo {author} {\bibfnamefont {Y.}~\bibnamefont
  {Abe}} \emph {et~al.},\ }\href@noop {} {\bibfield  {journal} {\bibinfo
  {journal} {J. High Energy Phys.}\ }\textbf {\bibinfo {volume} {01}},\
  \bibinfo {pages} {163} (\bibinfo {year} {2016})}\BibitemShut {NoStop}%
\bibitem [{\citenamefont {Fallot}\ \emph {et~al.}(2012)\citenamefont {Fallot}
  \emph {et~al.}}]{neutrinos_PRL}%
  \BibitemOpen
  \bibfield  {author} {\bibinfo {author} {\bibfnamefont {M.}~\bibnamefont
  {Fallot}} \emph {et~al.},\ }\href@noop {} {\bibfield  {journal} {\bibinfo
  {journal} {Phys. Rev. Lett.}\ }\textbf {\bibinfo {volume} {109}},\ \bibinfo
  {pages} {202504} (\bibinfo {year} {2012})}\BibitemShut {NoStop}%
\bibitem [{\citenamefont {Algora}\ \emph {et~al.}(2010)\citenamefont {Algora}
  \emph {et~al.}}]{DecayHeat}%
  \BibitemOpen
  \bibfield  {author} {\bibinfo {author} {\bibfnamefont {A.}~\bibnamefont
  {Algora}} \emph {et~al.},\ }\href@noop {} {\bibfield  {journal} {\bibinfo
  {journal} {Phys. Rev. Lett.}\ }\textbf {\bibinfo {volume} {105}},\ \bibinfo
  {pages} {202501} (\bibinfo {year} {2010})}\BibitemShut {NoStop}%
\bibitem [{IAE(2015)}]{IAEA2015}%
  \BibitemOpen
  \enquote {\bibinfo {title} {{IAEA} report {INDC(NDS)}-0676},}\ \ (\bibinfo
  {year} {2015})\BibitemShut {NoStop}%
\bibitem [{\citenamefont {Moore}\ \emph {et~al.}(2013)\citenamefont {Moore}
  \emph {et~al.}}]{Moore_IGISOLIV}%
  \BibitemOpen
  \bibfield  {author} {\bibinfo {author} {\bibfnamefont {I.~D.}\ \bibnamefont
  {Moore}} \emph {et~al.},\ }\href@noop {} {\bibfield  {journal} {\bibinfo
  {journal} {Nucl. Instrum. and Methods B}\ }\textbf {\bibinfo {volume}
  {317}},\ \bibinfo {pages} {208} (\bibinfo {year} {2013})}\BibitemShut
  {NoStop}%
\bibitem [{\citenamefont {Tain}\ \emph
  {et~al.}(2015{\natexlab{b}})\citenamefont {Tain} \emph
  {et~al.}}]{DTAS_design}%
  \BibitemOpen
  \bibfield  {author} {\bibinfo {author} {\bibfnamefont {J.~L.}\ \bibnamefont
  {Tain}} \emph {et~al.},\ }\href@noop {} {\bibfield  {journal} {\bibinfo
  {journal} {Nucl. Instrum. and Methods A}\ }\textbf {\bibinfo {volume}
  {803}},\ \bibinfo {pages} {36} (\bibinfo {year}
  {2015}{\natexlab{b}})}\BibitemShut {NoStop}%
\bibitem [{\citenamefont {Eronen}\ \emph {et~al.}(2012)\citenamefont {Eronen}
  \emph {et~al.}}]{JYFLTRAP}%
  \BibitemOpen
  \bibfield  {author} {\bibinfo {author} {\bibfnamefont {T.}~\bibnamefont
  {Eronen}} \emph {et~al.},\ }\href@noop {} {\bibfield  {journal} {\bibinfo
  {journal} {Eur. Phys. J. A}\ }\textbf {\bibinfo {volume} {48}},\ \bibinfo
  {pages} {46} (\bibinfo {year} {2012})}\BibitemShut {NoStop}%
\bibitem [{\citenamefont {Guadilla}\ \emph {et~al.}(2018)\citenamefont
  {Guadilla} \emph {et~al.}}]{NIMA_DTAS}%
  \BibitemOpen
  \bibfield  {author} {\bibinfo {author} {\bibfnamefont {V.}~\bibnamefont
  {Guadilla}} \emph {et~al.},\ }\href@noop {} {\bibfield  {journal} {\bibinfo
  {journal} {Nucl. Instrum. and Methods A}\ }\textbf {\bibinfo {volume}
  {910}},\ \bibinfo {pages} {79} (\bibinfo {year} {2018})}\BibitemShut
  {NoStop}%
\bibitem [{\citenamefont {Zakari-Issoufou}\ \emph {et~al.}(2015)\citenamefont
  {Zakari-Issoufou} \emph {et~al.}}]{Zak_PRL}%
  \BibitemOpen
  \bibfield  {author} {\bibinfo {author} {\bibfnamefont {A.-A.}\ \bibnamefont
  {Zakari-Issoufou}} \emph {et~al.},\ }\href@noop {} {\bibfield  {journal}
  {\bibinfo  {journal} {Phys. Rev. Lett.}\ }\textbf {\bibinfo {volume} {115}},\
  \bibinfo {pages} {102503} (\bibinfo {year} {2015})}\BibitemShut {NoStop}%
\bibitem [{\citenamefont {Rice}\ \emph {et~al.}(2017)\citenamefont {Rice} \emph
  {et~al.}}]{Simon_PRC}%
  \BibitemOpen
  \bibfield  {author} {\bibinfo {author} {\bibfnamefont {S.}~\bibnamefont
  {Rice}} \emph {et~al.},\ }\href@noop {} {\bibfield  {journal} {\bibinfo
  {journal} {Phys. Rev. C}\ }\textbf {\bibinfo {volume} {96}},\ \bibinfo
  {pages} {014320} (\bibinfo {year} {2017})}\BibitemShut {NoStop}%
\bibitem [{\citenamefont {Guadilla}\ \emph {et~al.}(2017)\citenamefont
  {Guadilla} \emph {et~al.}}]{100Tc}%
  \BibitemOpen
  \bibfield  {author} {\bibinfo {author} {\bibfnamefont {V.}~\bibnamefont
  {Guadilla}} \emph {et~al.},\ }\href@noop {} {\bibfield  {journal} {\bibinfo
  {journal} {Phys. Rev. C}\ }\textbf {\bibinfo {volume} {96}},\ \bibinfo
  {pages} {014319} (\bibinfo {year} {2017})}\BibitemShut {NoStop}%
\bibitem [{\citenamefont {Cano-Ott}\ \emph
  {et~al.}(1999{\natexlab{a}})\citenamefont {Cano-Ott} \emph
  {et~al.}}]{TAS_pileup}%
  \BibitemOpen
  \bibfield  {author} {\bibinfo {author} {\bibfnamefont {D.}~\bibnamefont
  {Cano-Ott}} \emph {et~al.},\ }\href@noop {} {\bibfield  {journal} {\bibinfo
  {journal} {Nucl. Instrum. and Methods A}\ }\textbf {\bibinfo {volume}
  {430}},\ \bibinfo {pages} {488} (\bibinfo {year}
  {1999}{\natexlab{a}})}\BibitemShut {NoStop}%
\bibitem [{\citenamefont {Agostinelli}\ \emph {et~al.}(2003)\citenamefont
  {Agostinelli} \emph {et~al.}}]{GEANT4}%
  \BibitemOpen
  \bibfield  {author} {\bibinfo {author} {\bibfnamefont {S.}~\bibnamefont
  {Agostinelli}} \emph {et~al.},\ }\href@noop {} {\bibfield  {journal}
  {\bibinfo  {journal} {Nucl. Instrum. and Methods A}\ }\textbf {\bibinfo
  {volume} {506}},\ \bibinfo {pages} {250} (\bibinfo {year}
  {2003})}\BibitemShut {NoStop}%
\bibitem [{\citenamefont {Brady}(1989)}]{Brady_thesis}%
  \BibitemOpen
  \bibfield  {author} {\bibinfo {author} {\bibfnamefont {M.}~\bibnamefont
  {Brady}},\ }\href@noop {} {\emph {\bibinfo {title} {Evaluation and
  Application of Delayed Neutron Precursor Data}}}\ (\bibinfo  {publisher} {Los
  Alamos National Laboratory},\ \bibinfo {year} {1989})\BibitemShut {NoStop}%
\bibitem [{\citenamefont {Kawano}\ \emph {et~al.}(2008)\citenamefont {Kawano}
  \emph {et~al.}}]{Kawano_neutron_spec}%
  \BibitemOpen
  \bibfield  {author} {\bibinfo {author} {\bibfnamefont {T.}~\bibnamefont
  {Kawano}} \emph {et~al.},\ }\href@noop {} {\bibfield  {journal} {\bibinfo
  {journal} {Phys. Rev. C}\ }\textbf {\bibinfo {volume} {78}},\ \bibinfo
  {pages} {054601} (\bibinfo {year} {2008})}\BibitemShut {NoStop}%
\bibitem [{\citenamefont {Kratz}\ \emph {et~al.}(1982)\citenamefont {Kratz}
  \emph {et~al.}}]{Kratz_bn94Sr}%
  \BibitemOpen
  \bibfield  {author} {\bibinfo {author} {\bibfnamefont {K.-L.}\ \bibnamefont
  {Kratz}} \emph {et~al.},\ }\href@noop {} {\bibfield  {journal} {\bibinfo
  {journal} {Z. Phys. A}\ }\textbf {\bibinfo {volume} {306}},\ \bibinfo {pages}
  {239} (\bibinfo {year} {1982})}\BibitemShut {NoStop}%
\bibitem [{\citenamefont {Hoff}(1981)}]{Hoff_bn94Sr}%
  \BibitemOpen
  \bibfield  {author} {\bibinfo {author} {\bibfnamefont {P.}~\bibnamefont
  {Hoff}},\ }\href@noop {} {\bibfield  {journal} {\bibinfo  {journal} {Nuclear
  Phys. A}\ }\textbf {\bibinfo {volume} {359}},\ \bibinfo {pages} {9} (\bibinfo
  {year} {1981})}\BibitemShut {NoStop}%
\bibitem [{\citenamefont {Gabelmann}(1987)}]{Thesis_Gabelmann}%
  \BibitemOpen
  \bibfield  {author} {\bibinfo {author} {\bibfnamefont {H.}~\bibnamefont
  {Gabelmann}},\ }\href@noop {} {\emph {\bibinfo {title} {Untersuchung des
  Beta-Verzogerten Neutronenzerfalls Neutronenreicher Brom-, Rubidium- und
  Caesiumisotope}}}\ (\bibinfo  {publisher} {Ph.D. thesis, Johannes
  Gutenberg-Universitat, Mainz},\ \bibinfo {year} {1987})\BibitemShut {NoStop}%
\bibitem [{\citenamefont {Cano-Ott}\ \emph
  {et~al.}(1999{\natexlab{b}})\citenamefont {Cano-Ott} \emph
  {et~al.}}]{TAS_MC}%
  \BibitemOpen
  \bibfield  {author} {\bibinfo {author} {\bibfnamefont {D.}~\bibnamefont
  {Cano-Ott}} \emph {et~al.},\ }\href@noop {} {\bibfield  {journal} {\bibinfo
  {journal} {Nucl. Instrum. and Methods A}\ }\textbf {\bibinfo {volume}
  {430}},\ \bibinfo {pages} {333} (\bibinfo {year}
  {1999}{\natexlab{b}})}\BibitemShut {NoStop}%
\bibitem [{\citenamefont {Tain}\ and\ \citenamefont
  {Cano-Ott}(2007{\natexlab{a}})}]{TAS_algorithms}%
  \BibitemOpen
  \bibfield  {author} {\bibinfo {author} {\bibfnamefont {J.~L.}\ \bibnamefont
  {Tain}}\ and\ \bibinfo {author} {\bibfnamefont {D.}~\bibnamefont
  {Cano-Ott}},\ }\href@noop {} {\bibfield  {journal} {\bibinfo  {journal}
  {Nucl. Instrum. and Methods A}\ }\textbf {\bibinfo {volume} {571}},\ \bibinfo
  {pages} {728} (\bibinfo {year} {2007}{\natexlab{a}})}\BibitemShut {NoStop}%
\bibitem [{\citenamefont {Tain}\ and\ \citenamefont
  {Cano-Ott}(2007{\natexlab{b}})}]{TAS_decaygen}%
  \BibitemOpen
  \bibfield  {author} {\bibinfo {author} {\bibfnamefont {J.~L.}\ \bibnamefont
  {Tain}}\ and\ \bibinfo {author} {\bibfnamefont {D.}~\bibnamefont
  {Cano-Ott}},\ }\href@noop {} {\bibfield  {journal} {\bibinfo  {journal}
  {Nucl. Instrum. and Methods A}\ }\textbf {\bibinfo {volume} {571}},\ \bibinfo
  {pages} {719} (\bibinfo {year} {2007}{\natexlab{b}})}\BibitemShut {NoStop}%
\bibitem [{\citenamefont {Capote}\ \emph {et~al.}(2009)\citenamefont {Capote}
  \emph {et~al.}}]{RIPL-3}%
  \BibitemOpen
  \bibfield  {author} {\bibinfo {author} {\bibfnamefont {R.}~\bibnamefont
  {Capote}} \emph {et~al.},\ }\href@noop {} {\bibfield  {journal} {\bibinfo
  {journal} {Nucl. Data Sheets}\ }\textbf {\bibinfo {volume} {110}},\ \bibinfo
  {pages} {3107} (\bibinfo {year} {2009})}\BibitemShut {NoStop}%
\bibitem [{\citenamefont {Kopecky}\ and\ \citenamefont {Uhl}(1990)}]{PSF}%
  \BibitemOpen
  \bibfield  {author} {\bibinfo {author} {\bibfnamefont {J.}~\bibnamefont
  {Kopecky}}\ and\ \bibinfo {author} {\bibfnamefont {M.}~\bibnamefont {Uhl}},\
  }\href@noop {} {\bibfield  {journal} {\bibinfo  {journal} {Phys. Rev. C}\
  }\textbf {\bibinfo {volume} {41}},\ \bibinfo {pages} {1941} (\bibinfo {year}
  {1990})}\BibitemShut {NoStop}%
\bibitem [{\citenamefont {Raman}\ \emph {et~al.}(2001)\citenamefont {Raman},
  \citenamefont {{Nestor Jr}},\ and\ \citenamefont
  {Tikkanen}}]{DeformationPar_exp}%
  \BibitemOpen
  \bibfield  {author} {\bibinfo {author} {\bibfnamefont {S.}~\bibnamefont
  {Raman}}, \bibinfo {author} {\bibfnamefont {C.~W.}\ \bibnamefont {{Nestor
  Jr}}}, \ and\ \bibinfo {author} {\bibfnamefont {P.}~\bibnamefont
  {Tikkanen}},\ }\href@noop {} {\bibfield  {journal} {\bibinfo  {journal}
  {Atomic Data and Nuclear Data Tables}\ }\textbf {\bibinfo {volume} {78}},\
  \bibinfo {pages} {1} (\bibinfo {year} {2001})}\BibitemShut {NoStop}%
\bibitem [{\citenamefont {Fogelberg}\ \emph {et~al.}(1985)\citenamefont
  {Fogelberg} \emph {et~al.}}]{137I_Fog_fermi_gas}%
  \BibitemOpen
  \bibfield  {author} {\bibinfo {author} {\bibfnamefont {B.}~\bibnamefont
  {Fogelberg}} \emph {et~al.},\ }\href@noop {} {\bibfield  {journal} {\bibinfo
  {journal} {Phys. Rev. C}\ }\textbf {\bibinfo {volume} {31}},\ \bibinfo
  {pages} {2041} (\bibinfo {year} {1985})}\BibitemShut {NoStop}%
\bibitem [{\citenamefont {Goriely}\ \emph {et~al.}(2008)\citenamefont
  {Goriely}, \citenamefont {Hilaire},\ and\ \citenamefont {Koning}}]{Gorieli1}%
  \BibitemOpen
  \bibfield  {author} {\bibinfo {author} {\bibfnamefont {S.}~\bibnamefont
  {Goriely}}, \bibinfo {author} {\bibfnamefont {S.}~\bibnamefont {Hilaire}}, \
  and\ \bibinfo {author} {\bibfnamefont {A.~J.}\ \bibnamefont {Koning}},\
  }\href@noop {} {\bibfield  {journal} {\bibinfo  {journal} {Phys. Rev. C}\
  }\textbf {\bibinfo {volume} {78}},\ \bibinfo {pages} {064307} (\bibinfo
  {year} {2008})}\BibitemShut {NoStop}%
\bibitem [{\citenamefont {Goriely}\ \emph {et~al.}(2007)\citenamefont
  {Goriely}, \citenamefont {Samyn},\ and\ \citenamefont {Pearson}}]{Gorieli2}%
  \BibitemOpen
  \bibfield  {author} {\bibinfo {author} {\bibfnamefont {S.}~\bibnamefont
  {Goriely}}, \bibinfo {author} {\bibfnamefont {M.}~\bibnamefont {Samyn}}, \
  and\ \bibinfo {author} {\bibfnamefont {J.}~\bibnamefont {Pearson}},\
  }\href@noop {} {\bibfield  {journal} {\bibinfo  {journal} {Phys. Rev. C}\
  }\textbf {\bibinfo {volume} {75}},\ \bibinfo {pages} {064312} (\bibinfo
  {year} {2007})}\BibitemShut {NoStop}%
\bibitem [{\citenamefont {Fogelberg}\ and\ \citenamefont
  {Tovedal}(1980)}]{137I_Fog}%
  \BibitemOpen
  \bibfield  {author} {\bibinfo {author} {\bibfnamefont {B.}~\bibnamefont
  {Fogelberg}}\ and\ \bibinfo {author} {\bibfnamefont {H.}~\bibnamefont
  {Tovedal}},\ }\href@noop {} {\bibfield  {journal} {\bibinfo  {journal}
  {Nuclear Physics A}\ }\textbf {\bibinfo {volume} {345}},\ \bibinfo {pages}
  {13} (\bibinfo {year} {1980})}\BibitemShut {NoStop}%
\bibitem [{\citenamefont {Browne}\ and\ \citenamefont {Tuli}(2007)}]{NDS_A137}%
  \BibitemOpen
  \bibfield  {author} {\bibinfo {author} {\bibfnamefont {E.}~\bibnamefont
  {Browne}}\ and\ \bibinfo {author} {\bibfnamefont {J.~K.}\ \bibnamefont
  {Tuli}},\ }\href@noop {} {\bibfield  {journal} {\bibinfo  {journal} {Nuclear
  Data Sheets}\ }\textbf {\bibinfo {volume} {108}},\ \bibinfo {pages} {2173}
  (\bibinfo {year} {2007})}\BibitemShut {NoStop}%
\bibitem [{\citenamefont {Rasco}\ \emph {et~al.}(2017)\citenamefont {Rasco}
  \emph {et~al.}}]{MTAS_137I}%
  \BibitemOpen
  \bibfield  {author} {\bibinfo {author} {\bibfnamefont {B.~C.}\ \bibnamefont
  {Rasco}} \emph {et~al.},\ }\href@noop {} {\bibfield  {journal} {\bibinfo
  {journal} {Phys. Rev. C}\ }\textbf {\bibinfo {volume} {95}},\ \bibinfo
  {pages} {054328} (\bibinfo {year} {2017})}\BibitemShut {NoStop}%
\bibitem [{Sup()}]{Suplement}%
  \BibitemOpen
  \href@noop {} {\enquote {\bibinfo {title} {See {S}upplemental {M}aterial at
  url for $\beta$ intensity distributions},}\ }\BibitemShut {NoStop}%
\bibitem [{\citenamefont {Koning}\ \emph {et~al.}(2005)\citenamefont {Koning},
  \citenamefont {Hilaire}, \citenamefont {Duijvestijn}, \citenamefont {Haight},
  \citenamefont {Chadwick},\ and\ \citenamefont {(Eds.)}}]{TALYS}%
  \BibitemOpen
  \bibfield  {author} {\bibinfo {author} {\bibfnamefont {A.}~\bibnamefont
  {Koning}}, \bibinfo {author} {\bibfnamefont {S.}~\bibnamefont {Hilaire}},
  \bibinfo {author} {\bibfnamefont {M.}~\bibnamefont {Duijvestijn}}, \bibinfo
  {author} {\bibfnamefont {R.}~\bibnamefont {Haight}}, \bibinfo {author}
  {\bibfnamefont {M.}~\bibnamefont {Chadwick}}, \ and\ \bibinfo {author}
  {\bibfnamefont {T.~K.}\ \bibnamefont {(Eds.)}},\ }\href@noop {} {\bibfield
  {journal} {\bibinfo  {journal} {Proceedings of the International Conference
  on Nuclear Data for Science and Technology, ND2004, AIP}\ }\textbf {\bibinfo
  {volume} {769}},\ \bibinfo {pages} {1154} (\bibinfo {year}
  {2005})}\BibitemShut {NoStop}%
\bibitem [{\citenamefont {Ohm}\ \emph {et~al.}(1980)\citenamefont {Ohm} \emph
  {et~al.}}]{137I_Ohm}%
  \BibitemOpen
  \bibfield  {author} {\bibinfo {author} {\bibfnamefont {H.}~\bibnamefont
  {Ohm}} \emph {et~al.},\ }\href@noop {} {\bibfield  {journal} {\bibinfo
  {journal} {Z. Phys. A}\ }\textbf {\bibinfo {volume} {296}},\ \bibinfo {pages}
  {23} (\bibinfo {year} {1980})}\BibitemShut {NoStop}%
\bibitem [{\citenamefont {Basu}\ \emph {et~al.}(2010)\citenamefont {Basu} \emph
  {et~al.}}]{NDS_A95}%
  \BibitemOpen
  \bibfield  {author} {\bibinfo {author} {\bibfnamefont {S.~K.}\ \bibnamefont
  {Basu}} \emph {et~al.},\ }\href@noop {} {\bibfield  {journal} {\bibinfo
  {journal} {Nuclear Data Sheets}\ }\textbf {\bibinfo {volume} {111}},\
  \bibinfo {pages} {2555} (\bibinfo {year} {2010})}\BibitemShut {NoStop}%
\bibitem [{\citenamefont {Agramunt}\ \emph {et~al.}(2016)\citenamefont
  {Agramunt} \emph {et~al.}}]{NIM_BELEN}%
  \BibitemOpen
  \bibfield  {author} {\bibinfo {author} {\bibfnamefont {J.}~\bibnamefont
  {Agramunt}} \emph {et~al.},\ }\href@noop {} {\bibfield  {journal} {\bibinfo
  {journal} {Nucl. Instrum. and Methods A}\ }\textbf {\bibinfo {volume}
  {807}},\ \bibinfo {pages} {69} (\bibinfo {year} {2016})}\BibitemShut
  {NoStop}%
\bibitem [{\citenamefont {Abriola}\ \emph {et~al.}(2011)\citenamefont {Abriola}
  \emph {et~al.}}]{Pn_IAEA_2011}%
  \BibitemOpen
  \bibfield  {author} {\bibinfo {author} {\bibfnamefont {D.}~\bibnamefont
  {Abriola}} \emph {et~al.},\ }\href@noop {} {\bibfield  {journal} {\bibinfo
  {journal} {IAEA Consultants Meeting on Beta Delayed Neutron Evaluation,
  Summary Report, INDC(NDS)-0599}\ } (\bibinfo {year} {2011})}\BibitemShut
  {NoStop}%
\bibitem [{\citenamefont {Greenwood}\ \emph {et~al.}(1992)\citenamefont
  {Greenwood}, \citenamefont {Struttmann},\ and\ \citenamefont
  {Watts}}]{Greenwood_GS}%
  \BibitemOpen
  \bibfield  {author} {\bibinfo {author} {\bibfnamefont {R.~C.}\ \bibnamefont
  {Greenwood}}, \bibinfo {author} {\bibfnamefont {D.~A.}\ \bibnamefont
  {Struttmann}}, \ and\ \bibinfo {author} {\bibfnamefont {K.~D.}\ \bibnamefont
  {Watts}},\ }\href@noop {} {\bibfield  {journal} {\bibinfo  {journal} {Nucl.
  Instrum. and Methods A}\ }\textbf {\bibinfo {volume} {317}},\ \bibinfo
  {pages} {175} (\bibinfo {year} {1992})}\BibitemShut {NoStop}%
\bibitem [{\citenamefont {Kratz}\ \emph {et~al.}(1983)\citenamefont {Kratz}
  \emph {et~al.}}]{Kratz_95Rb_decay}%
  \BibitemOpen
  \bibfield  {author} {\bibinfo {author} {\bibfnamefont {K.-L.}\ \bibnamefont
  {Kratz}} \emph {et~al.},\ }\href@noop {} {\bibfield  {journal} {\bibinfo
  {journal} {Z. Phys. A}\ }\textbf {\bibinfo {volume} {312}},\ \bibinfo {pages}
  {43} (\bibinfo {year} {1983})}\BibitemShut {NoStop}%
\bibitem [{log()}]{logftNNDC}%
  \BibitemOpen
  \href@noop {} {\enquote {\bibinfo {title} {{ENSDF A}nalysis {P}rograms,
  {LOGFT}, {N}ational {N}uclear {D}ata {C}enter, {B}rookhaven {N}ational
  {L}aboratory},}\ }\bibinfo {howpublished}
  {\url{http://www.nndc.bnl.gov/nndcscr/ensdf_pgm/analysis/logft/unx/}}\BibitemShut
  {NoStop}%
\bibitem [{\citenamefont {Tengblad}\ \emph {et~al.}(1989)\citenamefont
  {Tengblad} \emph {et~al.}}]{Olof_beta}%
  \BibitemOpen
  \bibfield  {author} {\bibinfo {author} {\bibfnamefont {O.}~\bibnamefont
  {Tengblad}} \emph {et~al.},\ }\href@noop {} {\bibfield  {journal} {\bibinfo
  {journal} {Nucl. Phys. A}\ }\textbf {\bibinfo {volume} {503}},\ \bibinfo
  {pages} {136} (\bibinfo {year} {1989})}\BibitemShut {NoStop}%
\bibitem [{\citenamefont {Rudstam}\ \emph {et~al.}(1990)\citenamefont {Rudstam}
  \emph {et~al.}}]{Rudstam}%
  \BibitemOpen
  \bibfield  {author} {\bibinfo {author} {\bibfnamefont {G.}~\bibnamefont
  {Rudstam}} \emph {et~al.},\ }\href@noop {} {\bibfield  {journal} {\bibinfo
  {journal} {Atomic Data and Nuclear Data Tables}\ }\textbf {\bibinfo {volume}
  {45}},\ \bibinfo {pages} {239} (\bibinfo {year} {1990})}\BibitemShut
  {NoStop}%
\bibitem [{\citenamefont {Chadwick}\ \emph {et~al.}(2011)\citenamefont
  {Chadwick} \emph {et~al.}}]{ENDF}%
  \BibitemOpen
  \bibfield  {author} {\bibinfo {author} {\bibfnamefont {M.~B.}\ \bibnamefont
  {Chadwick}} \emph {et~al.},\ }\href@noop {} {\bibfield  {journal} {\bibinfo
  {journal} {Nucl. Data Sheets}\ }\textbf {\bibinfo {volume} {112}},\ \bibinfo
  {pages} {2887} (\bibinfo {year} {2011})}\BibitemShut {NoStop}%
\bibitem [{JEF()}]{JEFF}%
  \BibitemOpen
  \href@noop {} {\enquote {\bibinfo {title} {The {JEFF} nuclear data
  library},}\ }\bibinfo {howpublished}
  {\url{http://www.oecd‐nea.org/dbdata/jeff/}}\BibitemShut {NoStop}%
\end{thebibliography}%

\end{document}